\documentclass[10pt, aps, pra, preprintnumbers,nofootinbib]{revtex4-2}
\usepackage{graphicx}
\usepackage{dcolumn}
\usepackage{amsmath}
\usepackage{float}
\usepackage{makecell}
\usepackage{bm}
\usepackage{comment}
\usepackage{subcaption}
\usepackage{slashed}
\usepackage{gensymb}
\usepackage{tikz}
\usepackage[normalem]{ulem}
\usepackage[compat=1.1.0]{tikz-feynman}
\usepackage{ragged2e}
\usepackage[utf8]{inputenc}
\usepackage{tabularx}
\usepackage{array}
\usepackage{graphics}
\graphicspath{{}}
\usepackage{psfrag}
\usepackage{epsfig}
\usepackage{amssymb}
\usepackage{setspace}
\usepackage{rotating}
\usepackage{colortbl}
\usepackage{slashed}
\usepackage{enumitem}
\usepackage[T1]{fontenc}
\makeatletter
\usepackage{textcomp}
\usepackage[normalem]{ulem}
\usepackage[colorlinks=true,allcolors=blue]{hyperref}
\usepackage[capitalize]{cleveref}
\usepackage{accents}
\newcommand{\mybracketbar}[1]{\accentset{(-)}{#1}}

\definecolor{lightblue}{rgb}{0.68, 0.85, 0.9}
\definecolor{deepblue}{rgb}{0.0, 0.0, 0.55}

\definecolor{nicegreen}{rgb}{0., 0.75, 0.46}

\begin{document}

\begin{flushright}
MI-HET-861\\
CETUP2025-001
\end{flushright}

\title{Monophotons from Scalar Portal Dark Matter at Neutrino Experiments}

\author{Bhaskar Dutta}
\email{dutta@tamu.edu}
\affiliation{
Mitchell Institute for Fundamental Physics and Astronomy,
Department of Physics and Astronomy, Texas A\&M University, College Station, TX 77843, USA
}
\author{Debopam Goswami}
\email{debopam22@tamu.edu}
\affiliation{
Mitchell Institute for Fundamental Physics and Astronomy,
Department of Physics and Astronomy, Texas A\&M University, College Station, TX 77843, USA
}
\author{Aparajitha Karthikeyan}
\email{aparajitha\_96@tamu.edu}
\affiliation{
Mitchell Institute for Fundamental Physics and Astronomy,
Department of Physics and Astronomy, Texas A\&M University, College Station, TX 77843, USA
}

\date{\today}

\begin{abstract}
In this work, we investigate monophoton signatures arising from dark matter via a $2\to 3$ scattering process $\chi + N \to \chi + N + \gamma$ that is mediated by a virtual scalar and a Standard Model photon. Since the final-state photon carries a large fraction of the initial dark matter's energy, this process offers a compelling handle for probing scalar portal dark matter scenarios. Their distinctive energy, angular, and timing distributions allow for effective separation of signal from neutrino-induced backgrounds. We analyze several models featuring different couplings to the scalar mediator, with the scalar photon coupling serving as the common detection channel. To distinguish between the models, we further examined their distinct spatial distributions. We considered the flux of dark matter produced both at the target and absorber of neutrino facilities such as the BNB, NuMI, and LBNF, and investigated the sensitivities at the ongoing SBND, ICARUS-NuMI, and future DUNE ND detectors. We further investigated the differences in the DM fluxes arising from various production mechanisms, as well as the distinctions between the target and absorber contributions. Our results demonstrate that the sensitivities at the considered experiments, especially DUNE ND, can place significantly improved constraints on viable parameter space in various scenarios. 

\end{abstract}

\maketitle

\section{\label{sec:level1}Introduction}

Over the past century, a remarkable amount of progress has been made in the arena of dark matter (DM)~\cite{Freese:2017idy}, from establishing its existence through various astrophysical evidence~\cite{babcock1939rotation, Zwicky:1933gu, Zwicky:1937zza, Markevitch:2001ri, Markevitch:2004qk, Neto:2007vq, Maccio:2008pcd, Springel:2008cc}, to pioneering various experimental pursuits~\cite{XENON:2019zpr, DarkSide:2018ppu, SuperCDMS:2020aus, PandaX:2023xgl, SENSEI:2024yyt} to find its particle nature. Weakly Interacting Massive Particles (WIMPs) have been long sought after as potential DM candidates, as the cold and massive thermal relic abundance is elegantly set by annihilation via weak interaction mediators. Despite its elegance, the lack of a positive WIMP signature~\cite{XENON:2018voc, XENON:2019gfn, XENON:2020fgj, DarkSide:2018ppu, SuperCDMS:2020aus} motivates us to pursue other candidates. 

One such attractive alternative is sub-GeV DM, which requires new portals or mediators lighter than electroweak mediators~\cite{Batell:2009di, deNiverville:2011it, deNiverville:2012ij, Kahn:2014sra, deNiverville:2015mwa,  Berlin:2018bsc,Dutta:2019fxn} to acquire the required thermal relic cross section~\cite{Lee:1977ua,Wolfram:1978gp,Steigman:1979kw}. Many mechanisms have been proposed and employed in pursuit of sub-GeV DM and their interactions, such as missing energy~\cite{Banerjee:2019pds, Andreev:2024lps, NA64:2024klw},  electron and nuclear recoils~\cite{LSND:2001akn,MiniBooNE:2017nqe,COHERENT:2021pvd, CCM:2021leg,Rott:2018rlw,SENSEI:2020dpa,Lattaud:2022jnq,PandaX:2024muv,XENON:2024wpa,SuperCDMS:2020aus,LZ:2023lvz,SuperCDMS:2024yiv}, nuclear deexcitation lines~\cite{Dutta:2023fij, Dutta:2024kuj}, dark trident~\cite{deGouvea:2018cfv, Adrian:2022nkt}, dark matter internal pair production (DIPP)~\cite{Dutta:2024nhg}, etc. The relevance of these detection mechanisms depends primarily on the Lorentz structure of the mediators and the beyond the standard model (BSM) scenario in which they appear. In this context, spin-1 mediators have received great attention due to well-motivated anomaly-free gauge extensions~\cite{Foot:1990mn, He:1990pn, He:1991qd, Heeck:2011wj}, as well as the efficiency in both production and detection at beam dump and direct detection experiments.

Scalar mediators have also been naturally implicated in spontaneous symmetry-breaking models~\cite{Dutta:2018qei, Dutta:2019fxn}, through extensions to the Higgs sector~\cite{Patt:2006fw}, 
modification of mass terms in neutrino oscillations through scalar non-standard interactions (sNSIs)~\cite{Dutta:2022fdt, Dutta:2024hqq, Chauhan:2023faf, Dev:2021axj}, implications on the magnetic moment of muons~\cite{Muong-2:2021ojo, Muong-2:2021vma, Dutta:2023fnl, Cesarotti:2023udo, Blinov:2024gcw, Dutta:2020scq}, etc. Several key features of scalar-mediated DM make them more distinctive than commonly sought-after vector-mediated scenarios. 
Firstly, the annihilation cross section for fermionic dark matter via scalar mediators is $p$-wave~\cite{Kumar:2013iva}. This therefore renders scalar portal DM models to be (CMB)-safe, i.e., preventing them from injecting energy into the photon plasma during recombination~\cite{Planck:2015fie, Slatyer:2009yq}. Therefore, scalar portal models do not require DM to be a pseudo-Dirac or complex scalar, such as in vector portal models. Secondly, scalar mediators have the additional freedom to couple to two photons, which is otherwise forbidden for vector mediators. Therefore, phenomenology-wise, scalar portal DM can give rise to interesting signatures through the $\text{DM}-\text{DM}-\gamma-\gamma$ effective vertex, both at astrophysical environments and at the detection front of terrestrial experiments.

In this paper, we will focus on the implications of the DM-photon interaction at the detection frontier. To this end, we propose a new detection mechanism for dark matter via scalar mediators, which utilizes the mediator's coupling to two photons to produce a single-photon final state. When dark matter interacts with a material, it can source off-shell scalars which can split into two photons, where one is exchanged with the nucleus and the other is produced on-shell. This can be viewed from the perspective of dark matter inducing an inverse Primakoff scattering by converting its scalar mediator to a photon. Based on recent studies~\cite{Dutta:2024nhg, Dutta:2025fgz, Dev:2025czz}, we find that light energetic DM prefers to share its energy with the third particle stemming from the support of a stationary nucleus rather than the nucleus itself. Kinematically, we find that this process leads to high-energy monophoton signals, without the need for minimum thresholds to make this process observable. This is particularly advantageous in comparison to recoil-based experiments, where the electron/nuclear recoil energies are lower. 
\par 

In order to demonstrate this detection mechanism, we will choose five benchmark Liquid Argon Time Projection Chamber (LArTPC) detectors: Coherent CAPTAIN-Mills (CCM200)~\cite{CCM:2021jmk, CCM:2021leg}, Short-Baseline Near Detector (SBND)~\cite{MicroBooNE:2015bmn}, ICARUS-NuMI~\cite{ICARUS:2004wqc}, MicroBooNE~\cite{MicroBooNE:2016pwy}, and the near detector at Deep Underground Neutrino Experiment (DUNE ND)~\cite{DUNE:2021tad, DUNE:2016hlj}. In addition, we also investigate this process in MiniBooNE~\cite{MiniBooNE:2008hfu}, which uses a mineral oil-based detection medium. These experiments are chosen to cover a wide range of beam energies, starting from 800~MeV CCM200 up to 120~GeV DUNE ND and ICARUS-NuMI. We will also consider a range of scalar portal models that not only include a coupling to photons, but also other Standard Model (SM) fermions. This will enable a study of the rich phenomenology of various production mechanisms at neutrino facilities as well as unique single-photon detection signatures. We will also show how the DM-induced single photon signatures can be distinguished from typical neutrino-induced backgrounds based on the timing distributions at all the above-mentioned detectors.

The paper is organized as follows: We will begin with a brief overview of an umbrella of scalar portal dark matter models and the related phenomenology at neutrino experiments in Section.~\ref{sec:models}. We then discuss these models in the context of our benchmark experiments in Sec.~\ref{sec:experiments}. Then, we will present our results in Sec.~\ref{sec:results} and discuss the salient features of the single photon signature. The results will include an analysis of the flux of produced dark matter, its spatial distribution, timing, and energy, and angular spectra of photons at all the experiments, followed by Section.~\ref{sec:sensitivity}, discussing the sensitivity plots. We will summarize our results in Sec.~\ref{sec:conclusion}.

\section{\label{sec:models}Models and Lagrangian}

We follow an effective field theory approach to study the phenomenology of fermionic dark matter with a scalar mediator, where the latter is a mediator between the SM and DM sectors. The Lagrangian of interest in our analysis is:

\begin{equation}\label{1.1}
\mathcal{L} \supset ig_D\bar{\chi}\chi\phi + \frac{1}{2}ig_{\phi \gamma \gamma}\phi F^{\mu \nu}F_{\mu \nu} + iy_f \bar{f}f\phi
\end{equation}

where $y_{f}$ and $g_D$ are the generalized Yukawa couplings of the scalar with SM fermions and dark matter, respectively. 
The dimension-5 operator contains a coupling $g_{\phi\gamma\gamma}$, which carries a dimension of inverse mass ($\text{GeV}^{-1}$). 
The $g_{\phi\gamma\gamma}$ coupling can appear from loops containing top, bottom, or heavy vector-like quarks~\cite{Kim:1979if}, which are much heavier than the energy scale corresponding to our choice of sub-GeV to GeV scale neutrino energies in these experiments. 
Therefore, from a phenomenological standpoint, we will treat all three couplings $g_{\phi\gamma\gamma},~y_f, g_D$ as independent parameters.
We can assume that the neutrinos, like all the other charged leptons, appearing in the Yukawa interactions are Dirac particles such that $y_f \bar{f}f = y_f(\bar{f}_Rf_L+\bar{f}_Lf_R)$. One could also consider the neutrinos to have Majorana interactions with the scalar, e.g., $\phi\nu^c_L\nu_L$ where $\phi$ is a triplet scalar. In the second case, $\nu_R$ is not needed. Detailed UV complete models for light $\phi$ containing interactions with neutrinos and other fermions based on the extensions of the SM Higgs sector have been constructed in Refs.~\cite{Dutta:2022fdt,Dutta:2020scq}. 
\par

We focus on minimal scenarios and categorize our study into two classes. The first class sets all Yukawa couplings to zero, such that the photon coupling entirely dictates the phenomenology. The second class builds upon the first by introducing one SM fermion coupling at a time, while fixing the photon coupling $g_{\phi\gamma\gamma}$ to its maximal allowed value. Within this class, we consider four benchmark scenarios: neutrinophilic, electro/muon-philic, and quarkphilic. Although the dark matter coupling $g_D$ enters implicitly through quantities such as cross sections and decay widths, we assume that $g_{D} \sim \mathcal{O}(1)$, which is much larger than the scalar couplings to fermions ($y_f$) and photons ($g_{\phi\gamma\gamma}$). This effectively renders the scalar invisible. As a result, the observable phenomenology relevant to dark matter detection becomes largely insensitive to the precise value of $g_D$.

\subsection{Production of Dark Matter}\label{sec:production}

Since dark matter does not directly couple to the SM particles, but via the scalar portal, we first investigate the production mechanisms for the scalar. Our benchmark experiments operate in proton-on-target facilities where high-energy protons impinge on a target, followed by a large decay volume. Therefore, there is a rich flux of photons, electrons, neutral, and charged mesons. These serve as potential sources of scalars, and consequently dark matter, at these facilities. We will now discuss the production mechanisms by segregating them based on the scalar's coupling to the SM sector, that is, to photons, neutrinos, charged leptons, and quarks. We will limit our discussion on production mechanisms that are dominant for our sub-GeV to GeV scale benchmark neutrino experiments.

\begin{figure}[!htbp]
    \centering
    \subfloat[]
    {
    \begin{tikzpicture}
    \tikzset{every node/.style={font=\large}}  
        \begin{feynman}
            \vertex (m) at (0,1);
            \vertex (a) at (-2,1) {$\gamma$};
            \vertex (b) at (2, 1) {$\textcolor{violet}{\phi}$};
            \vertex (c) at (-2, -1.3) {$N$};
            \vertex (d) at (2, -1.3) {$N$};
            \vertex (n) at (0, -1.3);
            \node at (0.3, 0.0) {$\gamma$};
            \diagram*{
            (a) -- [photon, very thick] (m) -- [violet, scalar, dotted, ultra thick] (b),
            (m) -- [boson, very thick](n),
            (c) -- [double,double distance=0.5ex,thick,with arrow=0.5,arrow size=0.2em](n) -- [double,double distance=0.5ex,thick,with arrow=0.5,arrow size=0.2em](d)
            };
        \end{feynman}
    \end{tikzpicture}
    \label{fig:Primakoff}
    }
    \hspace{1cm}
    \subfloat[]
    {
    \begin{tikzpicture}
    \tikzset{every node/.style={font=\large}}  
        \begin{feynman}
            \vertex (a) at (-2,0) {$\pi^{+}/K^+$ };
            \vertex (b) at (0,0);
            \vertex (c) at (0.6, -0.6);
            \vertex (d) at (1.5,1.5) {$\ell^{+}$};
            \vertex (e) at (1.5,-1.5) {$\nu_{\ell}$};
            \vertex (f) at (1.7,0.4) {$\textcolor{violet}{\phi}$};
            \diagram*{
            (a) -- [scalar, very thick](b), (d) -- [fermion, very thick](b),
            (b) -- [fermion, very thick](e),
            (c) -- [violet, scalar, ultra thick] (f),
            };
        \end{feynman}
    \end{tikzpicture}
    \label{fig:3bodyneutrino}
    }
    \hspace{1cm}
    \subfloat[]
    {
    \begin{tikzpicture}
    \tikzset{every node/.style={font=\large}}  
        \begin{feynman}
            \vertex (a) at (-2,0) {$\pi^{+}/K^+$ };
            \vertex (b) at (0,0);
            \vertex (c) at (0.6, 0.6);
            \vertex (d) at (1.5,1.5) {$\ell^{+}$};
            \vertex (e) at (1.5,-1.5) {$\nu_{\ell}$};
            \vertex (f) at (1.7,-0.4) {$\textcolor{violet}{\phi}$};
            \diagram*{
            (a) -- [scalar, very thick](b), (d) -- [fermion, very thick](b),
            (b) -- [fermion, very thick](e),
            (c) -- [violet, scalar, ultra thick] (f),
            };
        \end{feynman}
    \end{tikzpicture}
    \label{fig:3bodylepton}
    }
    \vspace{0.2cm}
    \subfloat[]
    {
    \begin{tikzpicture}
    \tikzset{every node/.style={font=\large}}  
        \begin{feynman}
            \vertex (a) at (-0.5,0);
            \vertex (b) at (-1.5, -1.5) {$u$};
            \vertex (c) at (-1.5, 1.5) {$\bar{s}$};
            \vertex (d) at (1.5, 0.0);
            \vertex (e) at (2.5, 1.5) {$\bar{d}$};
            \vertex (f) at (2.5, -1.5) {$u$};
            \vertex (g) at (-1.0, -1.0);
            \vertex (h) at (-0.3, -1.5) {$\textcolor{violet}{\phi}$};
            \node at (0.5, 0.4) {$W^{+}$};
            \node at (-1.7, 0.0) {$K^+$};
            \node at (2.7, 0.0) {$\pi^+$};
            \diagram*{
            (b) -- [fermion, very thick](a);
            (a) -- [fermion, very thick](c);
            (a) -- [photon, very thick](d);
            (e) -- [fermion, very thick](d);
            (d) -- [fermion, very thick](f);
            (g) -- [violet, scalar, ultra thick](h);
            };
        \end{feynman}
    \end{tikzpicture}
    \label{fig:Kptopipupcoupling}
    }
    \hspace{0.05cm}
    \subfloat[]
    {
    \begin{tikzpicture}
    \tikzset{every node/.style={font=\large}}  
        \begin{feynman}
                \vertex[blob, minimum size=1cm] (a) at (0,-1) {};
                \vertex (c) at (-2.0, -2.0) {$p$};
                \vertex (e) at (-2.0, 0.0) {$p$};
                \vertex (s) at (-1, -0.5);
                \vertex (b) at (2, -1.0);
                \vertex (d) at (2.0, -2.0);
                \vertex (g) at (2.0, 0.0);
                \vertex (h) at (-0.2, 0.6) {$\textcolor{violet}{\phi}$};
                \diagram*{
            (c) -- [fermion, very thick](a), 
            (e) -- [fermion, very thick](a),
            (a) -- [very thick](b),
            (a) -- [very thick](d),
            (a) -- [very thick](g),
            (s) -- [violet, scalar, ultra thick] (h)
            };
    \end{feynman}
    \draw [decorate, decoration={brace, amplitude=10pt}, thick]
    (2.1,0.1) -- (2.1,-2.1) node[midway,xshift=15pt] {\large $f$};
    \end{tikzpicture}
    \label{fig:protonbremscalar}
    }
    \hspace{0.05cm}
    \subfloat[]
    {
    \begin{tikzpicture}
    \tikzset{every node/.style={font=\large}}  
        \begin{feynman}
            \vertex (a) at (-2,0.5){$\bar{s}$};
            \vertex (b) at (-0.8,0.5);
            \vertex (c) at (0.8,0.5);
            \vertex (d) at (2,0.5){$\bar{d}$};
            \vertex (e) at (-2,-1.5){$u$};
            \vertex (f) at (2,-1.5) {$u$};
            \vertex (g) at (0.5, 1.5) {$\textcolor{violet}{\phi}$};
            \vertex (h) at (0, 0.5) ;
            \node at (0, 0.25) {$\bar{t}$};
            \node at (-2.2, -0.4) {$K^+$};
            \node at (2.2, -0.4) {$\pi^+$};
            \diagram*{
            (a) -- [fermion, very thick](b) -- [fermion, very thick, label = {[yshift =1cm]$\bar{t}$}](c) -- [fermion,very thick](d),
            (b) -- [boson, half right, looseness = 1.5](c),
            (e) --[fermion, very thick](f),
            (h) -- [violet, scalar, ultra thick] (g)
            };
        \end{feynman}
    \end{tikzpicture}
    \label{fig:Kptopiptopcoupling}
    }
    
    \captionsetup{justification=Justified, singlelinecheck=off}
    \caption{Feynman diagrams illustrate the production of (a) photophilic scalars via photon, (b) neutrinophilic and (c) electro/muon-philic scalars, both via three-body decays of charged mesons, (d) up-philic scalars via kaon two-body decays, and (e) via proton bremsstrahlung, and (f) top-philic scalars via one-loop kaon two-body decays.}
    \label{fig:scattering}
\end{figure}
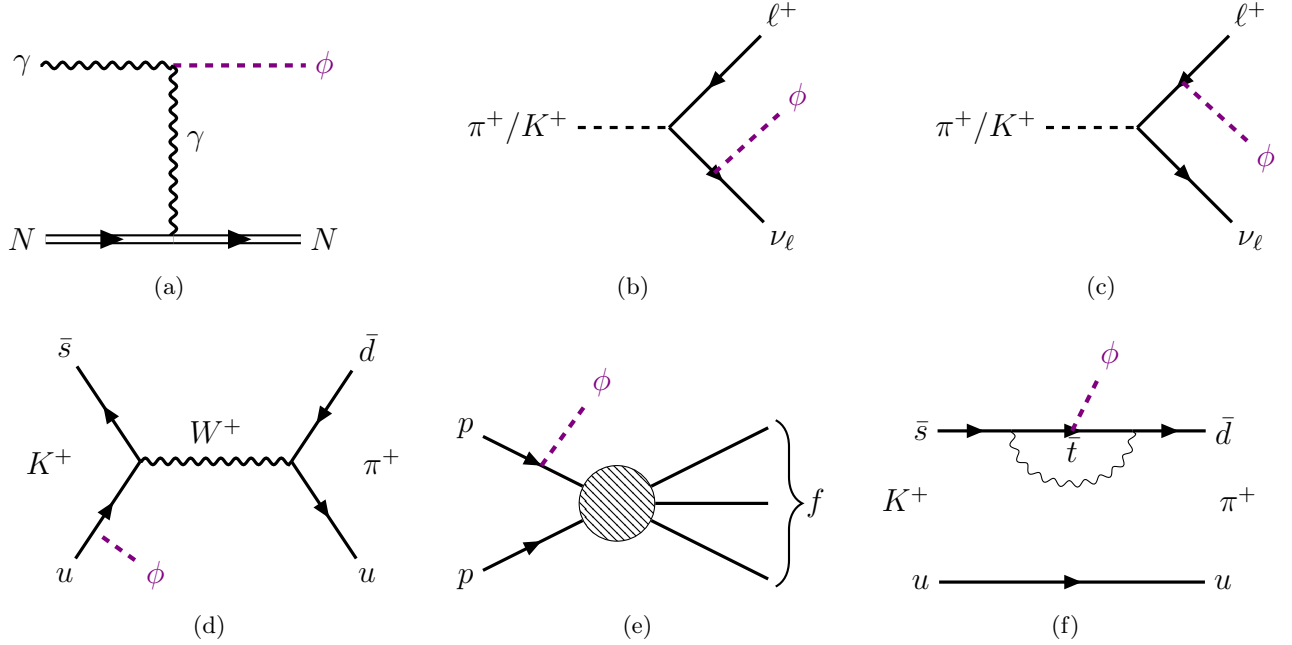

\medskip

\noindent \textbf{Photophilic scalars:} Scalars that couple to photons via the dimension-5 operator can be produced through Primakoff scattering when photons interact in a material, as depicted in the Feynman diagram in Fig.~\ref{fig:Primakoff}. This is an efficient production mechanism for two main reasons. Firstly, in the sub-GeV and GeV experiments of our interest, the cross section is coherently enhanced by a factor of $Z^2$ from the nuclear form factor. Secondly, since the photon propagator drives the process towards minimal momentum transfer to the nucleus, the produced scalar is both forward with respect to the initial photon's direction and carries almost all of its energy ($E_\gamma$). In the forward limit, for a scalar with momentum $p_\phi \simeq E_{\gamma}$, the cross section is approximately given as~\cite{PhysRev.81.899, Avignone:1988bv, Creswick:1997pg, SOLAX:1997lpz, Dent:2020jhf}: 

\begin{equation}
    \sigma_{\text{prim}}(m_{\phi}, E_\gamma)\simeq \sigma_{\text{prim}}(p_\phi) = \frac{Z^2\alpha_{\text{em}}g_{\phi\gamma\gamma}^2}{2}\Bigg( \frac{2r_0^2 p_{\phi}^2 + 1}{4r_0^2 p_{\phi}^2} \ln (1 + 4r_0^2 p_{\phi}^2) -1 \Bigg)
\end{equation}

where, $Z$ is the atomic number of the material, $r_0$ is the screening length, $m_\phi$ is the scalar mass and $\alpha_{\text{em}}$ is the fine structure constant. From the cross section, we calculate the probability of a scalar Primakoff for a given scalar mass and photon energy as follows:

\begin{equation}
    P_{\text{prim}}(m_\phi, E_\gamma) = \frac{\sigma_{\text{prim}}(m_\phi, E_\gamma)}{\sigma_{\text{prim}}(m_\phi, E_\gamma) + \sigma_{\text{SM}}(E_\gamma)} \simeq \frac{\sigma_{\text{prim}}(m_\phi, E_\gamma)}{\sigma_{\text{SM}}(E_\gamma)}
\end{equation}

where $\sigma_{SM}(E_\gamma)$ represents the total cross-section for a standard model photon with energy $E_\gamma$, interacting with any material as given in the Photon Cross Section Database Ref.~\cite{NIST:2019}.

\medskip

\noindent \textbf{Neutrinophilic scalars:} In this study, we will assume that the neutrinophilic scalar is flavor-universal. These scalars can be produced copiously through three-body decays of charged mesons, $\texttt{m}^{\pm} \to l^{\pm}+ \mybracketbar{\nu_l} +\phi$. The phase space availability of the third decay product alleviates helicity suppression~\cite{Carlson:2012pc, Barger:2011mt, Laha:2013xua} as observed in its SM two-body decay counterpart, making the three-body decay an enhanced production mechanism. Neutrinophilic scalars can emanate from the neutrino leg of the meson decay, as depicted in the Feynman diagram in Fig.~\ref{fig:3bodyneutrino}. The probability of producing a scalar can be found by calculating the decay width of the three-body decay and then finding its fraction with respect to the SM decay width of the charged mesons. We refer the reader to the appendix~\ref{app:threebody} for more details on the three-body decay width calculation. For the sub-GeV and GeV scale neutrino energies in these experiments of our interest, the charged mesons that are dominantly produced are pions and kaons only, i.e. $\texttt{m}^{\pm} = \pi^{\pm}~\text{and}~K^{\pm}$, and also $l=e,\mu$. 
Since the horns in these experiments enhance the charged meson flux by focusing them parallel to the beamline, this production mechanism is very efficient in the above neutrino facilities.
\medskip

\noindent \textbf{Electro/Muon-philic scalars:} Much like the neutrinophilic scalar scenario, the three-body decay process detailed above can also source the electro/muon-philic scalars, where the scalar can emanate from its charged lepton counterpart. This process, as depicted in the Feynman diagram of Fig.~\ref{fig:3bodylepton}, faces the same phase space enhancement as that for the neutrinophilic scalar. Unlike the neutrinophilic scenarios, we will treat the electron and muon flavors independently, i.e., assume flavor-nonuniversality.

Electrophilic scalars have additional sources through electromagnetic interactions of electrons, positrons, and photons with target electrons/nuclei. A few of these sources are Compton-like scattering $\gamma+e^-\to\phi+e^-$~\cite{AristizabalSierra:2020rom}, associated production $e^+ + e^- \to \phi + \gamma$~\cite{Nardi:2018cxi}, electron-positron annihilation $e^+ + e^- \to \phi$~\cite{Nardi:2018cxi}, and electron bremsstrahlung $e^{\pm}+N\to e^{\pm} + N +\phi$~\cite{Tsai:1989vw, Tsai:1986tx}. Muonphilic scalars can also be produced via muon bremsstrahlung, i.e., $\mu^{\pm} + N \to \mu^{\pm}+N+\phi$~\cite{PhysRevD.55.1233, Rella:2022len, Chen:2018vkr}. However, we find that these production mechanisms are subdominant as they yield much softer scalars as compared to the meson decays, especially at GeV-scaled neutrino facilities. 

\medskip

\noindent \textbf{Quarkphilic scalars:} Scalars that couple to quarks are mostly produced from hardonic interactions. For example, an up-philic scalar can be produced through two-body decays of Kaons ($K^+ \to \pi^+ + \phi$) such as that depicted in Fig.~\ref{fig:Kptopipupcoupling}. Since this two-body decay does not face any helicity suppression, it is more enhanced than the three-body decay process, which can occur through the scalar emanating from the meson leg. In addition to the meson decays, an up-philic scalar can also be produced through proton bremsstrahlung~\cite{Foroughi-Abari:2021zbm} ($p+N\to p+N+\phi$), such as that shown in the Feynman diagram in Fig.~\ref{fig:protonbremscalar}. Proton bremsstrahlung produces the scalar in the forward direction and transfers a large portion of the initial proton's energy to the scalar. Therefore, proton bremsstrahlung enhances the flux of up-philic scalars at proton-on-target facilities. Down-philic scalars can also be produced, analogous to the up-philic processes, through kaon two-body decays and proton bremsstrahlung after accounting for the appropriate multiplicity of down quarks in the mesons and protons, respectively. We remark that the $K^+ \to \pi^+ + \phi$ process can also occur via loops containing heavier quarks, such as charm, strange, top, bottom, through a one-loop diagram as shown in Fig.~\ref{fig:Kptopiptopcoupling} for an example top-philic scalar. However, for the sake of demonstration, we will limit our analysis to the up-philic scenario only.

\medskip

There exist other scalar scenarios where the scalar can couple to more than one SM fermion in addition to photons, such as Higgs Portal Scalars~\cite{Patt:2006fw}, the ones appearing in a $U(1)_{T_{3R}}$ model~\cite{Dutta:2019fxn}, etc. The coupling of Higgs portal scalars to leptons is equal to the product of the SM Higgs coupling and the mixing. On the other hand, $U(1)_{T_{3R}}$ scalar couplings are only constrained by the experimental data. The phenomenology of such scenarios would include a combination of the various production mechanisms discussed above. They will also call for additional production mechanisms, such as those from one-loop factors or a cross-term in the meson decays. Since our goal is to demonstrate the effect of dark matter coupled to scalar portals, the above-mentioned combinations are beyond the scope of this study.

\medskip

\noindent\textbf{Dark Matter from scalars:} DM can be produced via the scalars, which act as a portal between the SM and DM. We will be interested in the regime where $m_{\phi} > 2 m_{\chi}$. This enables the scalar to decay into dark matter via a two-body decay mechanism, $\phi \to \chi \bar{\chi}$, as depicted in the Feynman diagram in Fig.~\ref{fig:phidecaytoDM}. The decay width for such a process is given by: 

\begin{equation}
\Gamma(\phi \to \chi\bar{\chi})=\frac{g_D^2m_\phi}{8\pi}\Bigg(1-\frac{4m_\chi^2}{m_\phi^2}\Bigg)^{3/2}
\end{equation}

Our assumption on the scalar couplings, $g_D \gg g_{\phi\gamma\gamma}, y_f$, ensures that any on-shell $\phi$ would promptly decay into dark matter with a 100\% branching ratio. We therefore adhere to the conventional choice of $\alpha_D = g_D^2/4\pi = 0.5$ and $m_{\phi} = 3 m_{\chi}$~\cite{COHERENT:2021pvd, deNiverville:2015mwa}. Using the chosen values we obtain $\Gamma(\phi \to \chi \bar{\chi})=1.57\times 10^{23}~s^{-1}$. This implies that the decay lengths are very short ($c\tau =1.9\times 10^{-15}~$m), resulting in prompt decay of scalars into dark matter.

\subsection{Detection of dark matter}\label{sec:detection}

\begin{figure}[t]
    \centering
    \subfloat[]
    {
    \begin{tikzpicture}
    \tikzset{every node/.style={font=\large}}  
        \begin{feynman}
                    \vertex (a) at (0.0, 0.0);
                    \vertex (b) at (-2.0, 0.0) {$\textcolor{violet}{\phi}$};
                    \vertex (c) at (1.5, 1.5) {$\textcolor{blue}{\chi}$};
                    \vertex (d) at (1.5, -1.5) {$\textcolor{blue}{\bar{\chi}}$};
                    \diagram*{
                    (b) -- [scalar, violet, ultra thick](a),
                    (a) -- [fermion, blue, very thick](c),
                    (d) -- [fermion, blue, very thick](a)
                    };
                \end{feynman}
    \end{tikzpicture}
    \label{fig:phidecaytoDM}
    }
    \hspace{1cm}
    \subfloat[]
    {
    \begin{tikzpicture}
    \tikzset{every node/.style={font=\large}}  
        \begin{feynman}
                \vertex (a) at (0.0, 0.0); 
                \vertex (b) at (0.8, 0.0); 
                \vertex (c) at (2.0, 0.0) {$\gamma$}; 
                \vertex (d) at (0.0, -0.8) ; 
                \vertex (e) at (0.0, -1.6); 
                \vertex (f) at (0.0, 0.8); 
                \vertex (g) at (0.0, 1.6); 
                \vertex (h) at (2.0, 1.6) {$\textcolor{blue}{\chi}$}; 
                \vertex (i) at (-2.0, 1.6) {$\textcolor{blue}{\chi}$};
                \vertex (j) at (2.0, -1.6) {$N$};
                \vertex (k) at (-2.0, -1.6) {$N$};
                \node at (-0.3, -0.8) {$\gamma$};
                \node at (-0.3, 0.8) {$\textcolor{violet}{\phi}$};
                \diagram*{
            (a) -- [photon, very thick](c), 
            (a) -- [scalar, violet, ultra thick](g),
            (a) -- [photon, very thick](e),
            (e) -- [double,double distance=0.5ex, very thick,with arrow=0.5,arrow size=0.2em] (j),
            (k) -- [double,double distance=0.5ex, very thick,with arrow=0.5,arrow size=0.2em] (e),
            (i) -- [fermion, blue, very thick](g), 
            (g) -- [fermion, blue, very thick](h)
            };
                \end{feynman}
    \end{tikzpicture}
    \label{fig:2to3detection}
    }
    \captionsetup{justification=Justified,singlelinecheck=false}
            \caption{Feynman diagrams depicting (a) the two-body decay of scalar mediator into dark matter, that occurs with 100\% probability, and (b) the $2 \rightarrow 3$ process for detecting DM. Here, the incoming DM particle scatters with the target nuclei in the detector via scalar and photon propagators to produce a photon in the final state.}
    \end{figure}
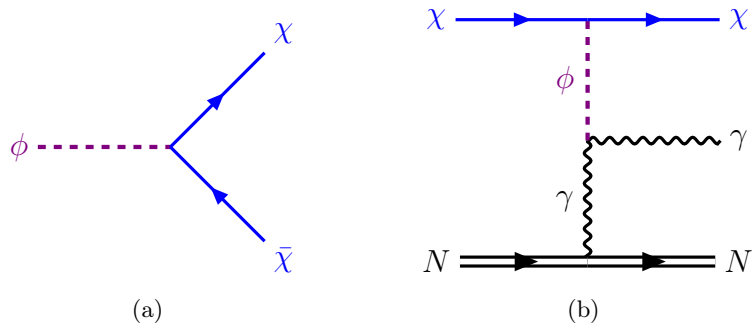

In the absence of any direct coupling to SM and lighter DM states, the single-component DM can only be detected by scattering mechanisms involving photons. Since all our scenarios of interest include the scalar's coupling to di-photons (via $g_{\phi\gamma\gamma}$), DM can produce a single photon final state via a $2\to 3$ scattering process $\chi + N \to \chi + N + \gamma$ (see Fig.~\ref{fig:2to3detection}). This process is facilitated by a virtual scalar that mediates the interaction between dark matter and the photons. Additionally, the process is supported by the nucleus through the exchange of a virtual photon. This, like the Primakoff process, leads to a coherent enhancement by a factor of $Z^2$, where $Z$ is the atomic number of the nucleus. Due to a virtual photon being exchanged with the nucleus, the final state photon carries a large portion of the initial DM's energy, and is also produced in the forward direction.

For scenarios involving an electrophilic or quarkphilic scalar, DM can also be detected through electron or nuclear recoils, respectively ($\chi+e^-/N \to \chi + e^-/N$). Although the cross sections for these $2\to2$ processes are larger than the $ 2\to3$ process discussed above, the recoil processes are not efficient detection mechanisms at neutrino experiments. For example, the final state electrons from electron recoil greatly suffer from neutrino charged and neutral current backgrounds, such as $\nu + e^- \to \nu + e^-$, which yield topologies very similar to DM-electron recoil. The nuclear recoil also suffers from the neutrino-induced background.  Additionally, the nuclear recoil energies are much softer than the single-photon final state, making them harder to detect given typical LArTPC thresholds to hadronic activities.
Therefore, the high-energy single photons from the $2\to3$ process are much more distinguishable and suffer less from neutrino/neutron-induced backgrounds at our benchmark neutrino experiments. We present the details on the cross-section calculations and the kinematics of the final state in Appendix.~\ref{app:2to3}.

\section{\label{sec:experiments}Dark Matter at Benchmark Experiments}

\begin{figure}[!htbp]
    \centering
        \includegraphics[scale=0.65]{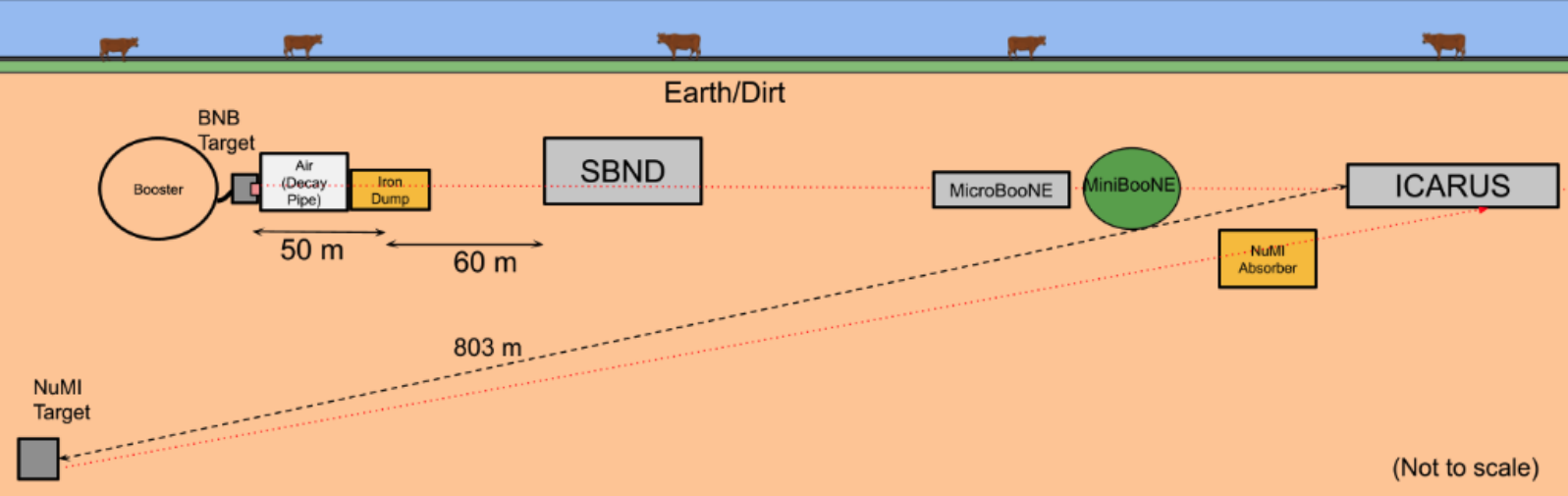}
    \captionsetup{justification=Justified, singlelinecheck=false}
    \caption{Schematic diagram (Ref.~\cite{PRISM}) of the SBN experimental program (not to scale), showing the layout of the Booster Neutrino Beam (BNB) line and the Neutrinos at the Main Injector (NuMI) line. The positions of the four detectors-SBND, MicroBooNE, MiniBooNE, and ICARUS are shown. SBND is located 110 meters downstream from the BNB target, while ICARUS is positioned 803 meters from the NuMI target. Red dotted lines show the directions of the BNB and NuMI beams. Both SBND BNB, and ICARUS-NuMI are situated off-axis with respect to their respective beam sources.}
    \label{fig:SBNSchematic}
\end{figure}

\begin{figure}[!htbp]
    \centering
        \includegraphics[scale=0.6]{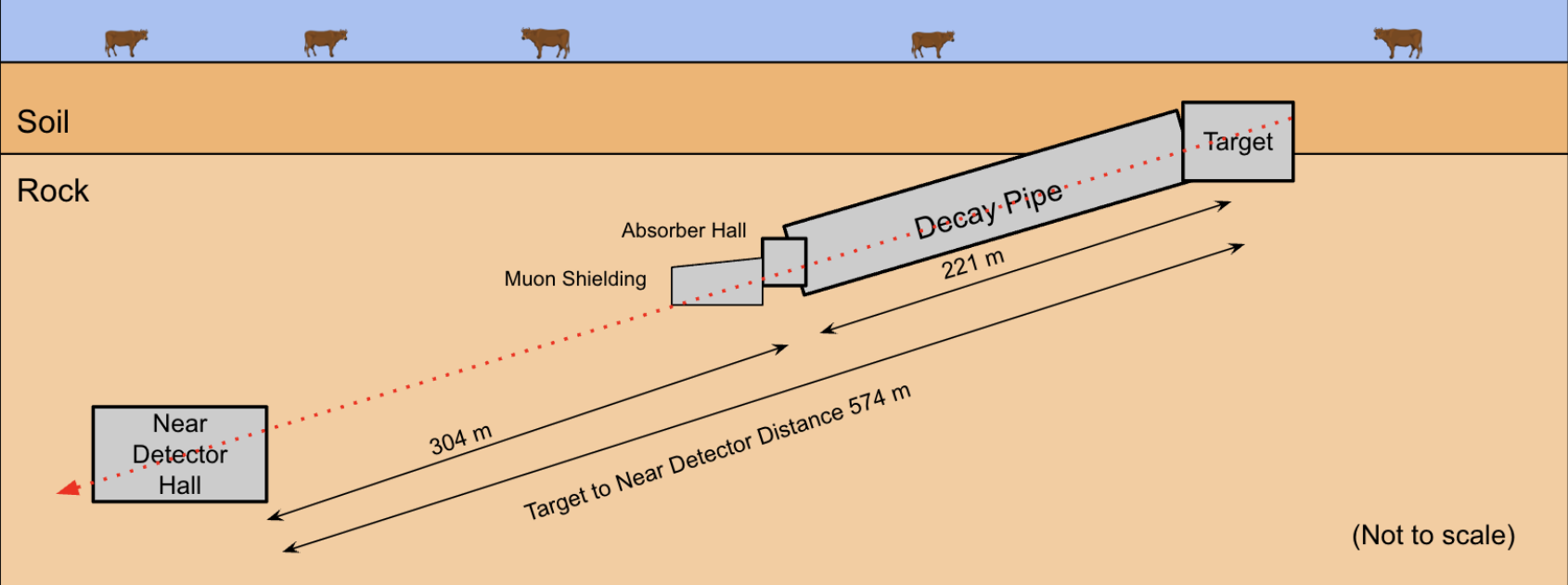}
    \captionsetup{justification=Justified, singlelinecheck=false}
    \caption{Schematic diagram of the DUNE near detector experimental setup (not to scale), showing the Long-Baseline Neutrino Facility (LBNF) beamline infrastructure. The decay pipe is 221 meters long. Downstream of this, the muon shielding and the absorber hall are also shown. The DUNE near detector is located 574 meters from the target. (Source: Fermilab)}
    \label{fig:DUNESchematic}
\end{figure}

\begin{figure}[!htbp]
    \centering
        \includegraphics[scale=0.45]{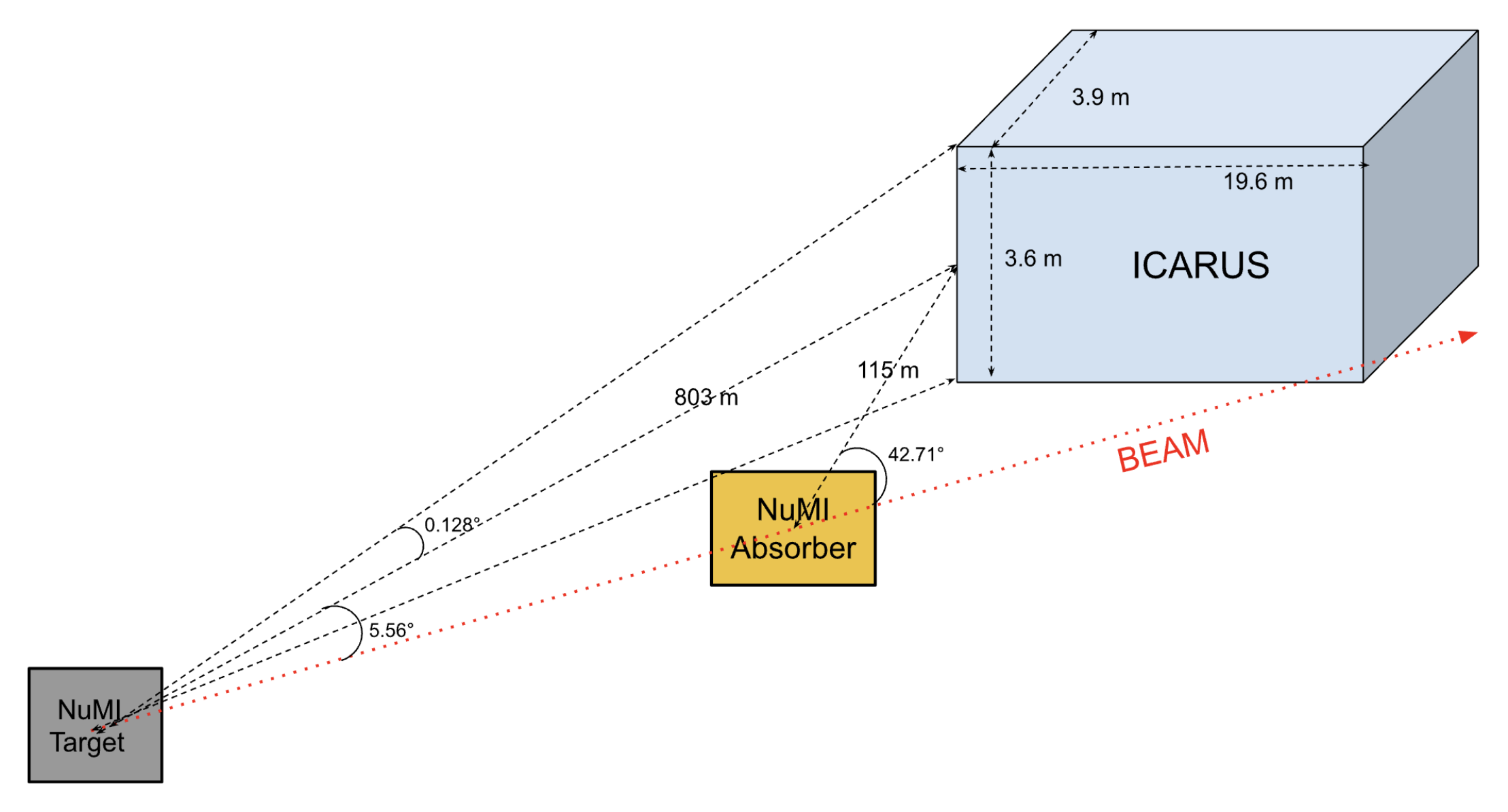}
    \captionsetup{justification=Justified, singlelinecheck=false}
    \caption{Schematic layout of the ICARUS detector in the NuMI beamline. The diagram illustrates the geometric configuration relative to the NuMI target and absorber. The ICARUS detector is positioned 803 meters from the target and 115 meters from the absorber, with off-axis angles of $5.56^\circ$ from the target and $42.71^\circ$ from the absorber, respectively. These off-axis placements significantly influence the detector's angular acceptance and sensitivity to various particle production mechanisms. The beam direction is shown as a red dashed line.}
    \label{fig:ICARUSSchematic}
\end{figure}

\begin{table*}[!htbp]
    \centering
    \renewcommand{\arraystretch}{1.2}
    \begin{tabular}{|c|c|c|c|c|c|c|c|}
        \hline
        Experiment & \makecell{Beam} & \makecell{Mass\\(tons)} & \makecell{Dimensions\\(m $\times$ m $\times$ m)} & \makecell{Distance\\(m)} & \makecell{Total\\POT} & 
        \makecell{Angle off-axis\\(degrees)}& \makecell{Threshold\\Energy\\(MeV)}\\
        \hline
        SBND & \makecell{8 GeV\\BNB} & 112.0 & $4 \times 4 \times 5$ & 110 & $10^{21}$ & 0.39 & 30\\
        (LAr, Ongoing) & & & & & & &  \\
        \hline
        ICARUS T600 & \makecell{120 GeV\\NuMI} & 476.0 & \makecell{$3.6 \times 3.9 \times$\\$19.6$} & 803 & $3 \times 10^{21}$ & 5.56 & 30\\
        (LAr, Ongoing) & & & & & & &  \\
        \hline
        DUNE ND & \makecell{120 GeV\\LBNF} & 50.0 & \makecell{$7 \times 3 \times 5$} & 574 & \makecell{$7.35 \times 10^{21}$} & 0 & 30\\
        (LAr, Upcoming) & & & & & & &  \\
        \hline
        CCM200 & \makecell{800 MeV\\LANSCE} & 10.0 & \makecell{Radius=1.25} & 20  & \makecell{$2 \times 10^{22}$} & 90 & 10 \\
        (LAr, Ongoing) & & & & & & &  \\
        \hline
        DUNE PRISM~\cite{Hasnip:2025gyi, Hasnip:2023ygr} & \makecell{120 GeV\\LBNF} & 50.0 & \makecell{$7 \times 3 \times 5$} & 574 & \makecell{$7.35 \times 10^{21}$} & 2.84 & 30\\
        (LAr, Upcoming) & & & & & & &  \\
        \hline
        MicroBooNE & \makecell{8 GeV\\BNB} & 86.8 & \makecell{$2.6 \times 2.3 \times 10.4$} & 470 & \makecell{$6.6 \times 10^{20}$} & 0 & 30\\
        (LAr, Completed) & & & & & & &  \\
        \hline
        MiniBooNE & \makecell{8 GeV\\BNB} & 818.0 & \makecell{Radius=5.0} & 537 & \makecell{$1.875 \times 10^{21}$} & 0 & 100\\
        (CH$_2$, Completed) & & & & & & &  \\
        \hline
    \end{tabular}
    \captionsetup{justification=Justified,singlelinecheck=false}
    \caption{List of key parameters for each experiment considered in this study. This includes the beam energy and source, active detector mass, physical dimensions, distance from the beam target, total number of protons on target (POT), off-axis angle, and the photon detection threshold energy. Experiments span ongoing, upcoming, and completed liquid argon (LAr) and CH$_2$-based detectors across the BNB, NuMI, LBNF, and LANSCE beamlines.} 
    \label{tab:ExperimentalDetails}
\end{table*}

In this section, we will discuss the experimental details and the flux of SM particles that are relevant for the various scalar scenarios. Schematic diagrams of all the SBN experiments, DUNE, and layout of the ICARUS detector in the NuMI beamline have been shown in Figs.~\ref{fig:SBNSchematic}, \ref{fig:DUNESchematic}, and \ref{fig:ICARUSSchematic} respectively.

The CCM200 detector is placed 20~m away from a tungsten target, which is delivered with 800~MeV protons from the LANSCE accelerator. The electromagnetic (EM) cascades and hadronic activities produce a large multiplicity of photons and electrons ($\mathcal{O}(0.1-1000)$ per POT in the MeV range), and approximately 0.1 $\pi^+$ per POT at rest.  EM particles and mesons have been utilized to study various BSM scenarios, such as axion-like particles~\cite{CCM:2021jmk}, meson-portal mediators~\cite{CCM:2023itc}, etc. The flux of $\pi^-$ is suppressed due to their absorption in the target through $\pi^- + p \to n + \gamma$~\cite{Dutta:2020vop}. Similarly, we realize that they can also serve as sources of various scalar portals. For example, photophilic and electrophilic scalars can be produced through the large flux of photons and electrons/positrons, respectively. In addition, charged pions can produce energetic electrophilic, muonphilic, and neutrinophilic scalars through the decay of pions at rest (PiDAR). Since the kaon production and proton bremsstrahlung at the 800~MeV facility is highly suppressed, there is a corresponding suppression in the flux of quark-philic scalars. Therefore, we will limit our analysis at CCM200 to photophilic, neutrinophilic, electrophilic, and muonphilic scalars only. For a total of $2\times 10^{22}~\text{POT}$, we will investigate the sensitivity of the 10-tonne LArTPC detector for scalar portal DM. 

SBND is a $4\times4\times5~\text{m}^3$ detector placed at 110 m along the BNB beamline from a 70~cm long Beryllium target, where 8~GeV protons are delivered. Along the same beamline, the MicroBooNE and MiniBooNE detectors are located 470~m and 537~m away from the BNB target, respectively. Due to the target geometry and beam energy, we find from \texttt{GEANT4}~\cite{Allison:2006ve, Allison:2016lfl, GEANT4:2002zbu} simulations that $\sim 18\%$ of the protons leak from the target, and travel up to the $4\times 4\times4.21~\text{m}^3$ iron dump that is located 50~m downstream. To maintain consistent terminology, we refer to the BNB iron dump as the BNB iron absorber throughout this work. The target and the iron absorber together act as a factory of electrons, positrons, and photons. Additionally, there is a large flux of $\pi^{\pm}~\text{and}~K^{\pm}$ produced at the target that are focused by magnetic horns. Since they are allowed to decay along the 50~m long decay pipe, they can source neutrinophilic, electrophilic, and muonphilic scalars via, for example, positively charged meson decays: $\pi^+/K^+ \to \mu^+\nu_\mu\phi/e^+\nu_e \phi$. The scalars produced from meson decays at the BNB facility, specifically electrophilic scalars, dominate over those produced from electromagnetic cascades, such as Compton-like, annihilation, and electron bremsstrahlung processes. Apart from these, we also find that the large decay volume and the 8~GeV facility allow for the production of up-philic quarks via $K^+ \to \pi^+ \phi$ in the decay volume, and proton bremsstrahlung at the target and absorber. We would like to stress the fact that while charged mesons dominate over electromagnetic cascade productions for the electrophilic scalars, the Primakoff production from photons at the target and absorber is the most dominant production of photophilic scalars, which, along with being high in energy and forward, are enhanced particularly by the large multiplicity of photons at the iron absorber. 

ICARUS is a $3.6 \times 3.9\times 19.6~\text{m}^3$ detector placed 803~m away from the NuMI source and positioned at a $5.56^\circ$ off-axis angle. The NuMI facility consists of a 120~GeV proton beam that impinges on a graphite target, which is followed by a 715-m-long decay volume along the beamline that is terminated by an iron absorber. Along with dark matter from Primakoff scattering at the target and three-body decays occurring in the decay pipe, ICARUS is also sensitive to dark matter from processes occurring at the absorber, which is only 110~m away from the detector. These processes include Kaon decay-at-rest (KDAR), Primakoff scattering, etc. Due to the off-forward positioning of the ICARUS detector, it can detect softer new physics particles that are produced in the 120~GeV facility. Similar to BNB, there is a smaller percentage, $6.2\%$, leakage of protons from the target to the absorber obtained using \texttt{GEANT4} simulations in the NuMI target configuration.

The proposed DUNE ND is a $7\times3\times5~\text{m}^3$ sized detector that is set to be located 574~m away from the graphite target at the LBNF facility. The LBNF sources 120~GeV protons to the target, which is followed by a 221~m long decay pipe that allows for the focused mesons to decay. The large flux of photons at the target and the focused mesons enable the production of photophilic, neutrinophilic, electron/muon-philic, and quarkphilic scalars. Additionally, the 120~GeV protons can radiate up-philic quarks. \texttt{GEANT4} simulations in LBNF target configuration, show only a small percentage of the beam protons, $4.25\%$, leak from the target. Therefore, the contribution from the absorber is not as significant as it is in the BNB facility.

We summarize the experimental details and specifications in Table.~\ref{tab:ExperimentalDetails}. In this analysis, we consider the neutrino mode of the BNB, NuMI, and DUNE ND facilities. In this mode, the magnetic horn configuration is such that the positively charged mesons are focused parallel to the beam. We simulate all the SM particle fluxes, such as mesons, photons, electrons, and positrons, using \texttt{GEANT4} simulation package. To simulate the horn effects, we utilize the fluxes generated by the \texttt{RKHorn} package~\cite{RKHorn2024}.

\begin{figure}[!htbp]
    \centering
    \begin{subfigure}{0.48\textwidth} 
        \centering
        \includegraphics[scale=0.34]{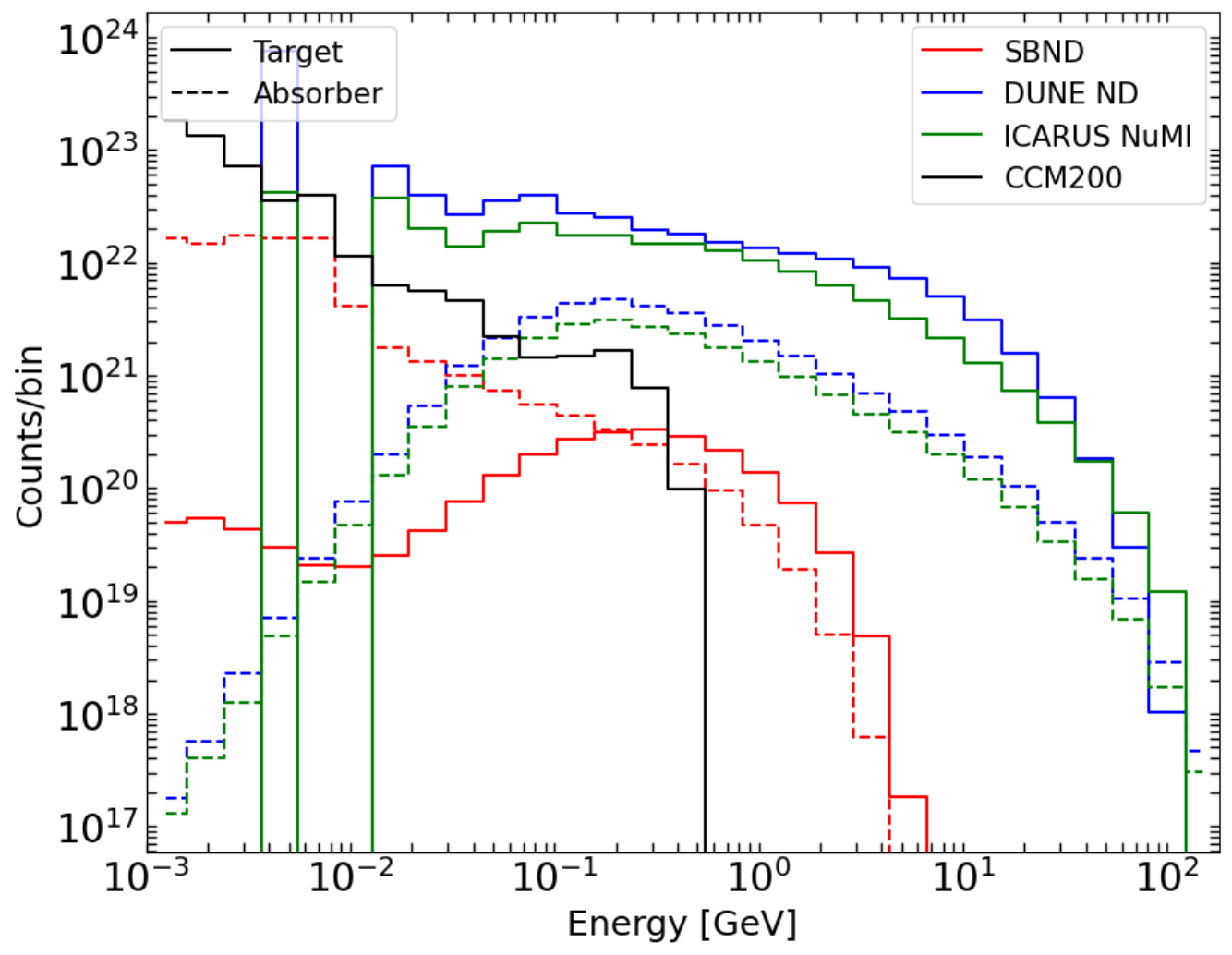}
        \caption{Photon Flux}
        \label{fig:PhotonFlux}
    \end{subfigure}
    \begin{subfigure}{0.48\textwidth} 
        \centering
        \includegraphics[scale=0.34]{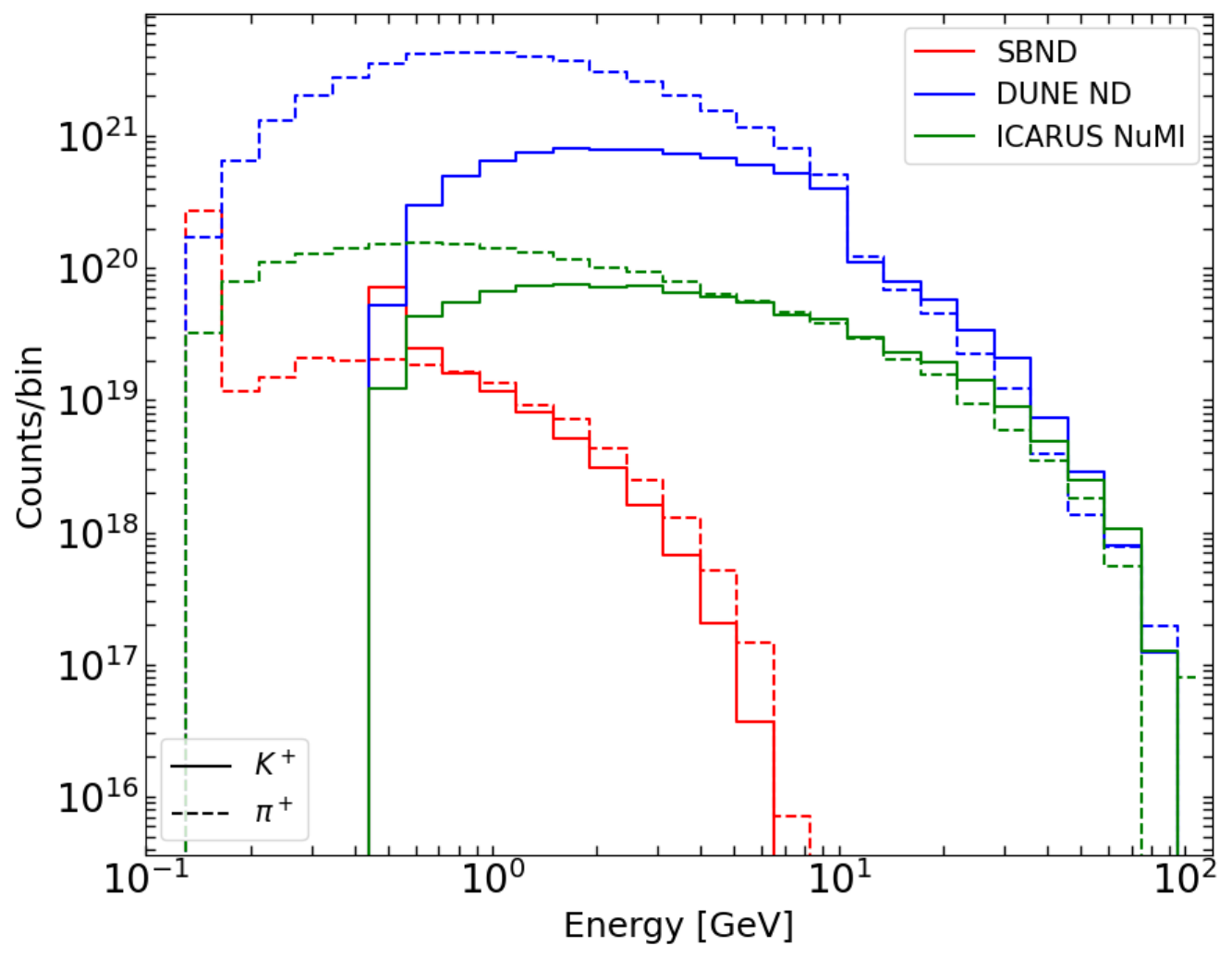}
        \caption{Meson Flux}
        \label{fig:MesonFlux}
    \end{subfigure}
    \captionsetup{justification=Justified, singlelinecheck=false}   
    \caption{Parent fluxes of scalars scaled to the total POT for the ongoing and upcoming experiments. (a) Photon fluxes that result in Primakoff production of photophilic scalars, and (b) Charged meson fluxes for neutrinophilic, electro/muon-philic, and up-philic scalars. Since, uniform binning is considered for the x-axis, the y-axis denotes counts/bin and not counts per bin width. Hence, in order to obtain the total number of photons/mesons one can sum over the values for each bin and not explicitly multiply with bin-widths.}
    \label{fig:Flux}
\end{figure}

\section{\label{sec:results}Results}

Due to differing experimental specifications, each detector yields distinct spectra and features for the various models discussed above. In this section, we systematically characterize these differences, which can be leveraged to enhance signal-to-background discrimination. Specifically, we focus on five key characteristics that vary across experiments, and also in the same experiment:
\begin{enumerate}[label=(\alph*)]
\item the energy spectrum of DM particles entering the detector,
\item the timing distribution of DM arrival, 
\item the number of DM produced as a function of mass,
\item the spatial distribution of DM particles at the face of the SBND detector, and
\item the energy and angular spectra of the final state photon and its comparison with the background for MicroBooNE
\end{enumerate}

We will focus most of our analysis on the ongoing and upcoming experiments, i.e., SBND, DUNE ND, ICARUS-NuMI, and CCM200. Throughout this section, we use consistent color coding for the experiments: SBND (red), DUNE ND (blue), ICARUS-NuMI (green), and CCM200 (black). Solid and dashed lines represent target and absorber contributions, respectively, unless otherwise noted.

A combined analysis of these observables provides a comprehensive understanding of how experimental geometry, beam energy, and production mechanisms influence signal characteristics.

\par Since, for all the following histogram plots (Figs.~\ref{fig:EnergySpectrum}, \ref{fig:TimingSpectrum}, \ref{fig:finalstatephotonspectra}, \ref{fig:MicroBooNEbkgspectra}), uniform binning has been considered for the x-axis, the y-axis denotes counts/bin and not counts per bin width. Hence, to obtain the total number, one can sum over the values in each bin rather than explicitly multiplying by the bin widths.

\subsection{Energy Spectra of DM}\label{sec:EnergySpectra}

\begin{figure}[!htbp]
    \centering
    \begin{subfigure}{0.45\textwidth} 
        \centering
        \includegraphics[scale=0.3]{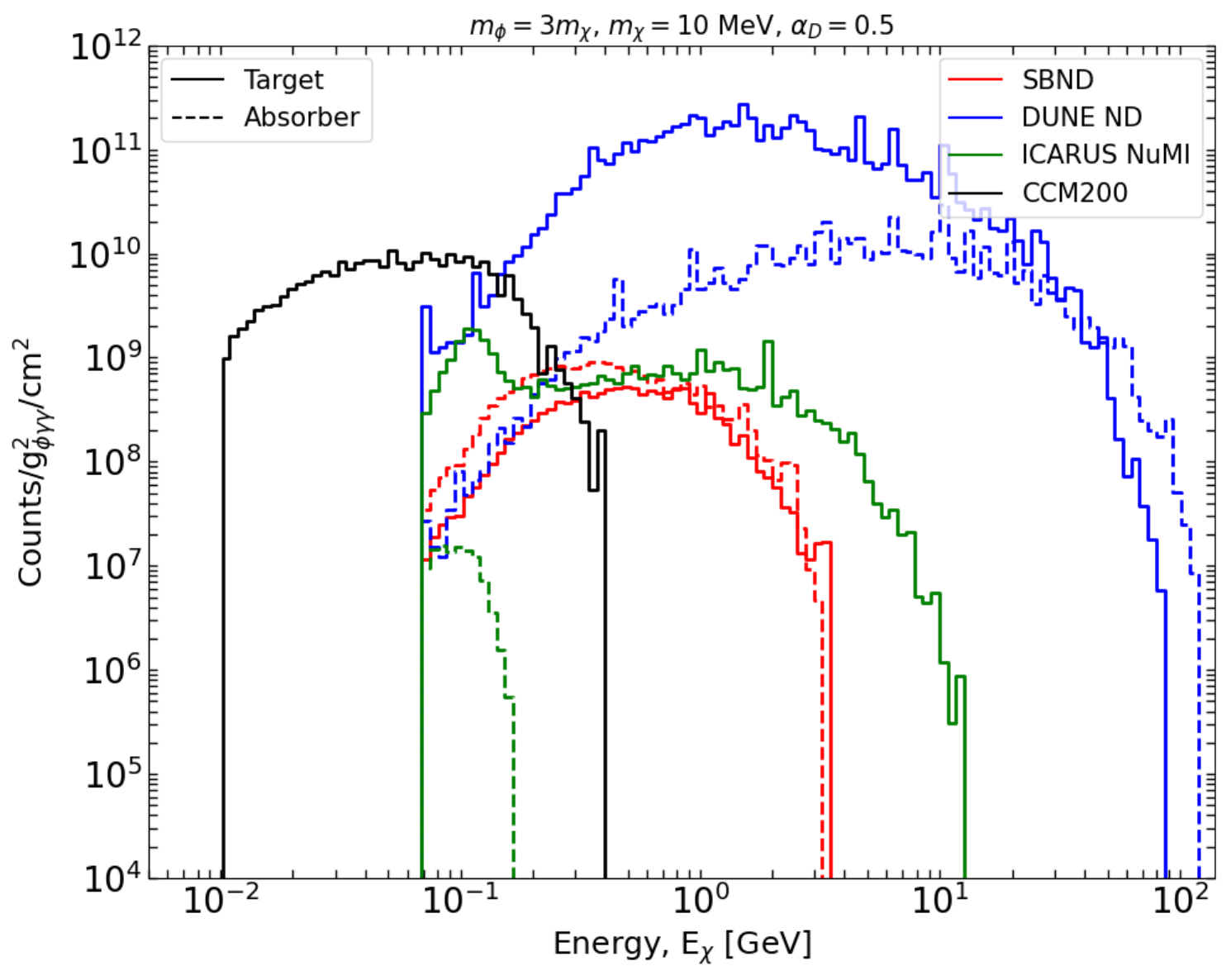}
        \caption{Photophilic model}
        \label{photonenergy}
    \end{subfigure}
    \begin{subfigure}{0.45\textwidth} 
        \centering
        \includegraphics[scale=0.3]{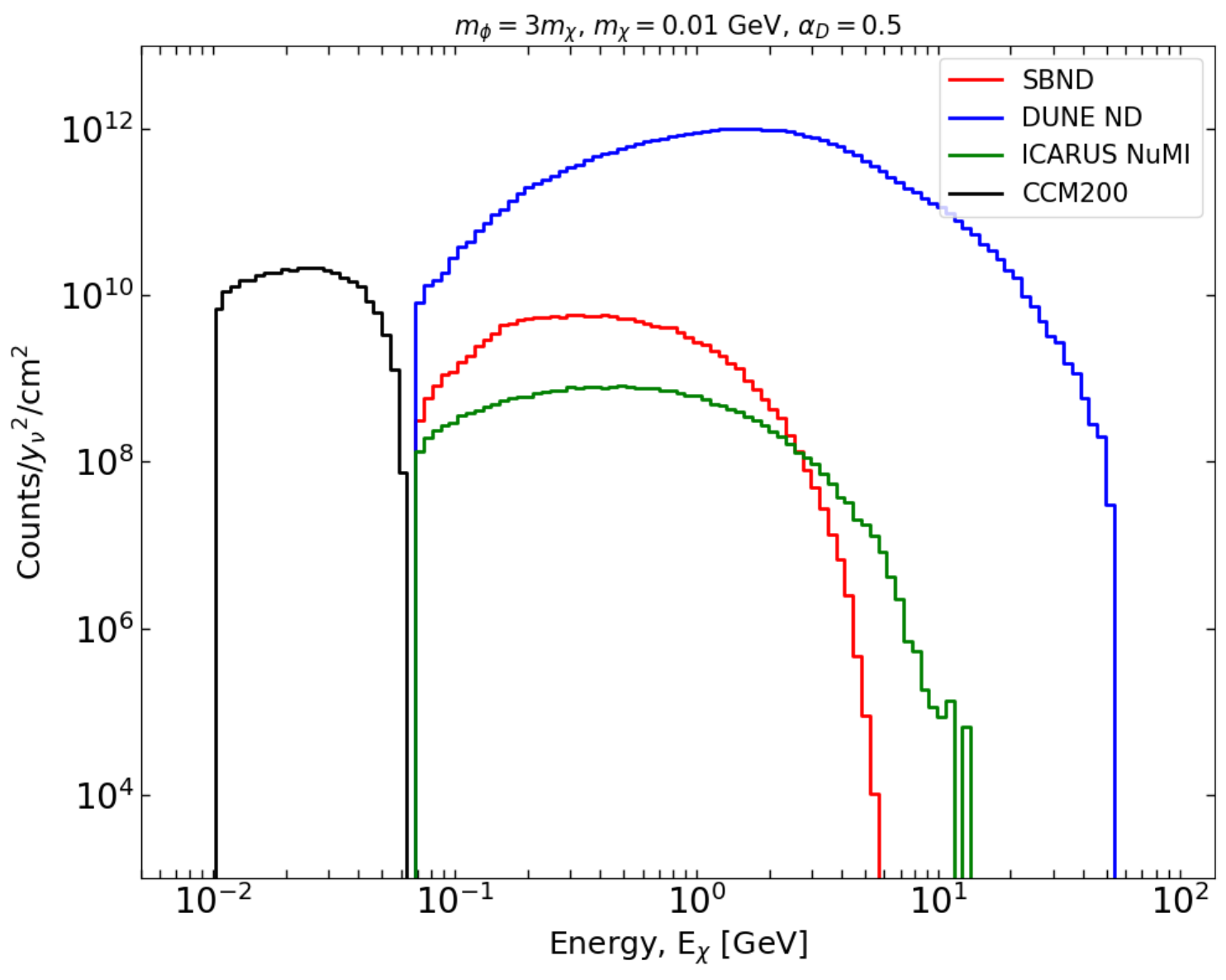}
        \caption{Neutrinophilic model}
        \label{neutrinoenergy}
    \end{subfigure}
    \begin{subfigure}{0.45\textwidth} 
        \centering
        \includegraphics[scale=0.3]{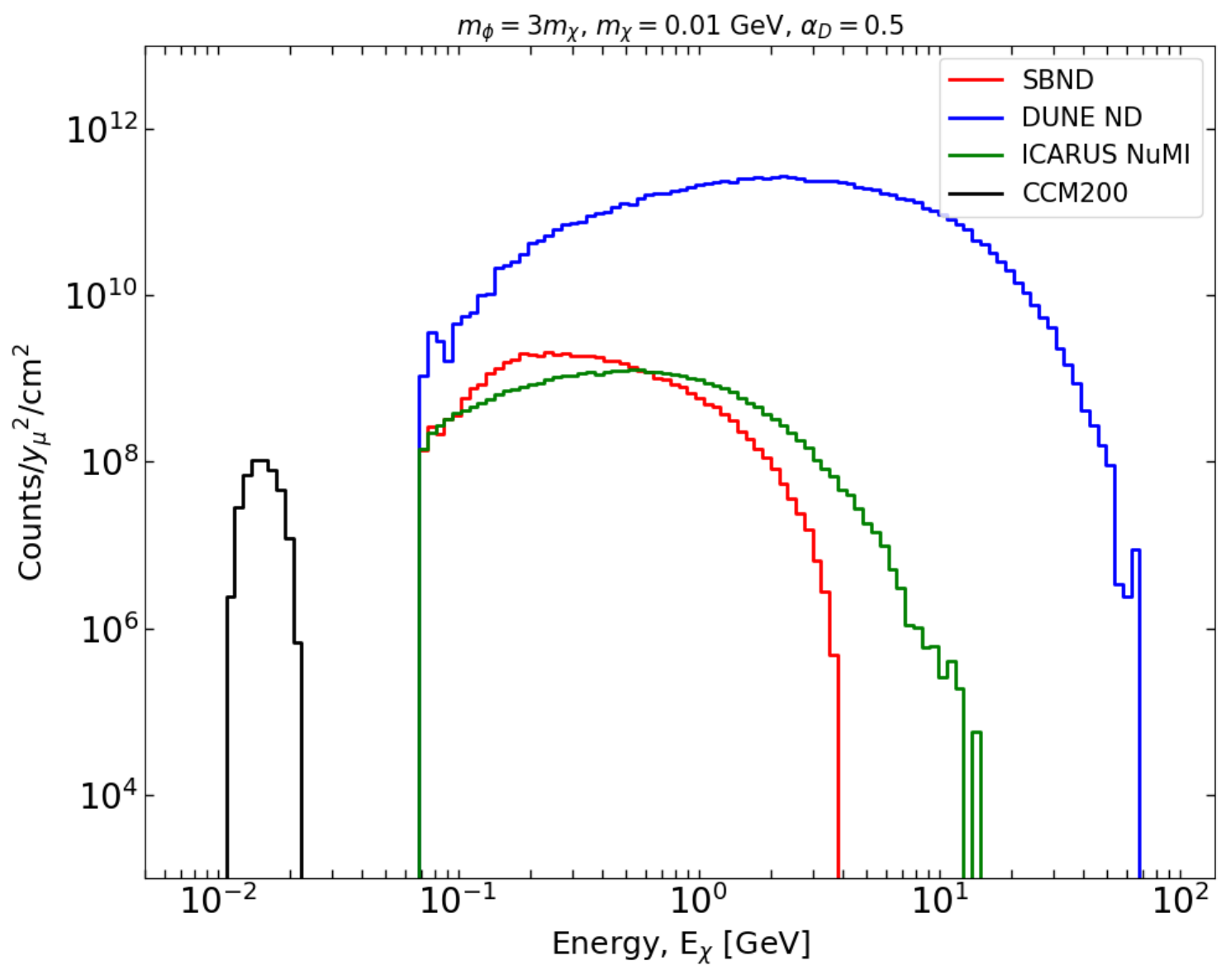}
        \caption{Muonphilic model}
        \label{muonenergy}
    \end{subfigure}
    \begin{subfigure}{0.45\textwidth} 
        \centering
        \includegraphics[scale=0.3]{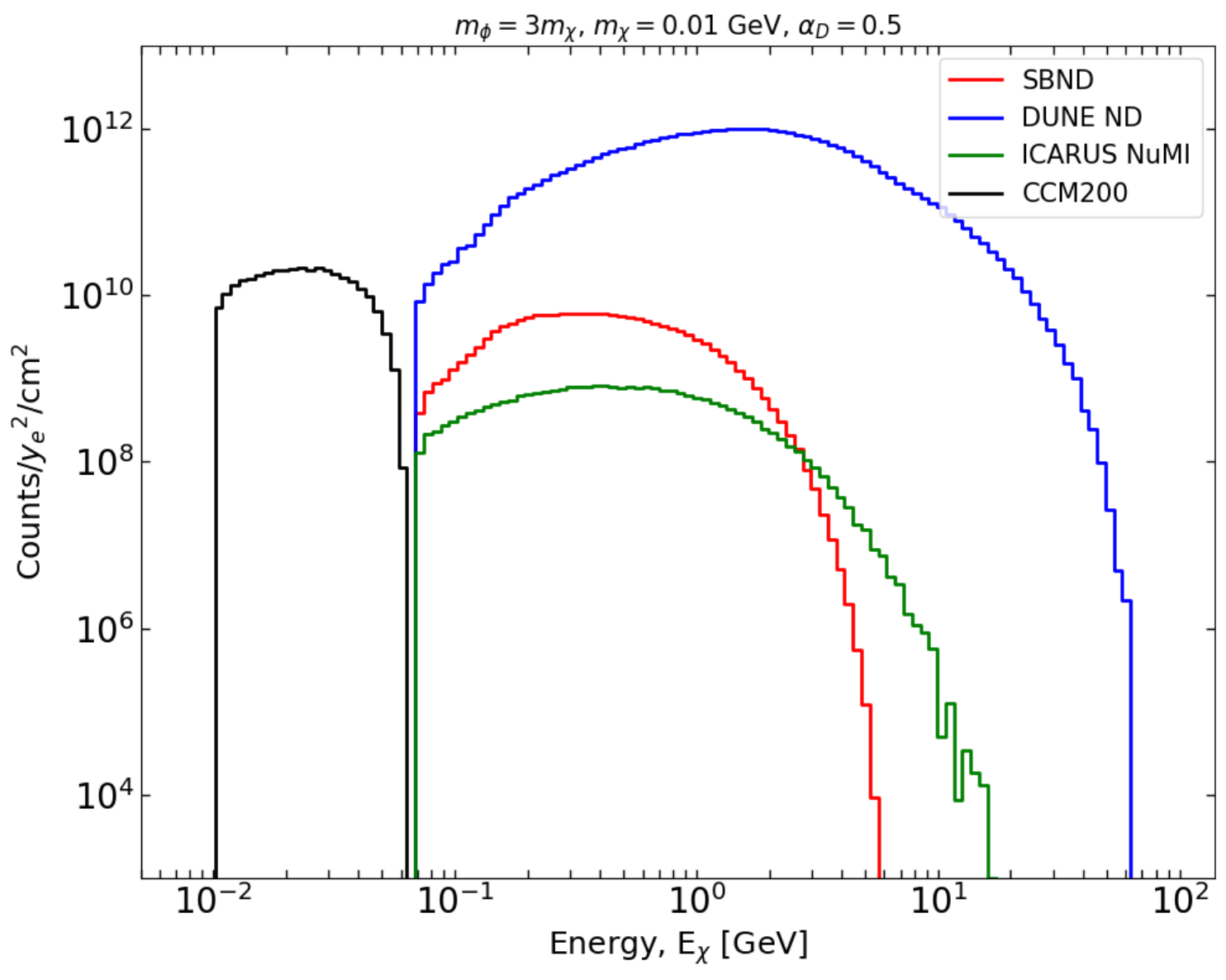}
        \caption{Electrophilic model}
        \label{electronenergy}
    \end{subfigure}
    \begin{subfigure}{0.45\textwidth} 
        \centering
        \includegraphics[scale=0.3]{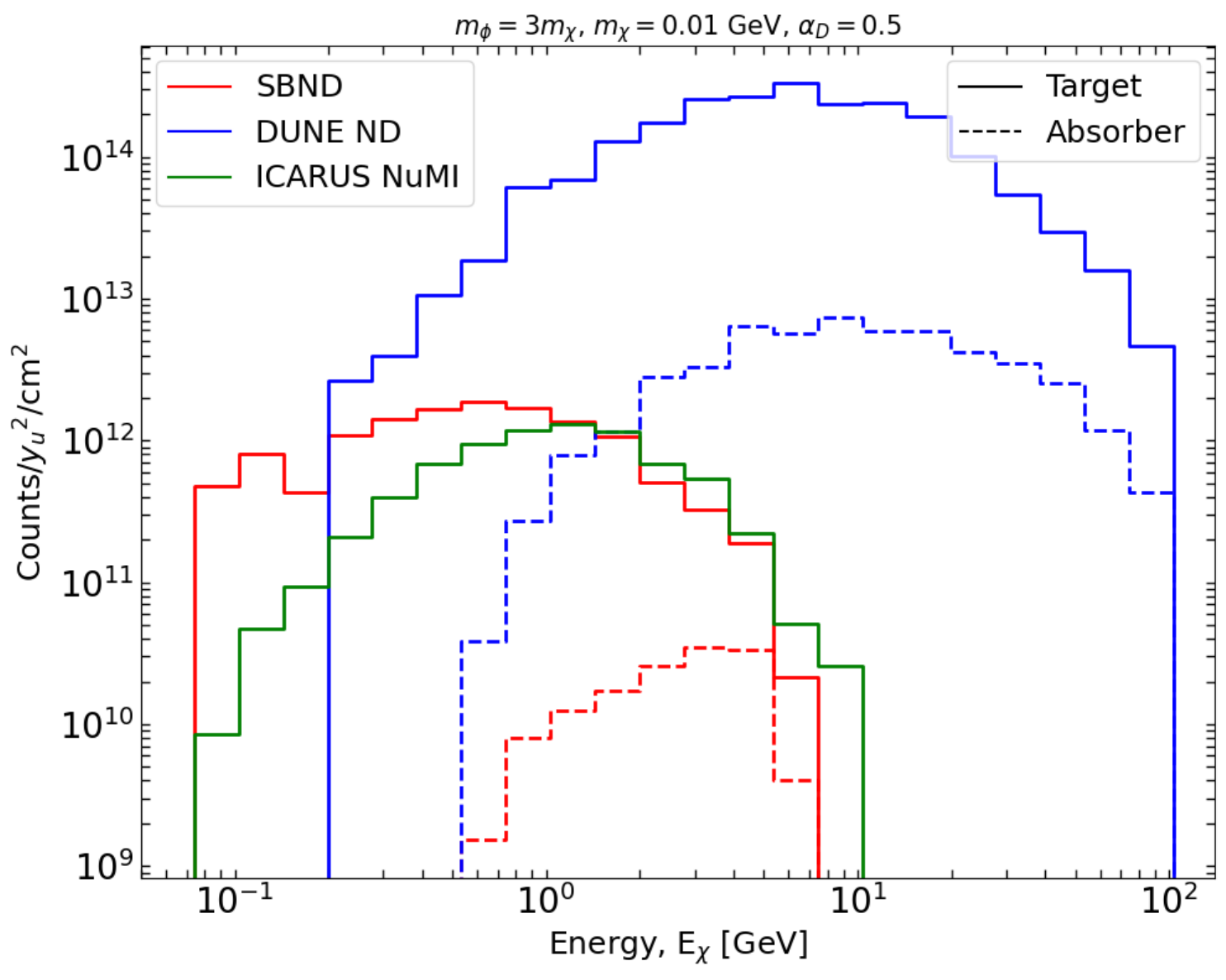}
        \caption{Upphilic model}
        \label{upquarkenergy}
    \end{subfigure}
    \captionsetup{justification=Justified, singlelinecheck=false}   
    \caption{Energy spectra of DM particles reaching the detectors for different production scenarios: (a) Photophilic, (b) Neutrinophilic, (c) Muonphilic, (d) Electrophilic, (e) and Up-philic. Curves represent different experiments: SBND (red), DUNE ND (blue), ICARUS-NuMI (green), and CCM200 (black). Solid and dashed lines (in panel a) show target and absorber contributions, respectively. Fluxes are normalized per unit coupling squared and account for total POT and detector geometry. A photon energy threshold of 1 MeV is applied for CCM200, while for SBND, DUNE ND, and ICARUS-NuMI, a dark matter (DM) kinetic energy threshold of 60 MeV is used.}
    \label{fig:EnergySpectrum}
\end{figure}

Figure.~\ref{fig:EnergySpectrum} shows the characteristic energy spectra of DM arriving at the front face of SBND, DUNE ND, ICARUS-NuMI, and CCM200 detectors, across different production models. We fix $m_\chi = 10$~MeV in all the energy spectra subplots.

To estimate the total signal (per unit coupling-squared), we integrate over the DM energy spectrum given in Fig.~\ref{fig:EnergySpectrum}, and weigh it by the number of target nuclei and the $2 \rightarrow 3$ cross-section. 
Since SBND, DUNE ND, and ICARUS require a minimum threshold of 30 MeV on the final state photon, we present the DM fluxes starting at 60~MeV. This threshold choice is consistent with the observation that the incoming DM almost equally shares its kinetic energy with the outgoing photon and outgoing DM, with a smaller share carried by the recoil nucleus, as shown in Fig.~\ref{fig:recoilenergyspectra} of Appendix~\ref{RecoilEnergy}.

Figure.~\ref{photonenergy} shows the energy spectra for the photophilic scenario, where the solid and dashed lines represent the DM flux from the target and absorber, respectively. The absorber contributions originate from the Primakoff production of scalar particles $\phi$ by the absorber-produced photons. For DUNE ND and ICARUS, these contributions are suppressed as compared to those from the target. SBND, however, shows a comparable absorber contribution, due to a larger leakage percentage, more dense absorber (iron versus aluminium), and a closer proximity while only being slightly off-axis. This feature also corroborates with the photon energy spectra as shown in Fig.~\ref{fig:PhotonFlux}. Additionally, the higher beam energies at DUNE ND and ICARUS enable more energetic DM flux compared to SBND and CCM200, although ICARUS suffers from reduced counts and lower energies due to its off-axis positioning. 

In Figs.~\ref{neutrinoenergy},~\ref{muonenergy}, and~\ref{electronenergy}, corresponding to neutrinophilic, muonphilic, and electrophilic scenarios, respectively, the scalars are produced through charged meson decays in the decay pipe, and consequently decay into DM. Here, absorber contributions are negligible because of their short interaction length as compared to the material thickness. While events from kaons decaying at rest (KDAR) can in principle have a contribution, it is subdominant due to their largely isotropic and softer DM distribution. This has also been pointed out in Ref.~\cite{Batell:2019nwo} in the context of Higgs Portal Scalars at ICARUS-NuMI. Figure.~\ref{upquarkenergy} shows the DM energy distribution for up-philic scalar mediator models, which includes contributions from two-body decay of kaons as well as proton bremsstrahlung. The absence of the CCM200 line is due to its negligible $K^+$ flux and insufficient beam energy to initiate the proton bremsstrahlung process.

Overall, DUNE ND exhibits both higher DM flux and harder energy spectra across all models, owing to its higher proton beam energy and on-axis positioning. Upon comparing the up-philic model with the remaining models for individual experiments, we observe that the higher energy peaks are obtained from proton bremsstrahlung as compared to meson decays. This is expected since the proton beam energy gets redistributed in the production of the mesons, whereas for the bremsstrahlung process, the whole beam energy is utilized.

\subsection{Number of DM particles vs mass}\label{sec:NumberSpectra}

\begin{figure}[!htbp]
    \centering
    \begin{subfigure}{0.45\textwidth} 
        \centering
        \includegraphics[scale=0.3]{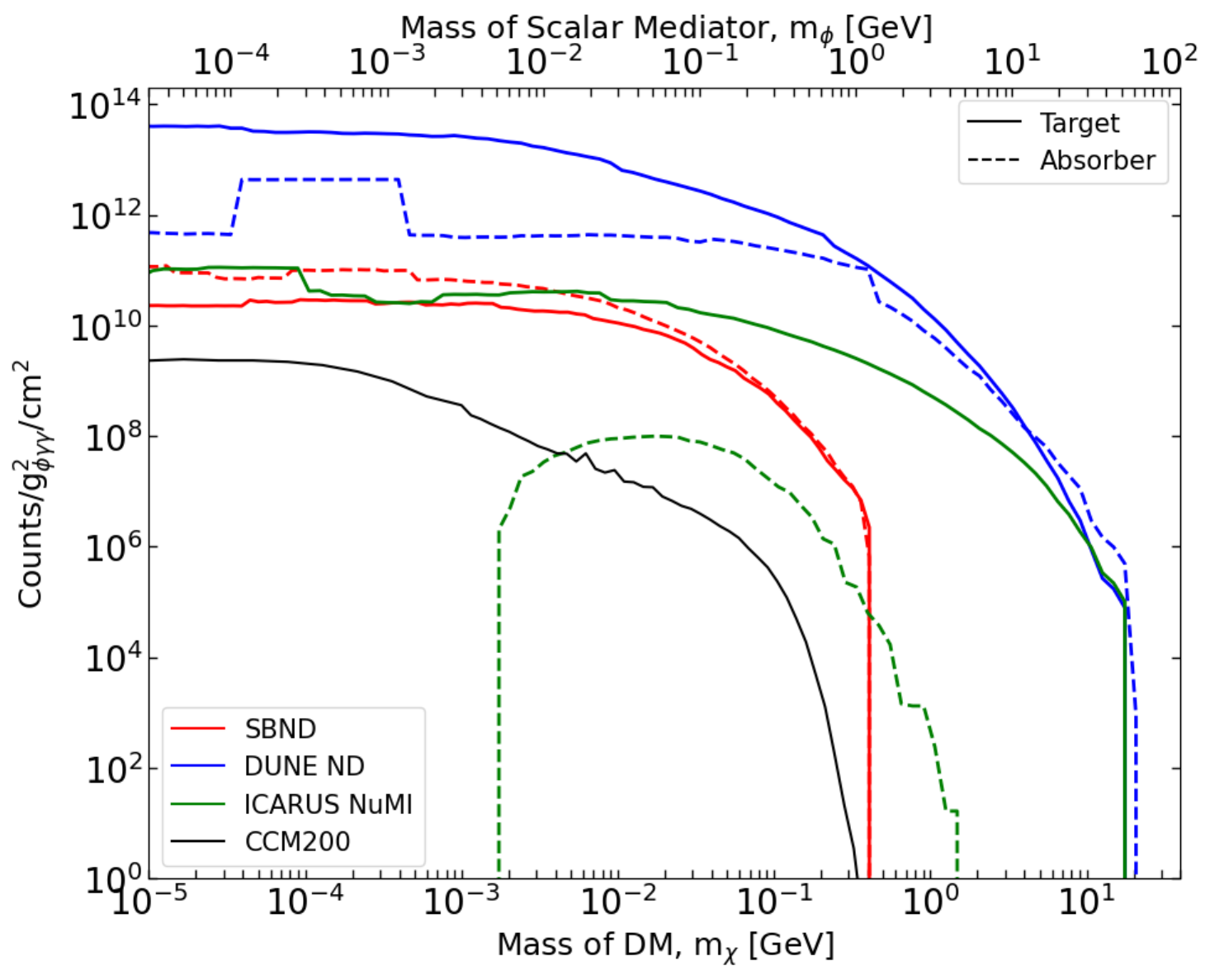}
        \caption{Photophilic model}
        \label{PhotonNumber}
    \end{subfigure}
    \begin{subfigure}{0.45\textwidth} 
        \centering
        \includegraphics[scale=0.3]{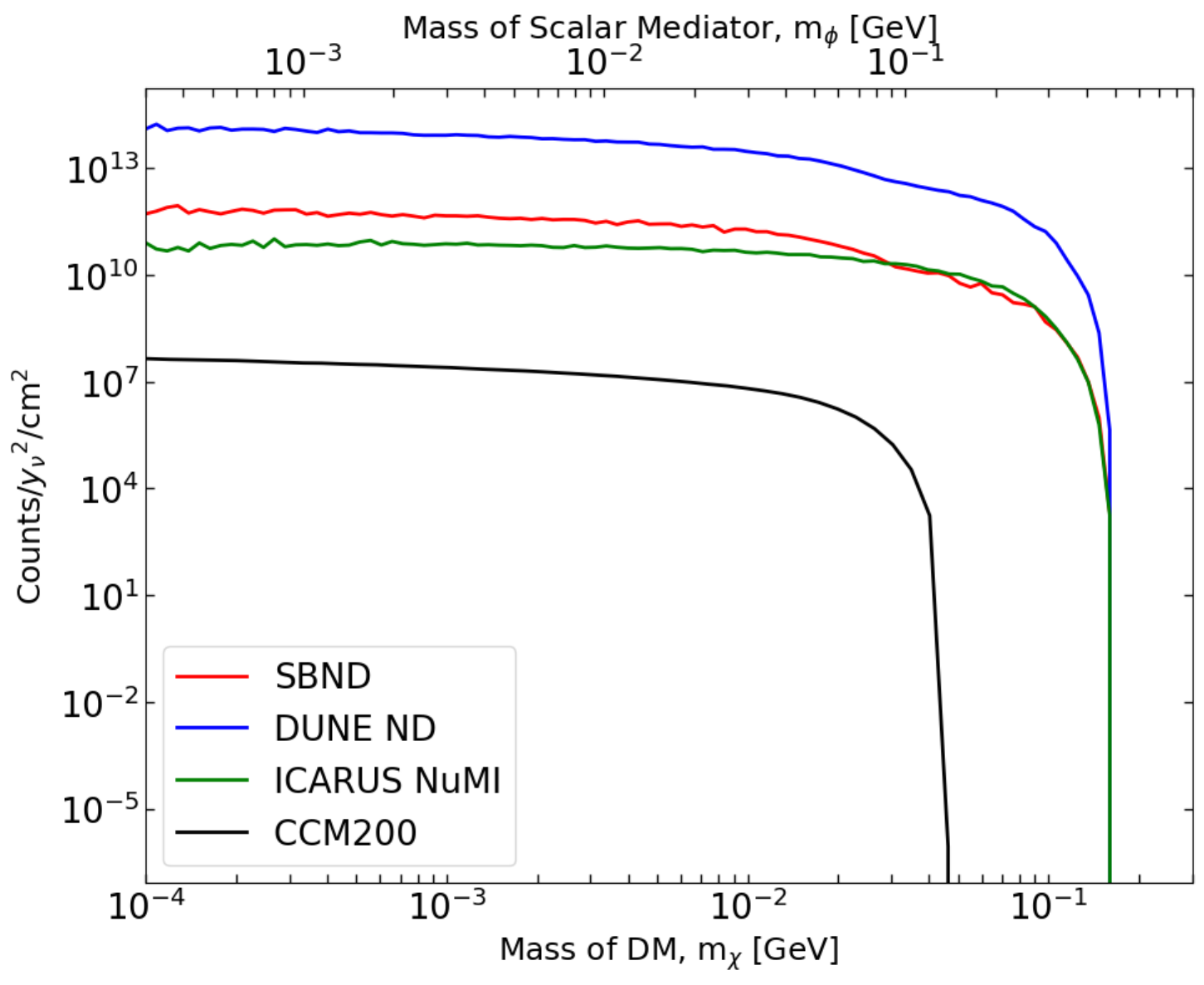}
        \caption{Neutrinophilic model}
        \label{NeutrinoNumber}
    \end{subfigure}
    \begin{subfigure}{0.45\textwidth} 
        \centering
        \includegraphics[scale=0.3]{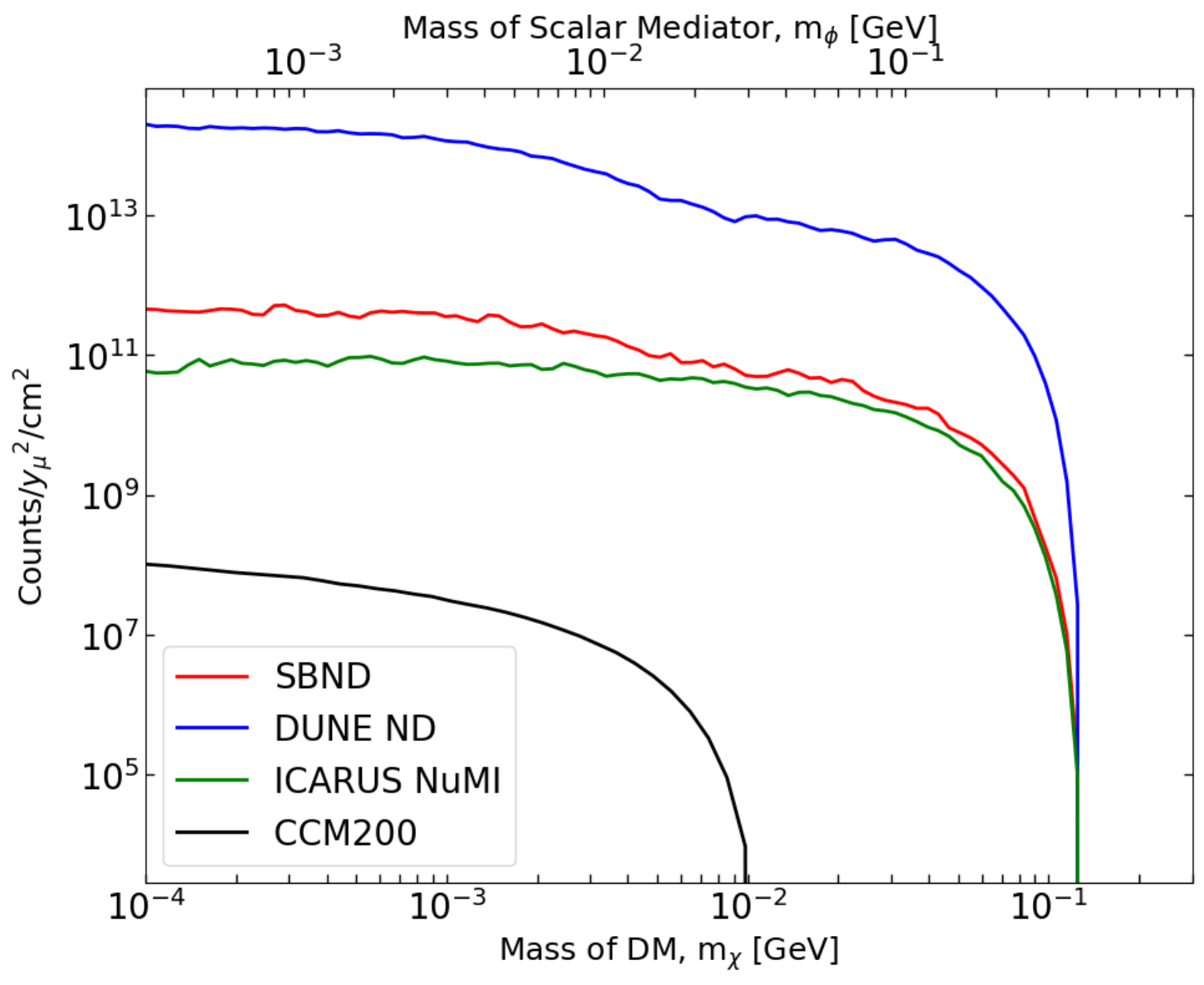}
        \caption{Muonphilic model}
        \label{MuonNumber}
    \end{subfigure}
    \begin{subfigure}{0.45\textwidth} 
        \centering
        \includegraphics[scale=0.3]{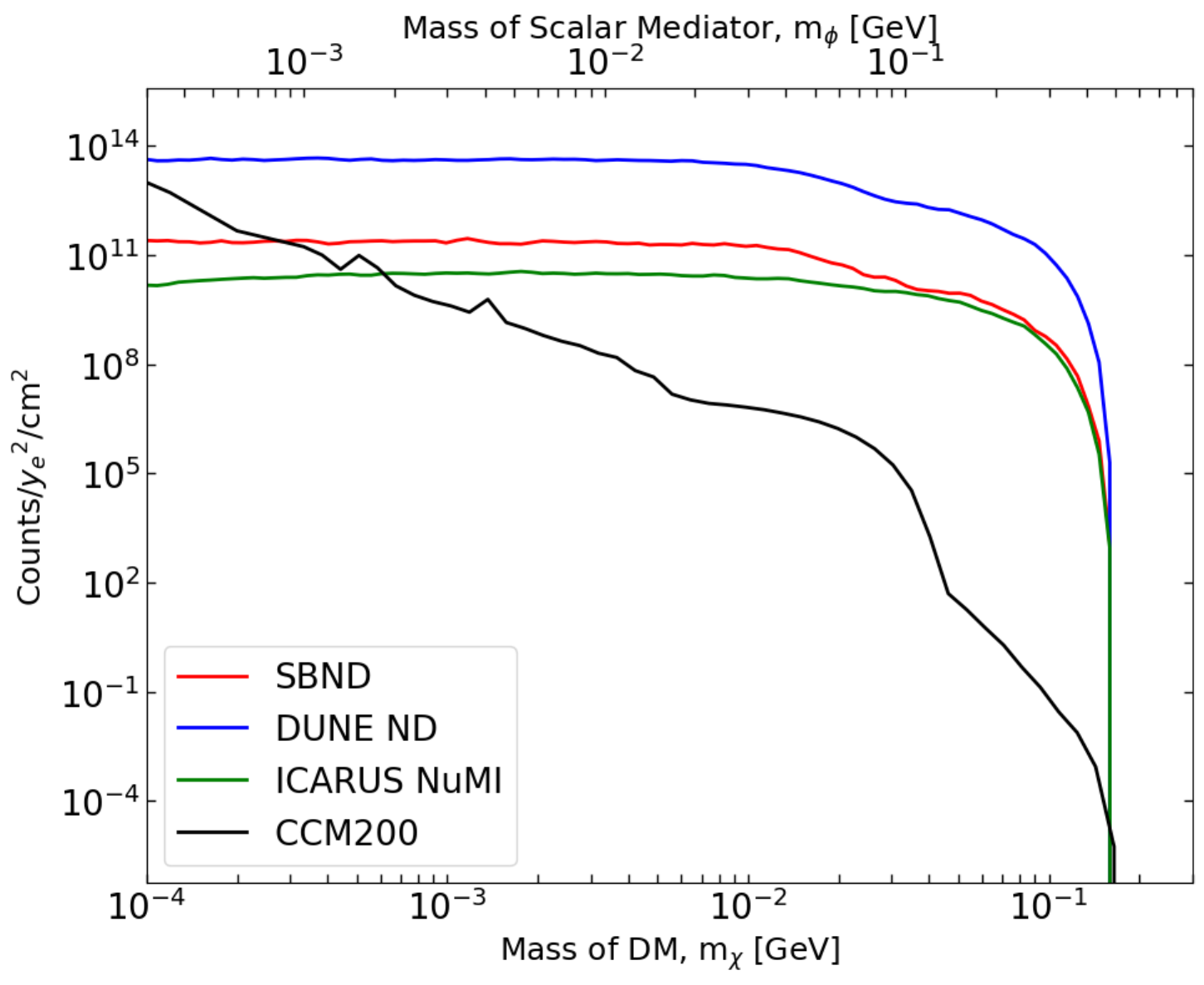}
        \caption{Electrophilic model}
        \label{ElectronNumber}
    \end{subfigure}
    \begin{subfigure}{0.45\textwidth} 
        \centering
        \includegraphics[scale=0.3]{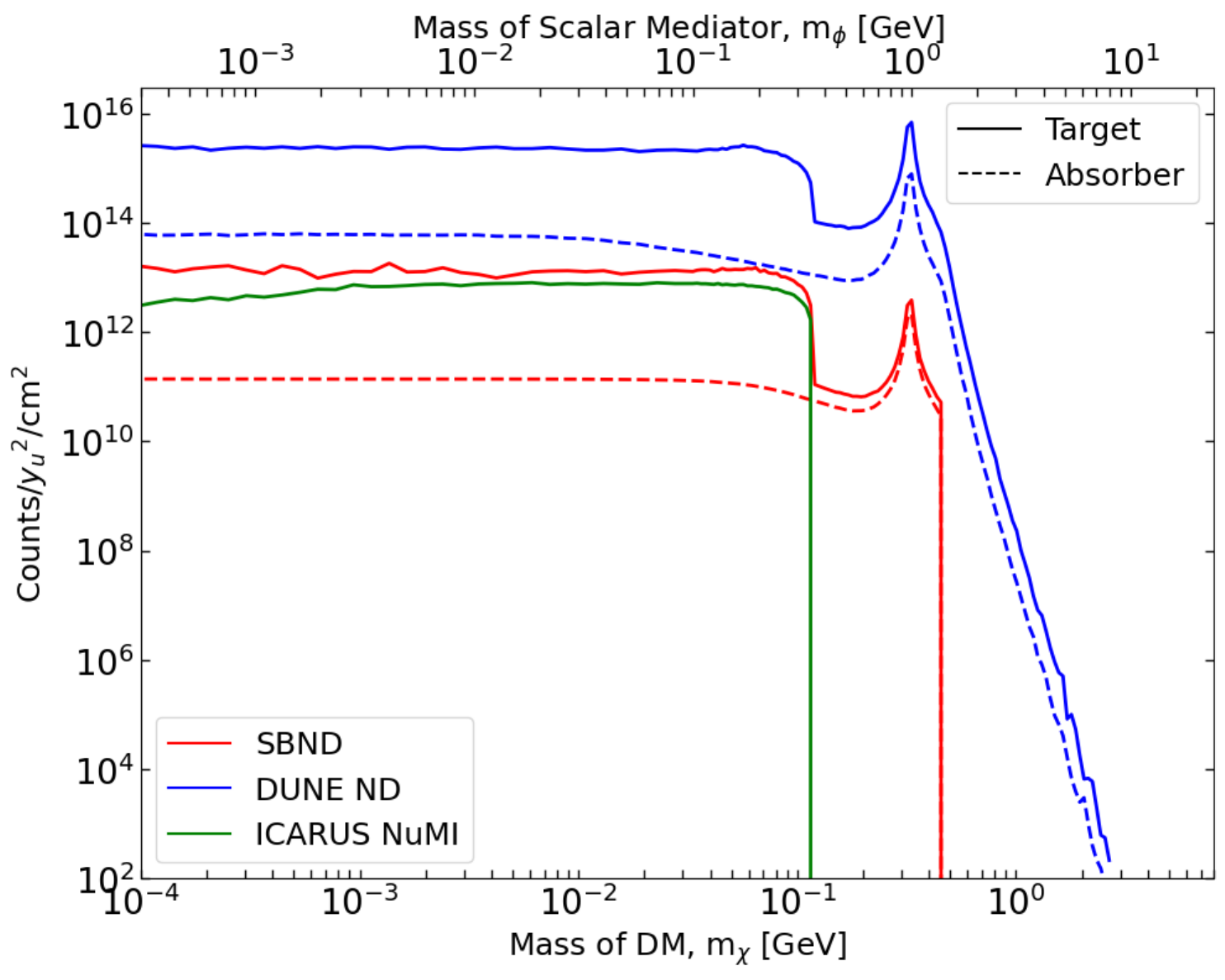}
        \caption{Up-philic model}
        \label{upquarkNumber}
    \end{subfigure}
    \captionsetup{justification=Justified, singlelinecheck=false}   
    \caption{Number spectra of DM particles entering the respective detectors as a function of DM mass (where $m_\chi = m_\phi /3$) for different models. The spectra are normalized per unit coupling squared for the corresponding model. Contributions from target and absorber are shown with solid and dashed lines, respectively, for SBND (red), DUNE ND (blue), and ICARUS-NuMI (green). CCM200 (black) only includes target contributions. Total POT and detector geometry have been accounted for in calculating the flux of DM.}
    \label{fig:NumberSpectrum}
\end{figure}

In this subsection, we will discuss Fig.~\ref{fig:NumberSpectrum}, which shows the number of DM particles entering each detector (counts per unit coupling squared per unit area) as a function of DM mass.
\par

In Fig.~\ref{PhotonNumber}, we find a correspondence between the number of DM reaching the detector and the photon energy spectra (as shown in Fig.~\ref{fig:PhotonFlux}): higher beam energies allow for production of higher energetic photons and thereby heavier scalars from the Primakoff process. Since DUNE ND and ICARUS-NuMI operate at 120~GeV, we can see that a larger number of heavier DM particles can reach the detector as compared to SBND. The ICARUS-NuMI absorber contribution is suppressed at both lower and higher masses. At lower masses, the off-forward positioning of the detector with respect to the beamline misses the lighter and therefore more forward dark matter particles. While the heavier scalars tend to be more off-forward, the limited photon statistics prevent any sizeable contribution to the DM counts. Since the photon flux is almost isotropic at the CCM200 facility, heavier masses are also able to contribute to the DM flux. 
\par

Figures.~\ref{NeutrinoNumber} and \ref{ElectronNumber} show results for scalars produced in three-body meson decays from the neutrino and electron legs, respectively. Unlike the Primakoff process, the maximum mass of the scalars that can be produced is governed by the mass difference between the initial and final state particles, rather than the parent's energy.  Therefore, a sharp drop in SBND, DUNE ND, and ICARUS occurs near $m_{\phi} \simeq m_{K^+}-m_e=0.49~$GeV. In the neutrinophilic case, CCM200 drops earlier at $m_{\phi} \simeq m_{\pi^+}-m_e=0.14~$GeV because of its negligible $K^+$ flux. In contrast, for the electrophilic scalar, CCM200 extends up to $m_{\phi} \simeq 0.5$~GeV due to contributions from additional channels such as Compton, annihilation, and electron bremsstrahlung. Figure.~\ref{MuonNumber} illustrates the muonphilic model, where the cutoff appears at $m_{\phi} \simeq m_{K^+}-m_\mu=0.39~$GeV for SBND, DUNE ND, and ICARUS-NuMI, and $m_{\phi} \simeq m_{\pi^+}-m_\mu \approx 0.03~$GeV for CCM200.
\par

In Fig.~\ref{upquarkNumber}, CCM200 yields no events due to negligible $K^+$ flux and insufficient beam energy to trigger proton bremsstrahlung as mentioned above. For ICARUS, the dominant contribution comes from $K^+ \rightarrow \pi^+ \phi$, as proton bremsstrahlung is predominantly forward and less accessible for an off-axis detector. In contrast, SBND, being comparatively forward than ICARUS, captures this contribution. DUNE benefits both from high beam energy and on-axis placement, resulting in the largest number of events. The sharp peak near $m_\phi=1~$GeV is characteristic of the $\phi$-meson resonance that manifests itself in the proton form factor in the bremsstrahlung process \cite{Foroughi-Abari:2020gju}.

\subsection{Timing Spectra of DM Induced Events}\label{sec:TimingSpectra}

\begin{figure}[!htbp]
    \centering
    \begin{subfigure}{0.45\textwidth} 
        \centering
        \includegraphics[scale=0.35]{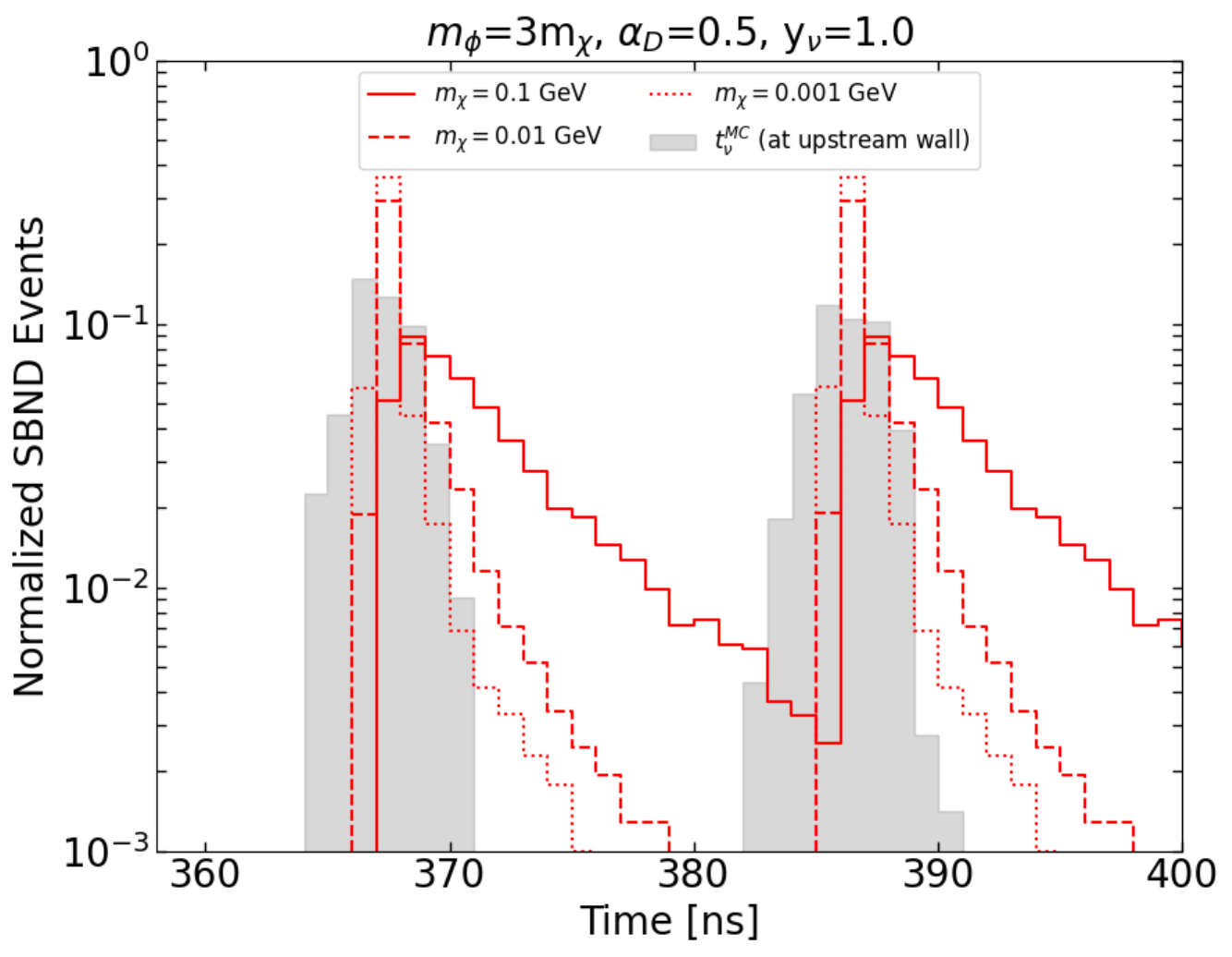}
        \caption{SBND}
        \label{SBNDTime}
    \end{subfigure}
    \begin{subfigure}{0.45\textwidth} 
        \centering
        \includegraphics[scale=0.35]{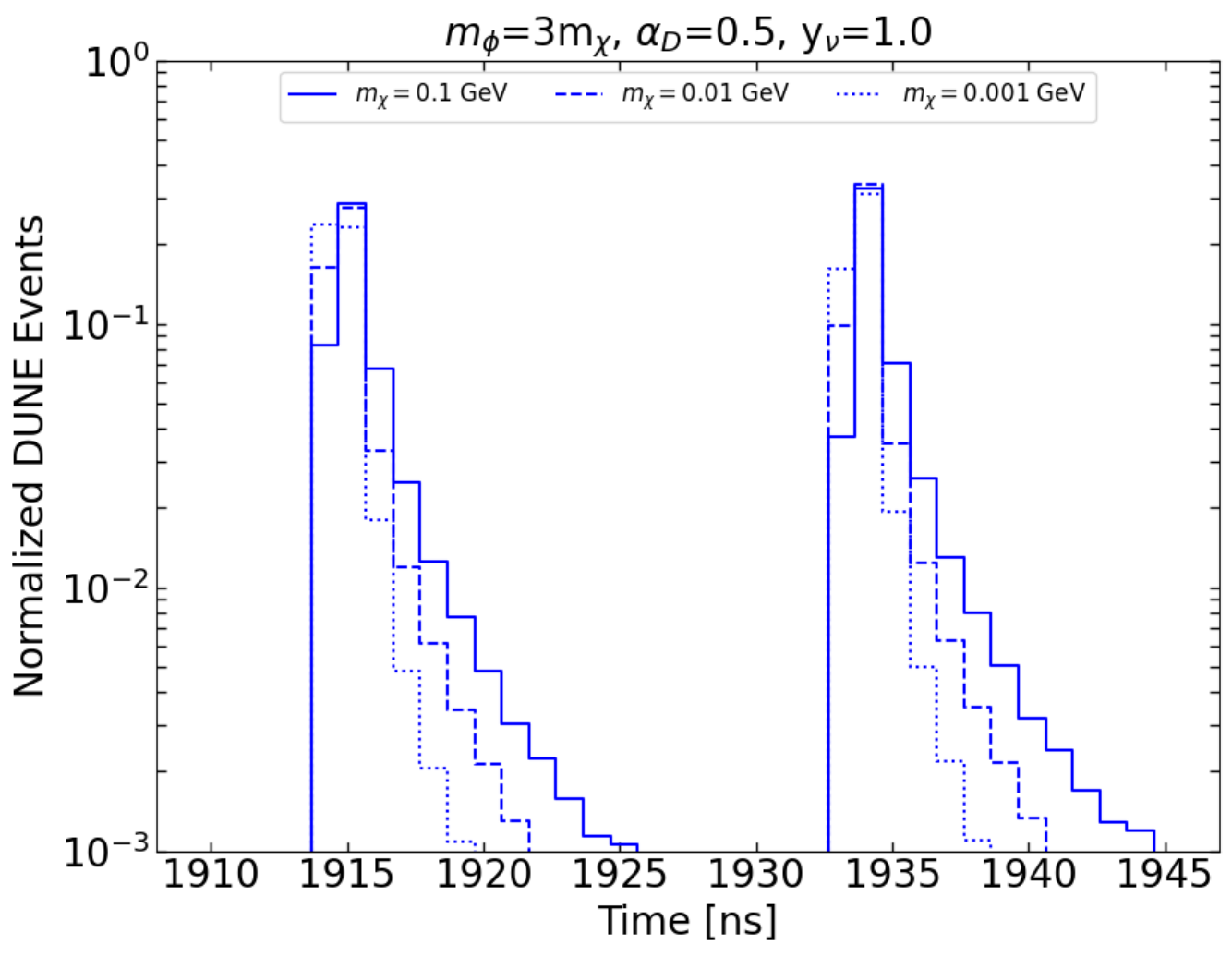}
        \caption{DUNE}
        \label{DUNETime}
    \end{subfigure}
    \begin{subfigure}{0.45\textwidth} 
        \centering
        \includegraphics[scale=0.35]{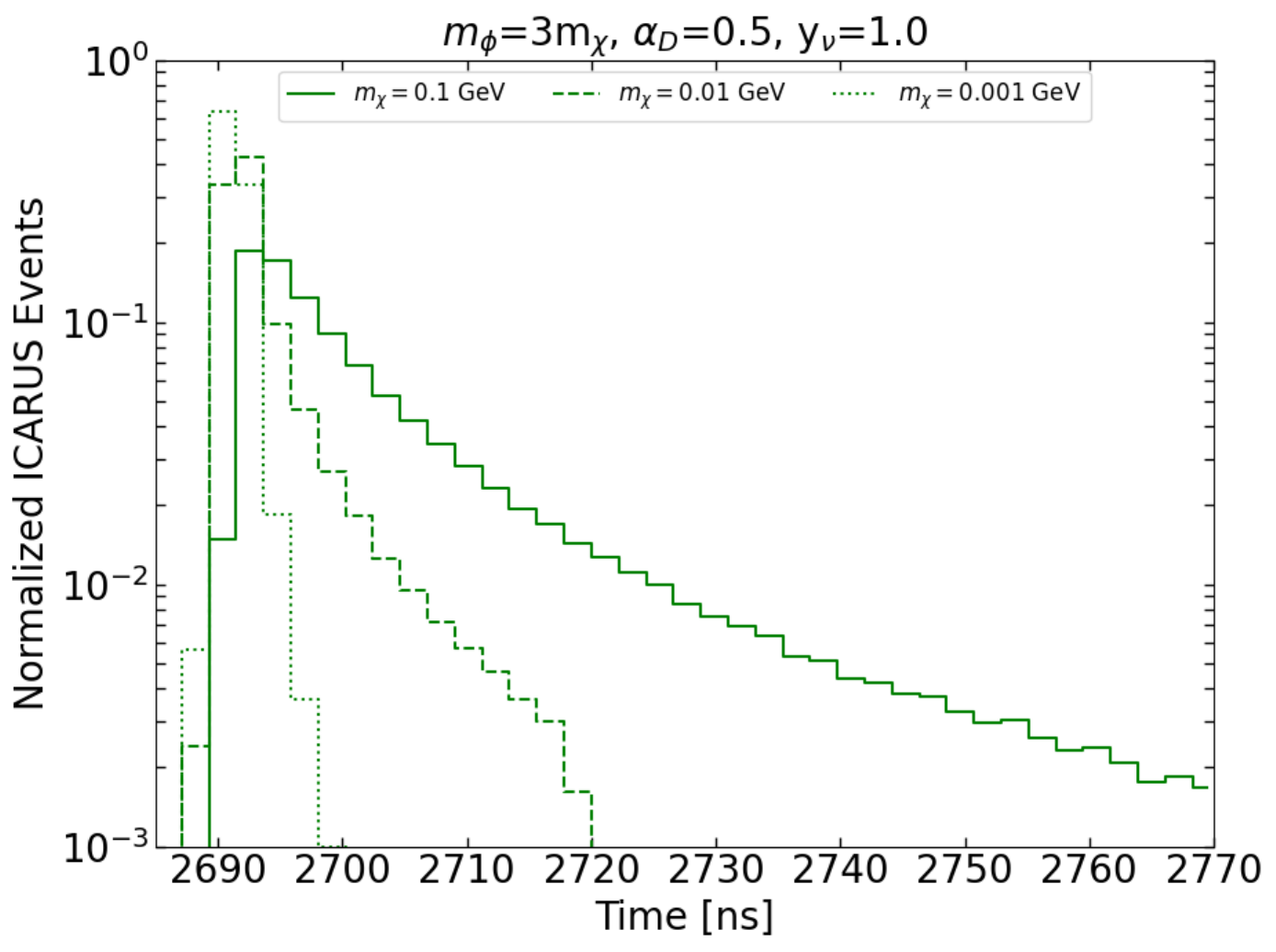}
        \caption{ICARUS}
        \label{ICARUSTime}
    \end{subfigure}
    \begin{subfigure}{0.45\textwidth} 
        \centering
        \includegraphics[scale=0.35]{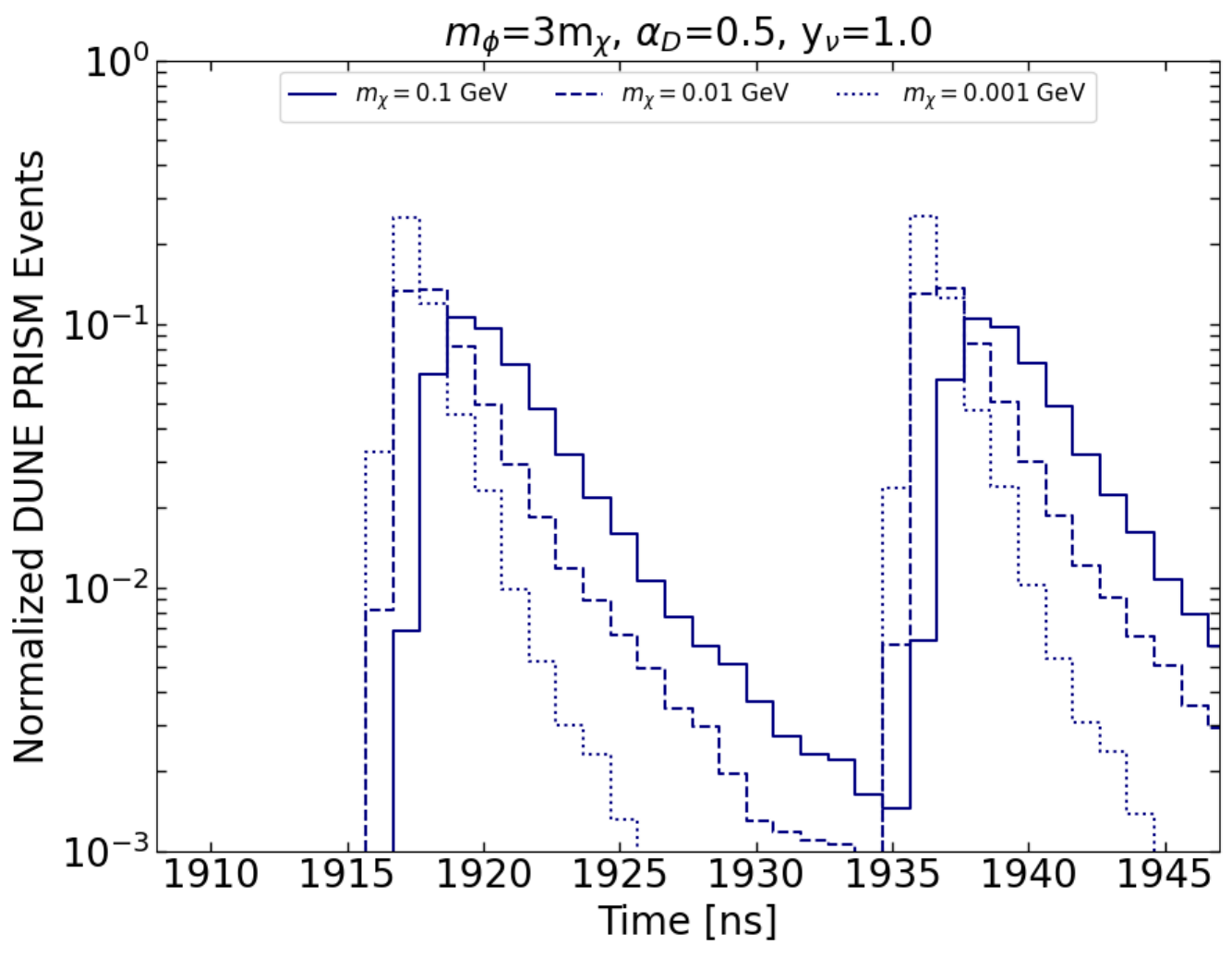}
        \caption{DUNE PRISM}
        \label{DUNEPRISMTime}
    \end{subfigure}
    \captionsetup{justification=Justified, singlelinecheck=false}   
    \caption{Timing spectra of DM-induced events in (a) SBND, (b) DUNE ND, (c) ICARUS-NuMI, and (d) DUNE PRISM. Solid, dashed, and dotted lines correspond to $m_\chi = 0.1$, $0.01$, and $0.001$ GeV, respectively. Only events producing a final-state photon above the detection threshold are shown. Time $t = 0$ marks the production of SM-charged mesons at the respective sources. The $x$-axis represents the total travel time to the detector. The shaded gray band in panel (a) shows the expected timing profile of BNB $\nu_\mu$ flux at SBND. Spectral width reflects both DM mass and detector geometry, with broader distributions for off-axis distant detectors and heavier DM. To enable a direct comparison across spectra, each distribution is normalized such that the total counts, summing up the contributions in each bin, result in unity.}
    \label{fig:TimingSpectrum}
\end{figure}

Figure.~\ref{fig:TimingSpectrum} displays the distribution of DM arrival times at SBND, DUNE ND (and DUNE PRISM), and ICARUS-NuMI for $m_\chi = 0.1$ GeV (solid), $0.01$ GeV (dashed), and $0.001$ GeV (dotted). In addition to the detectors discussed in the previous sections, we include timing results for DUNE PRISM (represented by navy blue), which is located $2.84^\circ$ off-axis with respect to the LBNF beamline. For comparison, we overlay the expected arrival time distribution of neutrino events at SBND, produced at the BNB~\cite{SBND:2024vgn, SearchesforBSM, SearchesforBSM1}. The mentioned references demonstrate the feasibility of nanosecond-level timing resolution at SBND, which is critical for separating BSM signals from SM neutrino backgrounds. 
\par 

In each timing spectrum, $t = 0$ is defined as the time at which the proton beam spill impinges on the respective targets. This also approximately coincides with the time at which SM charged mesons are produced. The proton beam at the BNB~\cite{SBND:2024vgn}, NuMI~\cite{numirookiebook}, and the LBNF~\cite{kevinwood} source has a pulsed structure, with a periodicity of $19~\text{ns}$.  
SM neutrinos, traveling nearly at the speed of light, generate early-arriving background events, making them temporally distinguishable from DM-induced events (in our case, the $2 \rightarrow 3$ process) arising from the heavier and therefore slower DM particles. This distinction is vital for separating DM signals from neutrino-induced processes, such as those discussed in Ref.~\cite{Gehrlein:2025tko}. This has also been demonstrated in the context of Heavy Neutral Leptons (HNLs), such as that in Ref.~\cite{Dutta:2025npn}.
\par

In these plots, we consider DM from neutrinophilic scalar mediators, as outlined in Sec.~\ref{sec:production}, where production occurs via Primakoff processes and 3-body decays of charged mesons; the latter being the dominant process. Heavier DM particles result in broader timing distributions because they are less relativistic and therefore take longer to reach the detector. In contrast, lighter DM particles are more energetic and arrive sooner, leading to narrower spectra.
\par

This effect is particularly evident in DUNE ND (Fig.~\ref{DUNETime}), where the spectrum is more concentrated due to its on-axis configuration, which suppresses off-forward DM contributions, especially those from heavier and hence slower particles with broader angular distributions. Conversely, ICARUS-NuMI (Fig.~\ref{ICARUSTime}) exhibits the broadest spectrum, consistent with its highly off-axis positioning: $5.56^\circ$ with respect to the target and $42.71^\circ$ with respect to the absorber, as illustrated in Fig.~\ref{fig:ICARUSSchematic}. A comparison between DUNE ND and DUNE PRISM (Fig.~\ref{DUNEPRISMTime}) further highlights this geometry dependence. DUNE PRISM is positioned 28.5 meters off-axis, resulting in a broader timing profile than DUNE ND.
\par

Due to the absence of official neutrino timing distributions for DUNE ND, DUNE PRISM, and ICARUS-NuMI, we do not include shaded background neutrino spectra similar to those used for SBND. This plot does not include the timing spectra at CCM200 as they lie within the prompt neutrino window. This enables the discrimination of signals from neutron-induced backgrounds, where the majority of the latter lies in the delayed window. Additionally, the photon from the $2\to 3$ scattering of DM is characteristically much higher in energy than any neutrino/neutron induced backgrounds at the 800~MeV facility. This serves as an additional handle for distinguishing DM-induced events from backgrounds at CCM200. 
\par

\subsection{Spatial Distributions of DM}

\begin{figure}[!htbp]
    \centering
    \begin{subfigure}{0.4\textwidth} 
        \centering
        \includegraphics[scale=0.3]{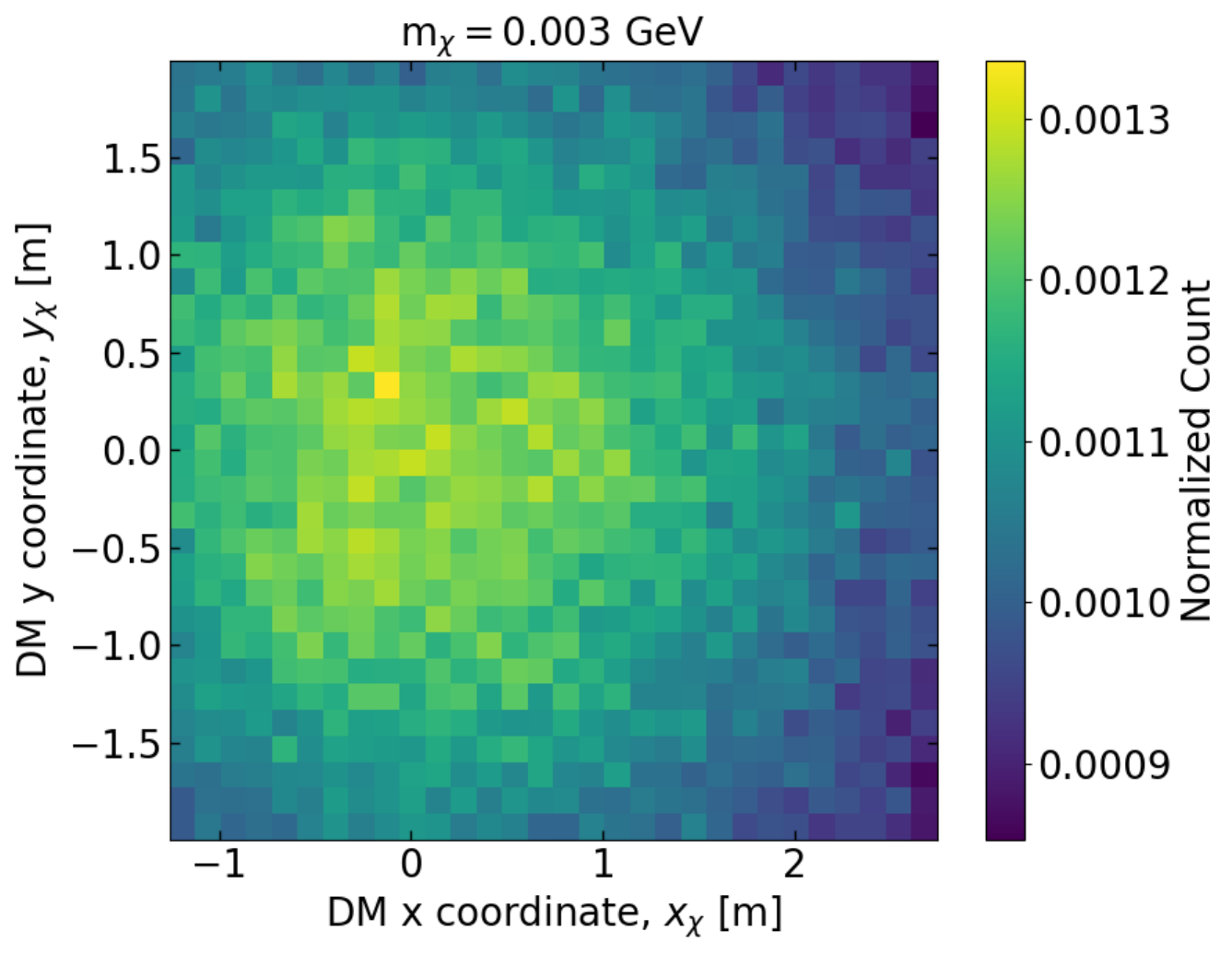}
        \label{meson0.003spatial}
    \end{subfigure}
    \begin{subfigure}{0.4\textwidth} 
        \centering
        \includegraphics[scale=0.3]{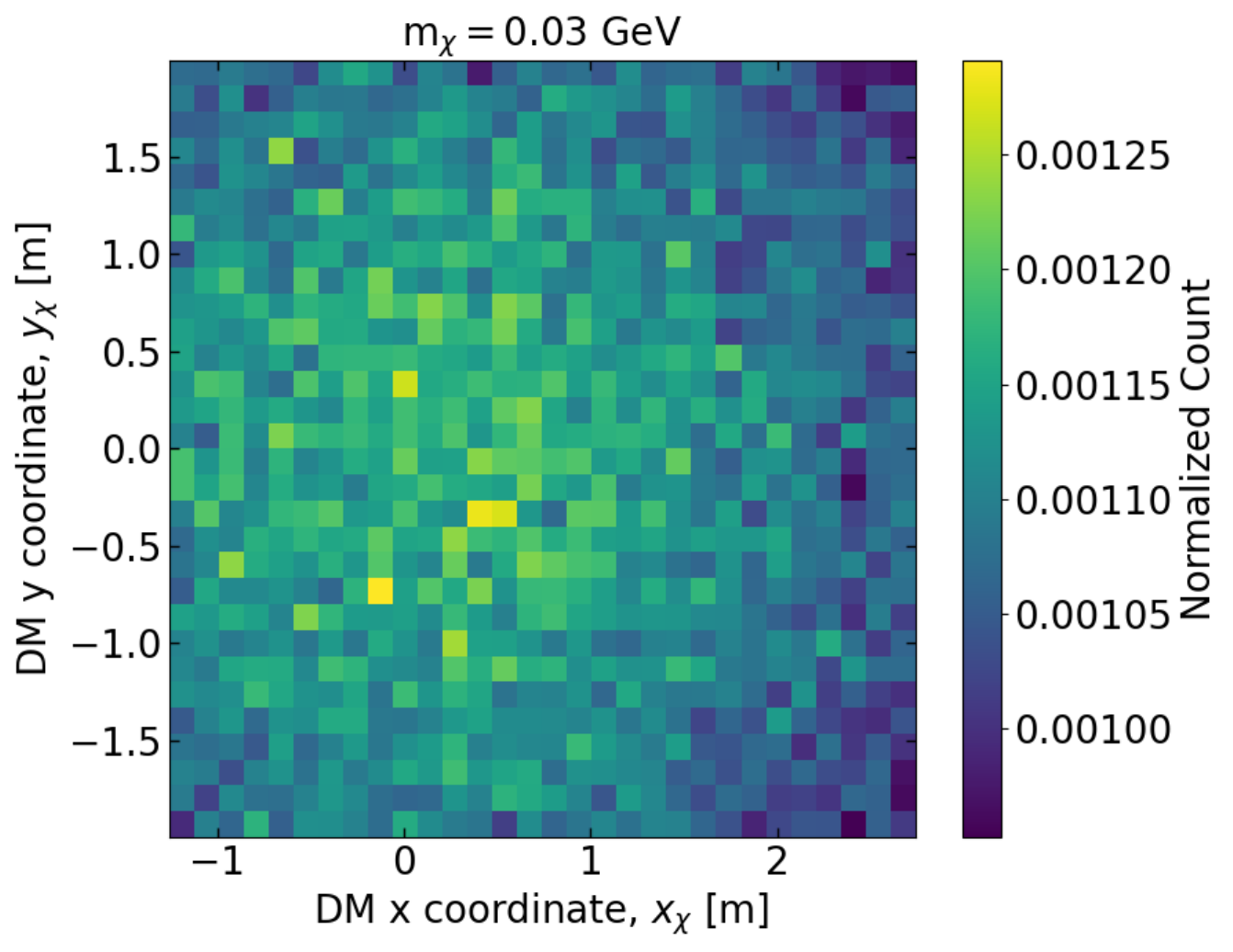}
        \label{meson0.03spatial}
    \end{subfigure}
    \begin{subfigure}{0.4\textwidth} 
        \centering
        \includegraphics[scale=0.3]{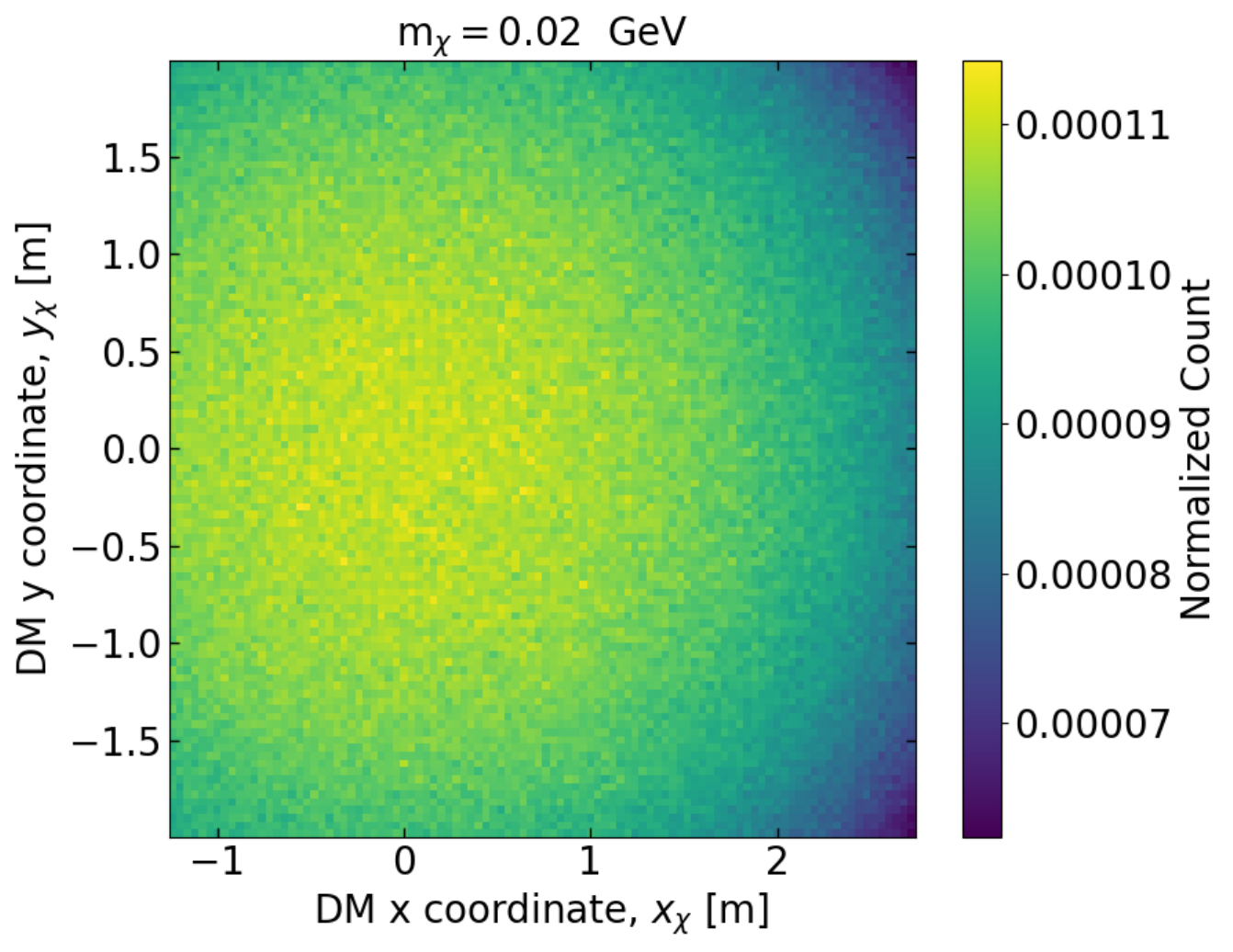}
        \label{brem0.02spatial}
    \end{subfigure}
    \begin{subfigure}{0.4\textwidth} 
        \centering
        \includegraphics[scale=0.3]{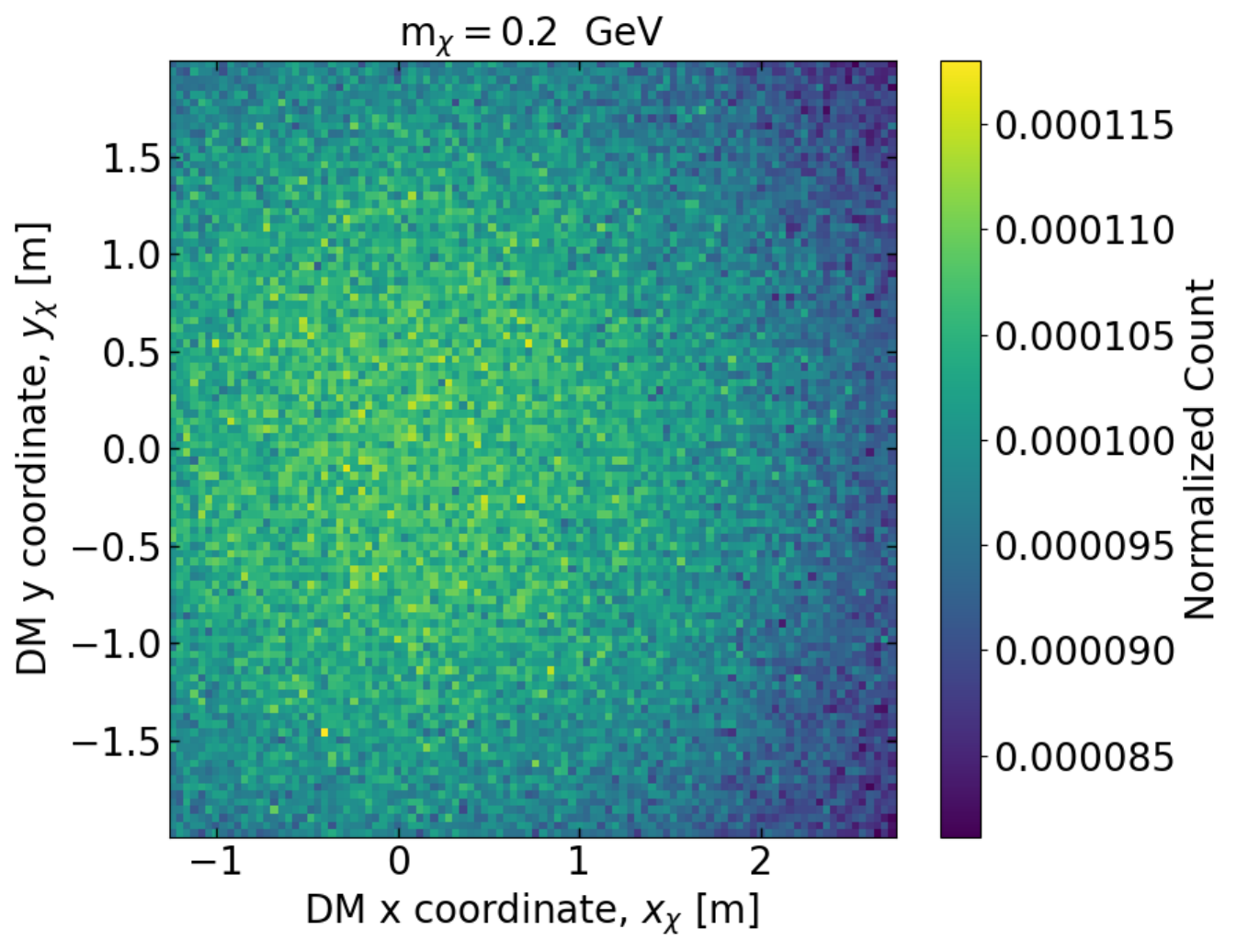}
        \label{brem0.2spatial}
    \end{subfigure}

    \captionsetup{justification=Justified, singlelinecheck=false}   
    \caption{Normalized spatial distribution of DM particles at the front face of the SBND detector produced from (Top row): Charged meson decays for neutrinophilic model and (Bottom row:) Proton bremsstrahlung for the upphilic model. The two figures in the top panel denote (Left:) $m_\chi=0.003~$GeV and (Right:) $m_\chi=0.03~$GeV DM masses. Similarly, the two plots in the bottom panel denote (Left:) $m_\chi=0.02~$GeV and (Right:) $m_\chi=0.2~$GeV DM masses.}
    \label{fig:SpatialDistribution}
\end{figure}

\begin{figure}[!htbp]
    \centering
    \begin{subfigure}{0.4\textwidth} 
        \centering
        \includegraphics[scale=0.3]{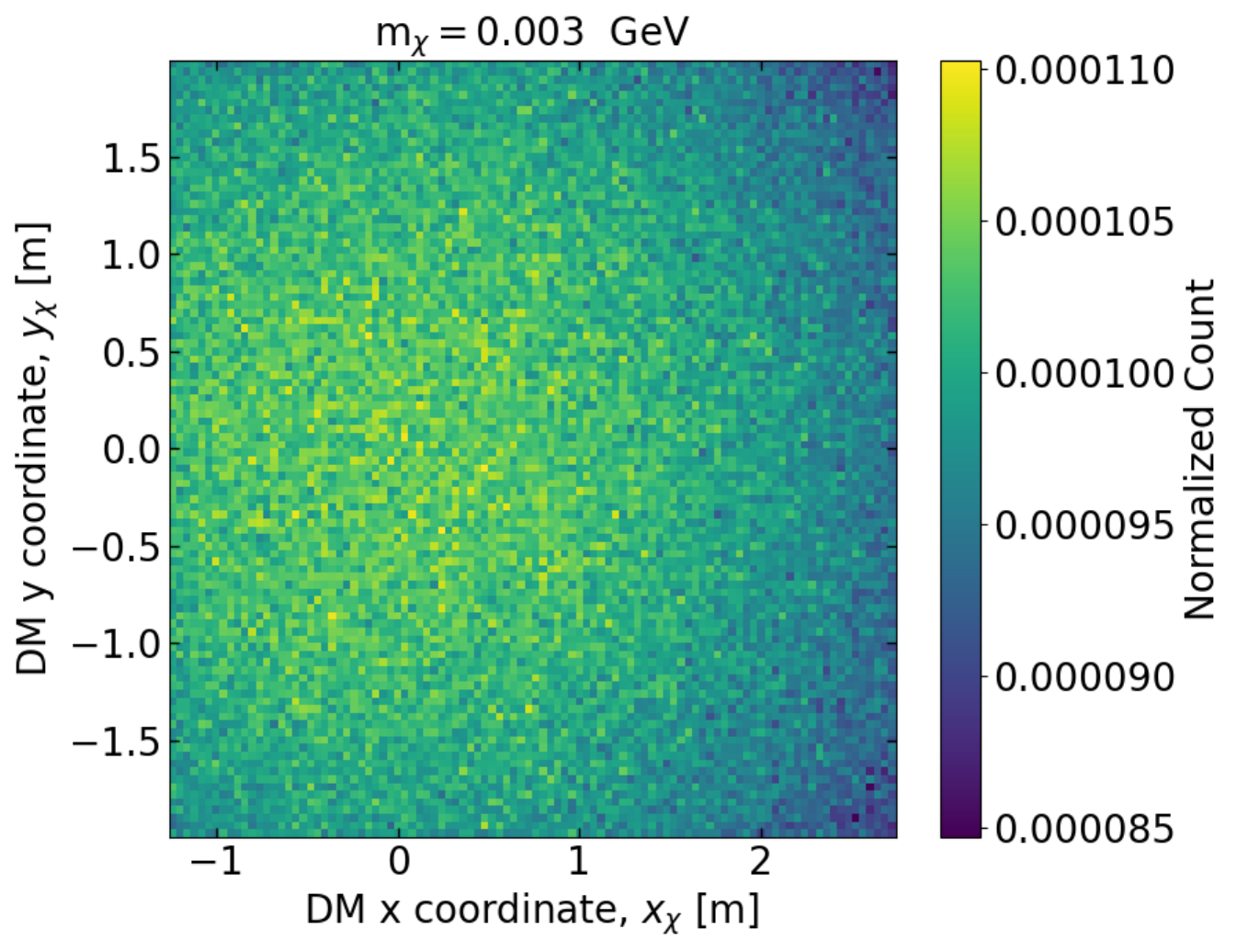}
        \label{brem0.003spatialtarget}
    \end{subfigure}
    \begin{subfigure}{0.4\textwidth} 
        \centering
        \includegraphics[scale=0.3]{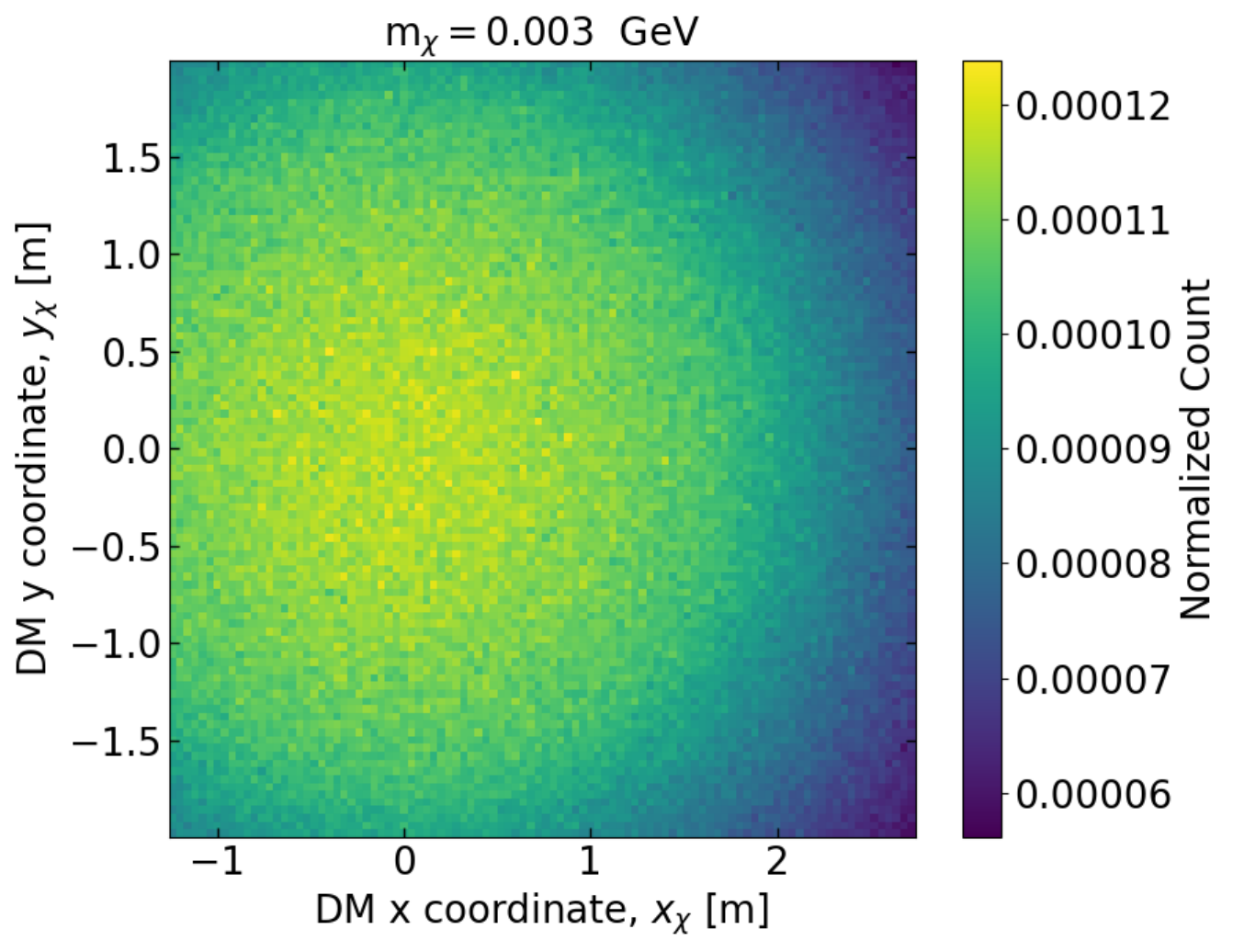}
        \label{brem0.003spatialdump}
    \end{subfigure}

    \captionsetup{justification=Justified, singlelinecheck=false}   
    \caption{Normalized spatial distribution of DM particles at the front face of the SBND detector, produced via proton bremsstrahlung in the target (left) and in the absorber (right). Both distributions correspond to $m_\chi=0.003~$GeV.}
    \label{fig:dumpvstarget}
\end{figure}

Figure.~\ref{fig:SpatialDistribution} shows the spatial distribution of DM particles at the SBND detector, where the color map represents the normalized counts for a given coordinate $(x,y)$ in meters. The top and bottom panels illustrate the neutrinophilic and up-philic scenarios, respectively. In the neutrinophilic scalar model, DM is produced via the three-body decay of charged mesons, whereas for the up-philic model, production occurs exclusively through proton bremsstrahlung. For the neutrinophilic case (top panel), the left and right subplots correspond to increasing DM masses of $0.003$ and $0.03$~GeV, respectively. Conversely, in the up-philic scenario (bottom panel), the left and right subplots correspond to DM masses of $0.02$ and $0.2$~GeV, respectively. The coordinate point (0 m, 0 m) in the plots represents the beam axis rather than the detector center, which is located at (0.74 m, 0 m). All the plots in this subsection are normalized to facilitate a clearer comparison of the distribution shapes rather than the absolute event counts. To gain insights into the total signal yields from each production mechanism, one may refer to the number count plots presented in Sec.~\ref{sec:NumberSpectra} or the sensitivity plots in Sec.~\ref{sec:sensitivity}.
\par

The two model scenarios highlight the distinct spatial distributions arising from their respective production mechanisms. While the mesons are focused by the magnetic horn, their subsequent decays into scalars and DM result in a broader angular spread of DM along the beamline. In contrast, the kinematics from proton bremsstrahlung result in more forward-peaked $Z'$, and hence DM, than meson decays. The higher energy of the protons ($8~$GeV), compared to the mesons (peaking around $500~$MeV), produces scalars and consequently DM that are more boosted and strongly collimated along the beamline. This effect is evident when comparing the right subplot of the top row with the left subplot of the bottom row, which corresponds to similar DM masses. We also notice that heavier DM is more widespread than lighter DM for a given production process, as seen when comparing the left and right subplots. This is because heavier scalars have a smaller boost factor ($E/m$) compared to lighter ones, which results in less boosted DM.

\par

Figure.~\ref{fig:dumpvstarget} presents the spatial distributions of DM particles at the front face of the SBND detector produced via proton bremsstrahlung in the target (left) and the absorber (right) scenarios. The absorber contribution is notably more forward-peaked compared to that from the target. This behavior is expected since the absorber is closer to the detector, leaving the DM particles less time to spread transversely.
\par

The spatial distribution thus provides an additional handle for distinguishing both the source (target vs absorber) and the production mechanism of the DM particles.

\subsection{Characteristic Energy and Angular Spectra of the Final State Photon}

\begin{figure}[!htbp]
    \centering
    \begin{subfigure}{0.45\textwidth} 
        \centering
        \includegraphics[scale=0.35]{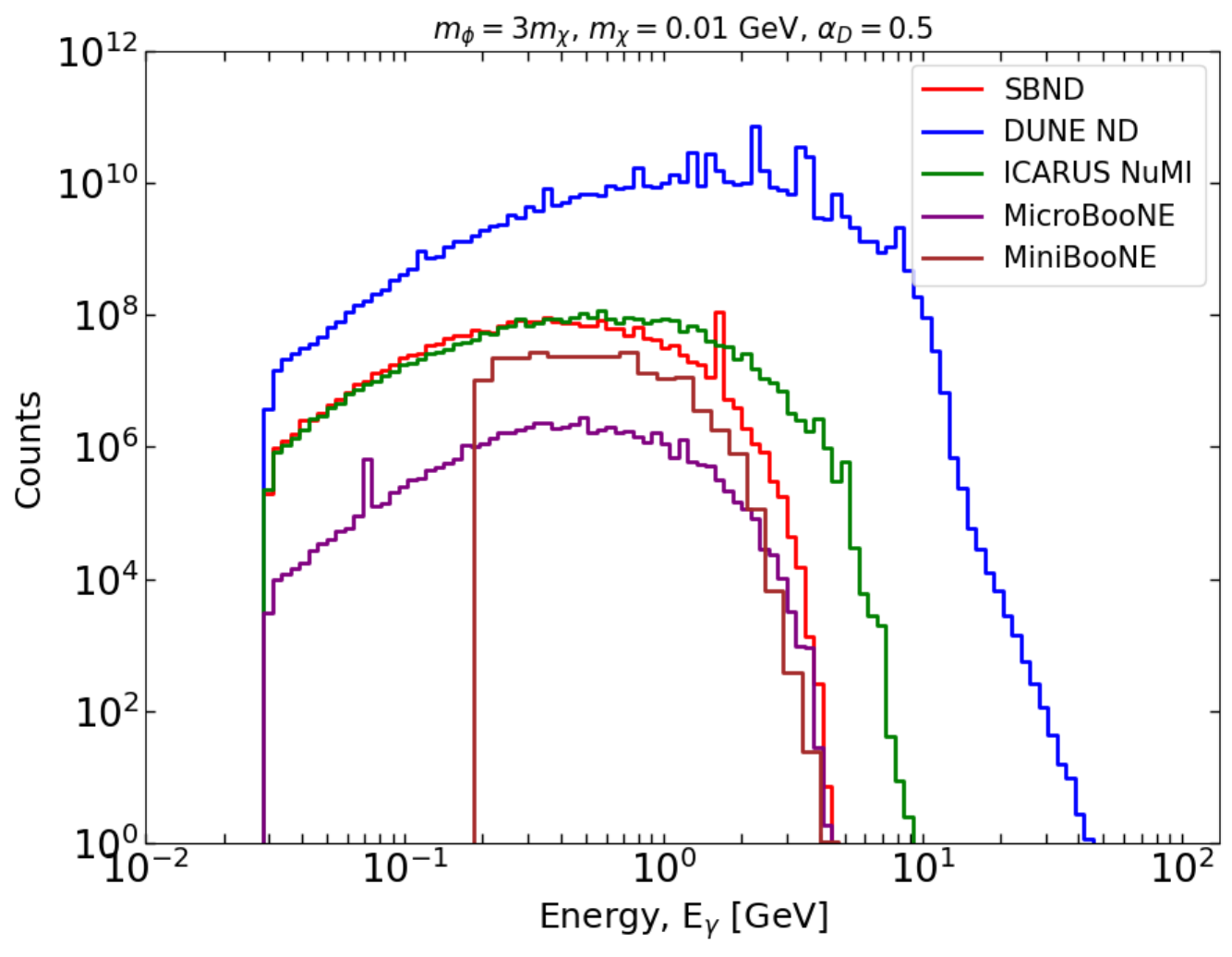}
        \caption{Energy Spectra}
        \label{energyspectra}
    \end{subfigure}
    \begin{subfigure}{0.45\textwidth} 
        \centering
        \includegraphics[scale=0.35]{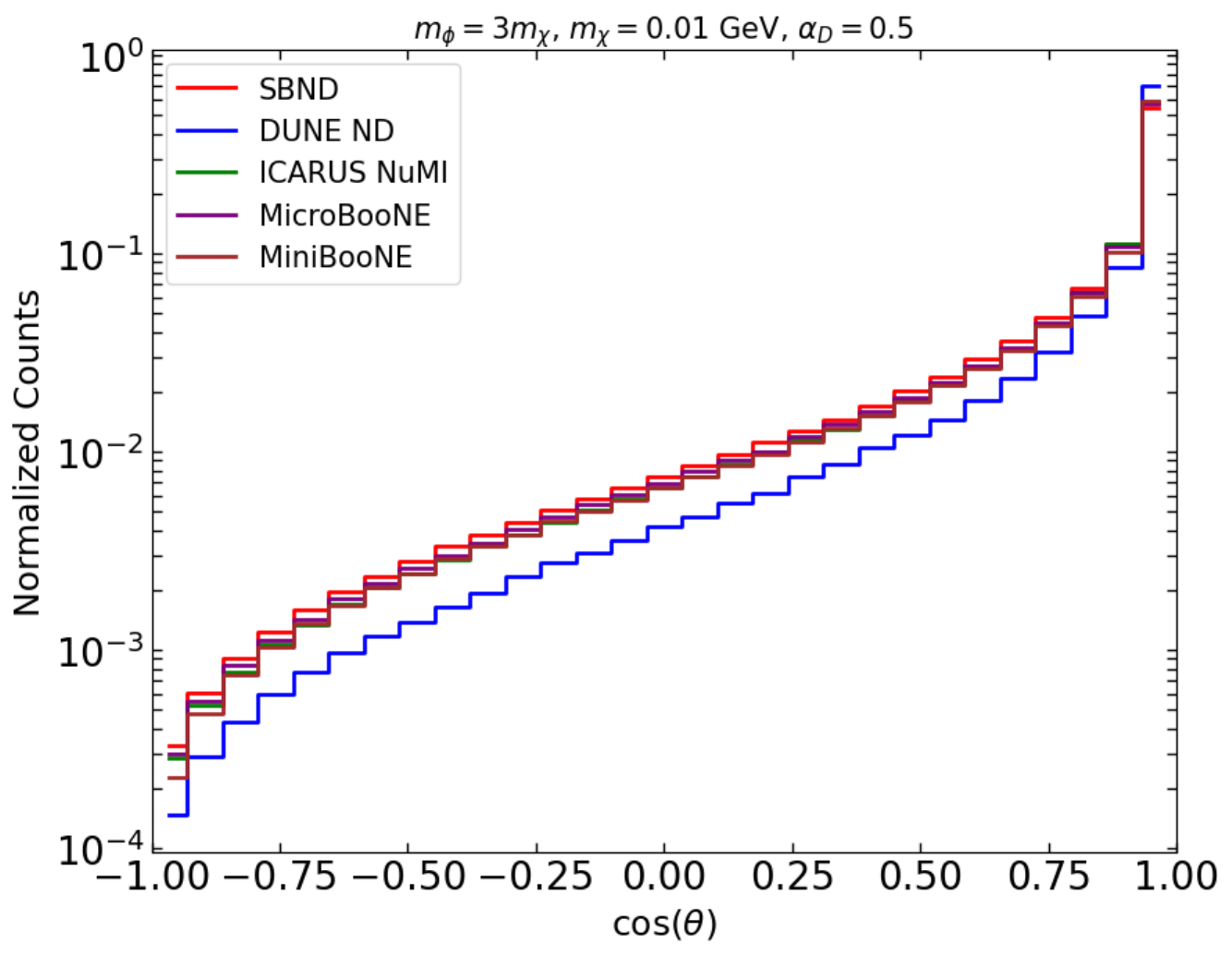}
        \caption{Angular Spectra}
        \label{angularspectra}
    \end{subfigure}
    \captionsetup{justification=Justified, singlelinecheck=false}   
    \caption{Comparison of the final-state (a) photon energy and (b) photon angular spectra across multiple experiments: SBND, DUNE ND, ICARUS-NuMI, MicroBooNE, and MiniBooNE, within the framework of the neutrinophilic model. The spectra are simulated for the benchmark parameters $m_\phi = 30$~MeV, $m_\chi = 10$~MeV, $g_{\phi \gamma \gamma} = 1.0$~GeV$^{-1}$, $y_\nu = 5 \times 10^{-3}$, and $\alpha_D = 0.5$. The photon energy $E_\gamma$ is shown on a logarithmic scale, highlighting differences in spectral shapes and event rates across detectors. The angular spectra are plotted with normalized counts to showcase the forwardness of the final state photon with respect to the beamline for each experiment.}
    \label{fig:finalstatephotonspectra}
\end{figure}

In this subsection, we examine how the characteristic energy and angular spectra of the final-state photon in the $2 \rightarrow 3$ scattering process, which constitutes our signal, vary across different experiments. The final state photon with a distinct energy and angular spectrum can be leveraged to distinguish DM interactions from SM background processes.
\par

Figure.~\ref{fig:finalstatephotonspectra} presents the (a) energy and (b) angular spectra of the final-state photon for the neutrinophilic model, after convoluting the DM spectra with the $2\to 3$ cross section. The color coding follows the conventions described earlier, with additional curves included for MicroBooNE (purple) and MiniBooNE (brown). As discussed earlier in Sec.~\ref{sec:EnergySpectra}, we can safely eliminate most of the backgrounds at CCM200 with a 1~MeV energy cut due to improved background rejections by shielding and Cherenkov light, i.e., 1/100th of that considered in Ref.~\cite{CCM:2021jmk}. Therefore, we do not show the CCM200 spectra in this plot. Here, we adopt the following benchmark parameters: $m_\phi = 3m_\chi = 30$~MeV, $g_{\phi \gamma \gamma} = 1.0$~GeV$^{-1}$, $y_\nu = 5 \times 10^{-3}$, and $\alpha_D = 0.5$. For this choice of parameters, we find that the peak photon energy lies around 600~MeV for SBND, MicroBooNE, and MiniBooNE, 1 GeV for ICARUS-NuMI, and 5~GeV for DUNE ND, as shown in Fig.~\ref{energyspectra}. Also, as shown in Fig.~\ref{angularspectra}, we observe that DUNE ND would observe most forward photons because of its distance and on-axis placement. Other experiments like ICARUS-NuMI, MicroBooNE, MiniBooNE, and SBND follow, providing slightly off-forward photons as compared to DUNE ND. Although one anticipates that ICARUS-NuMI, the most distant detector, would observe the most forward-going signal, its off-axis placement makes it more susceptible to softer DM, resulting in the photons' off-forward nature.
These characteristically high-energy and forward signals constitute a clean and distinguishable signal.
\par

\begin{figure}[!htbp]
    \centering
    \begin{subfigure}{0.45\textwidth} 
        \centering
        \includegraphics[scale=0.35]{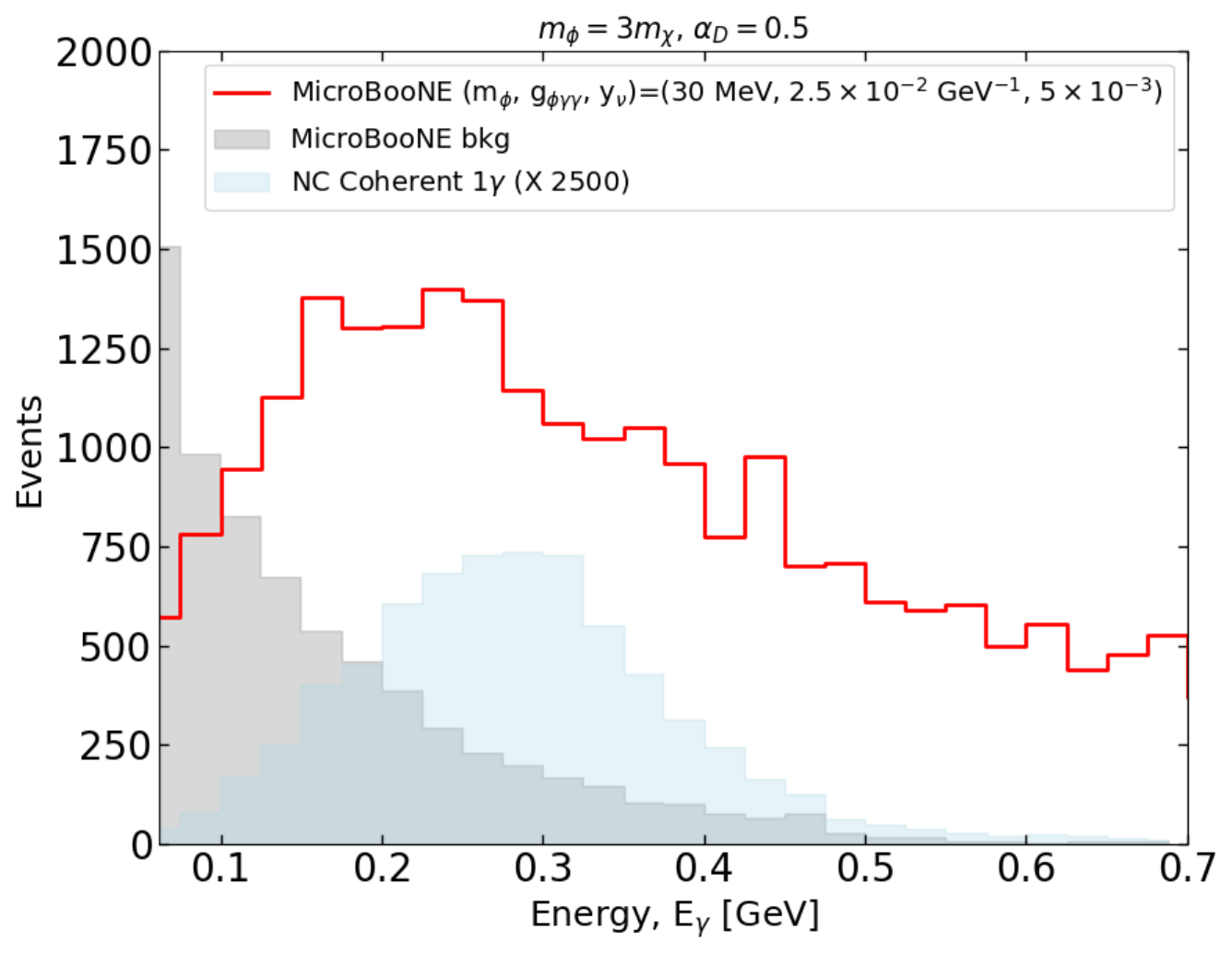}
        \caption{Energy Spectra}
        \label{energyspectraMicroBooNE}
    \end{subfigure}
    \begin{subfigure}{0.45\textwidth} 
        \centering
        \includegraphics[scale=0.35]{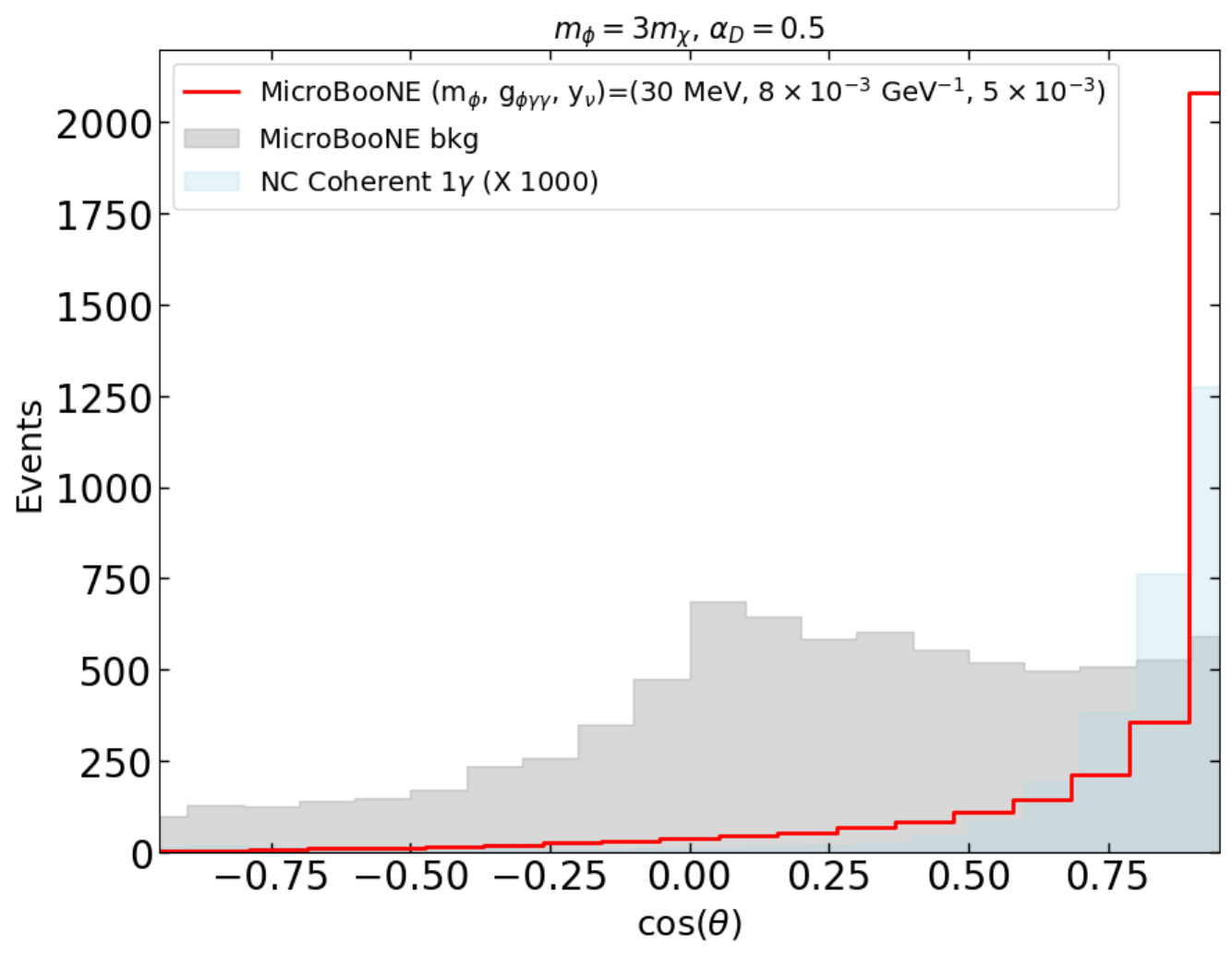}
        \caption{Angular Spectra}
        \label{angularspectraMicroBooNE}
    \end{subfigure}
    \captionsetup{justification=Justified, singlelinecheck=false}   
    \caption{(a) Energy and (b) Angular spectrum of the final-state photon in MicroBooNE (POT=$6.87 \times 10^{20}$) in the neutrinophilic model (red line), shown in comparison with the MicroBooNE background (gray shaded region) from Ref.~\cite{MicroBooNE:2025rsd}. The NC coherent $1\gamma$ background (light blue shaded region) is scaled by (a) $\times2500$ and (b) $\times 1000$ for visibility, respectively. The signal spectrum corresponds to the parameter choice $m_\phi = 30$~MeV, $m_\chi = 10$~MeV, $y_\nu = 5 \times 10^{-3}$, and $\alpha_D = 0.5$.}
    \label{fig:MicroBooNEbkgspectra}
\end{figure}

Figure.~\ref{fig:MicroBooNEbkgspectra} compares the energy and angular spectra of the signal photon in the neutrinophilic scalar model (red curve) along with the expected MicroBooNE background (shaded gray region)~\cite{MicroBooNE:2025rsd}. For enhanced visibility, the neutral current (NC) coherent $1\gamma$ background is shown separately as a light blue shaded region and multiplied by a factor of 2500 and 1000 in Figs.~\ref{energyspectraMicroBooNE} and~\ref{angularspectraMicroBooNE}, respectively. To roughly match the total background counts, we choose the following set of parameters for this plot: $m_\phi = 30$~MeV, $m_\chi = 10$~MeV, $y_\nu = 5 \times 10^{-3}$, and $\alpha_D = 0.5$. For Figs.~\ref{energyspectraMicroBooNE} and~\ref{angularspectraMicroBooNE}, we choose $g_{\phi \gamma \gamma}=2.5 \times 10^{-2}~\text{GeV}^{-1}$ and $g_{\phi \gamma \gamma}=8.0 \times 10^{-3}~\text{GeV}^{-1}$, respectively to show clearly the differences in the nature of the spectra. For lower values of $m_\phi$, the photon energy spectrum shifts toward lower energies. However, even for a scalar as light as $3~$MeV, we verified that the spectral peak remains well around 100 MeV, as can also be seen in Fig.~\ref{fig:recoilenergyspectra} for SBND.
\par 

Due to the unavailability of background analysis for the $1\gamma 0 p$ topology in ongoing and upcoming experiments, we present the available background energy spectra for MicroBooNE, which employs a similar detection mechanism. Similar analysis can be carried out for the ongoing and upcoming experiments. As seen in the figures, the signal photons are significantly more energetic and forward than those from the background. When combined with the timing spectrum of the $2 \rightarrow 3$ process, this provides a robust and clean handle to isolate DM signals from SM-induced backgrounds.

\section{Model Sensitivity \& Discussion}\label{sec:sensitivity}

\begin{figure}[!htbp]
    \centering
    \begin{subfigure}{0.45\textwidth} 
        \centering
        \includegraphics[scale=0.33]{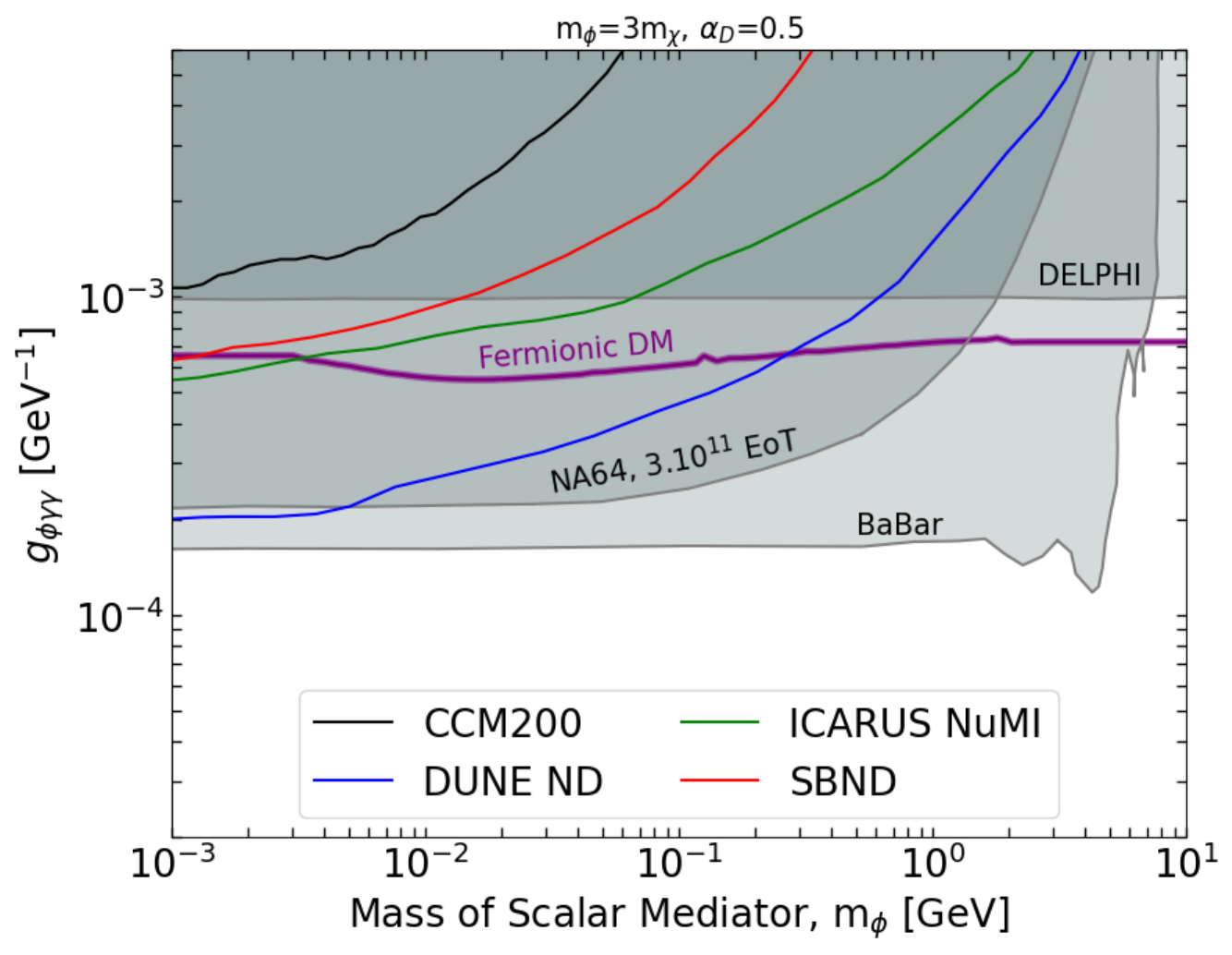}
        \caption{Photon Coupling}
        \label{photoncouplingsensitivity}
    \end{subfigure}
    \begin{subfigure}{0.45\textwidth} 
        \centering
        \includegraphics[scale=0.33]{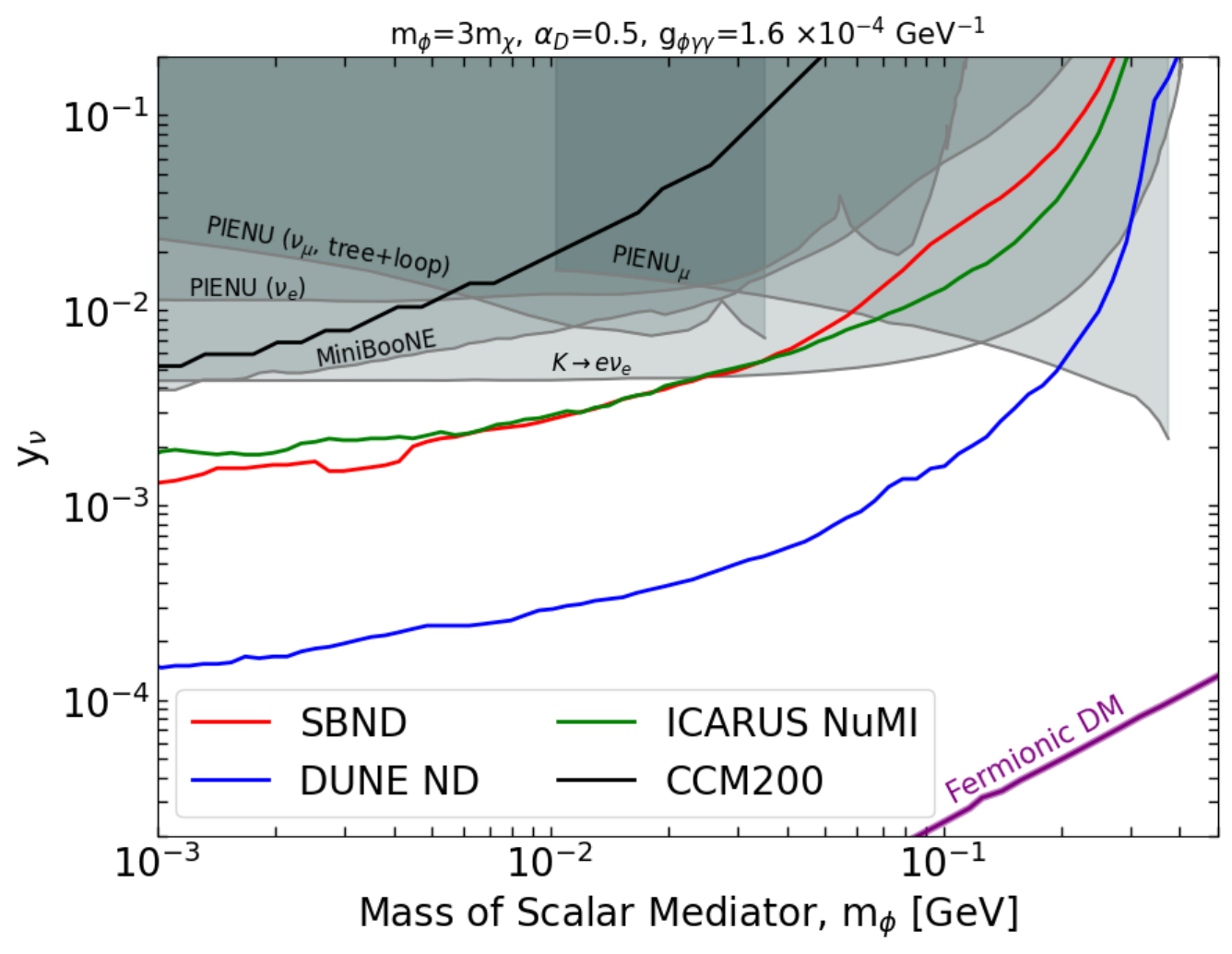}
        \caption{Neutrino Coupling}
        \label{neutrinocouplingsensitivity}
    \end{subfigure}
    \begin{subfigure}{0.45\textwidth} 
        \centering
        \includegraphics[scale=0.1]{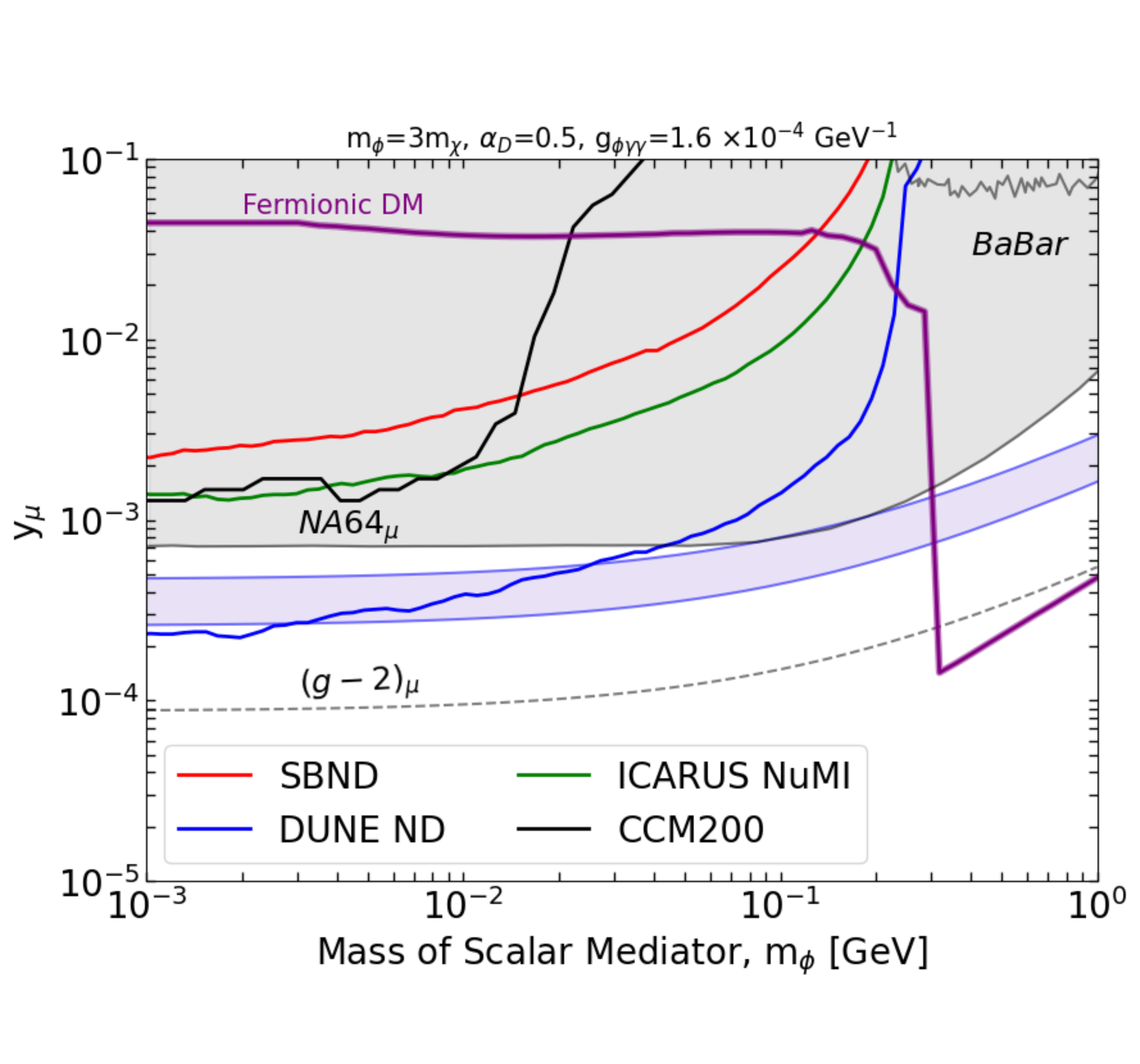}
        \caption{Muon coupling}
        \label{muoncouplingsensitivity}
    \end{subfigure}
    \begin{subfigure}{0.45\textwidth} 
        \centering
        \includegraphics[scale=0.1]{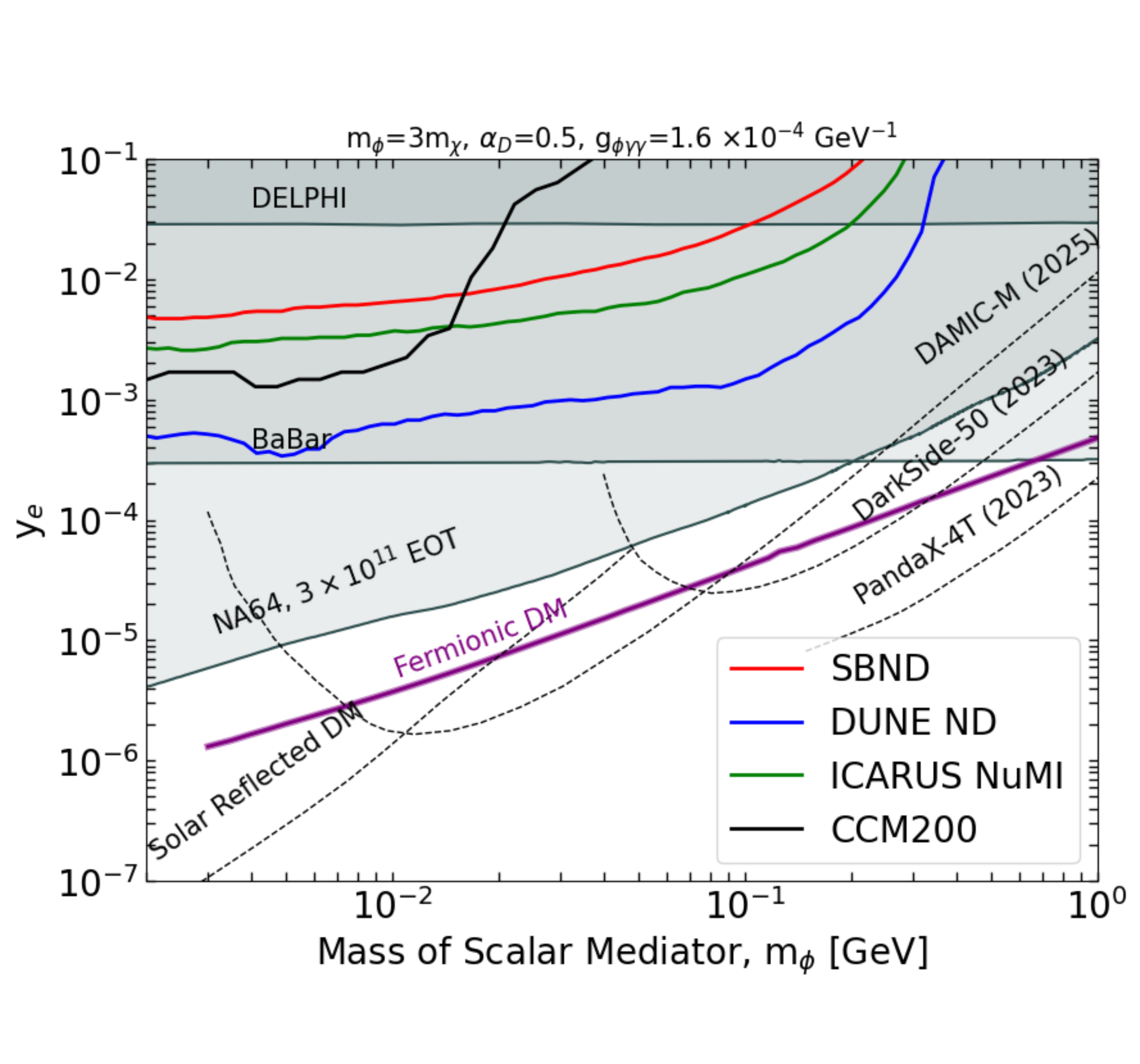}
        \caption{Electron Coupling}
        \label{electroncouplingsensitivity}
    \end{subfigure}
    \begin{subfigure}{0.45\textwidth} 
        \centering
        \includegraphics[scale=0.33]{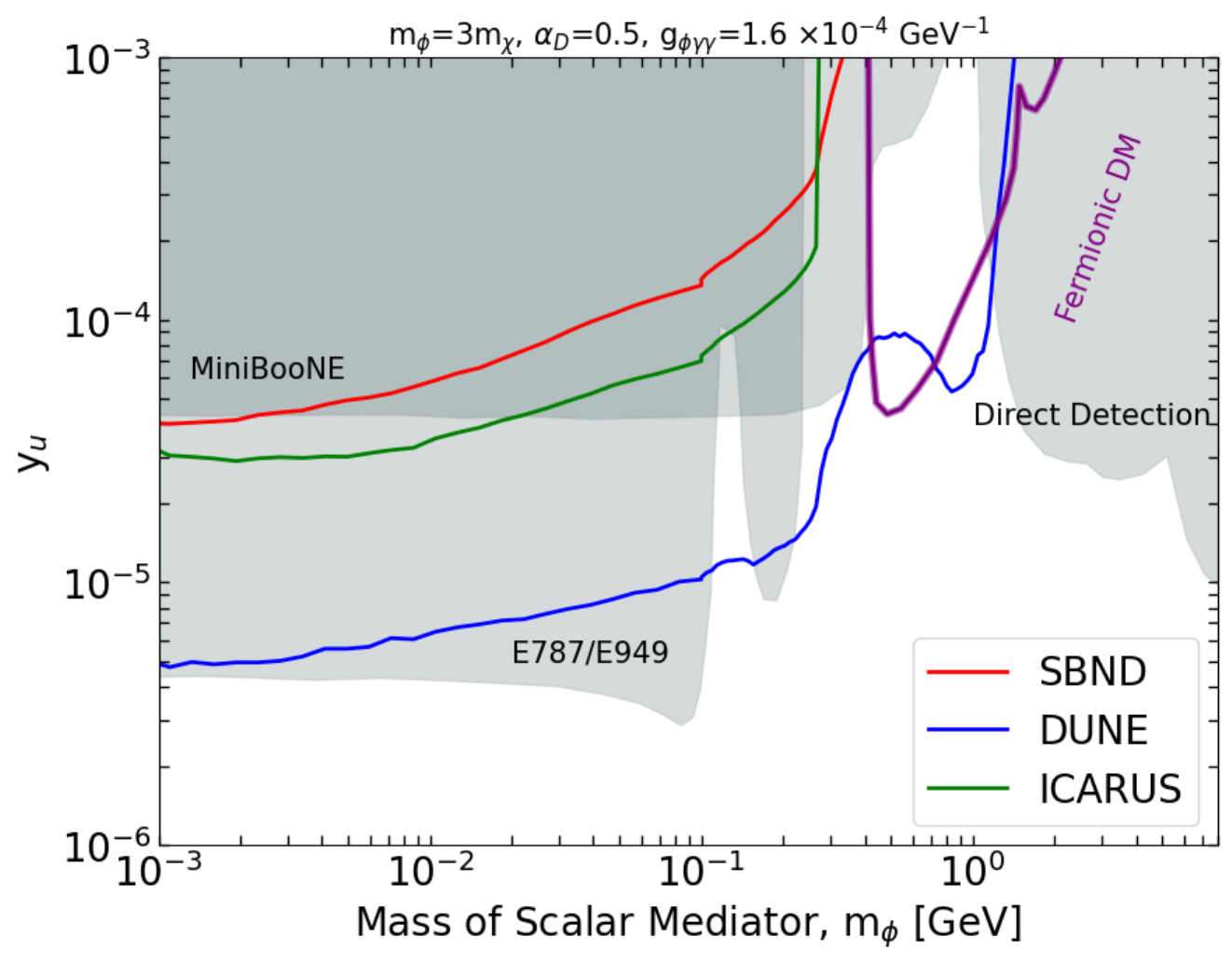}
        \caption{Up-quark Coupling}
        \label{upquarkcouplingsensitivity}
    \end{subfigure}
    \captionsetup{justification=Justified, singlelinecheck=false}   
    \caption{Projected sensitivities of SBND (red), DUNE ND (blue), ICARUS-NuMI (green), and CCM200 (black) to scalar mediator models under various coupling scenarios. Shaded gray areas indicate existing exclusions. Subpanels (a)-(e) correspond to: (a) photon coupling via Primakoff production, (b) neutrino coupling with $\phi$ emitted from the neutrino leg, (c) muon coupling from charged meson decay, (d) electron coupling, and (e) up-quark coupling via meson decay and proton bremsstrahlung. The magenta band in (c) represent the $(g-2)_\mu$ preferred region. The thick purple solid line shows the relic abundance for the respective considered model.}
    \label{fig:Sensitivity}
\end{figure}

In this section we analyze the sensitivity plots in Fig.~\ref{fig:Sensitivity}, which illustrate the regions of parameter space that each experiment can probe. While detection in all cases proceeds via the $2 \rightarrow 3$ process described in Section \ref{sec:detection}, the production mechanisms of the scalar mediator differ across subplots as described in Section \ref{sec:production}. 
\par

In each panel, the colored lines represent projected sensitivities of different experiments: SBND (red), DUNE ND (blue), ICARUS-NuMI (green), and CCM200 (black). For all experiments except CCM200 (where there is no absorber), contributions from both the target and absorber are summed to obtain the total signal for the respective experiment. Shaded gray regions represent existing experimental exclusions. The magenta bands in Fig.~\ref{muoncouplingsensitivity} indicate the 2$\sigma$(dashed) regions that favored the long-sought anomalous magnetic moment of the muon; regions above these bands are currently excluded~\cite{Muong-2:2025xyk, Aliberti:2025beg}. Across all couplings, DUNE ND provides the strongest sensitivity, followed by ICARUS-NuMI, SBND, and CCM200, although this ranking shows exceptions in certain mass ranges, which will be discussed below. As per our discussions in earlier sections, we expect to separate our signal from existing backgrounds in the experiments considered. Therefore, our sensitivities correspond to 10 signal events satisfying the energy threshold (given in Table.~\ref{tab:ExperimentalDetails}). Studies at the MicroBooNE detector show that LArTPC detectors have good efficiencies ($\sim 20\%$) for high-energy signals~\cite{MicroBooNE:2025ntu}. Though we have assumed 100\% detection efficiency for demonstration purposes, the 20\% detection efficiency will not change the sensitivity much. 
For $m_\phi\lesssim 1$ MeV,  $\phi$ can contribute to the effective number of relativistic species, $\Delta N_{eff}$~\cite{Escudero:2019gvw}. In our scenario, $\phi$ primarily decays into light dark matter, with subdominant channels into $\gamma\gamma$ and $\bar{\nu}\nu$. A full cosmological analysis of its impact on  $\Delta N_{eff}$ is beyond the scope of this paper. However, it is noteworthy that laboratory-based experiments can probe regions of parameter space that are also relevant for these cosmologically interesting scenarios.
\par

In Fig.~\ref{photoncouplingsensitivity}, the photophilic scalar mediator $\phi$ is produced via the Primakoff process. Existing constraints (at 90\% confidence level (C.L.)) on the photophilic invisible scalar~\cite{Darme:2020sjf} appear from missing energy searches at NA64~\cite{NA64:2020qwq}, and monophoton searches at BaBar~\cite{BaBar:2017tiz} and DELPHI~\cite{DELPHI:2008uka}. Since the number of events at benchmark experiments is proportional to $g_{\phi \gamma \gamma}^4$, as opposed to the existing bounds that depend on $g_{\phi \gamma \gamma}^2$, we find that their sensitivities lie within the parameters excluded by NA64, BaBar, and DELPHI. Nevertheless, DUNE ND and ICARUS-NuMI benefit from higher beam energies, allowing for more sensitivity to heavier mediators as compared to SBND and CCM200. The thick purple line in this figure denotes the parameters required to set the correct fermionic DM freeze-out annihilations for the observed relic abundance. For scalar masses below 8 GeV, we find that the unexplored parameter space would result in an overabundance of the present day DM, assuming that DM annihilates via a purely photophilic scalar mediator. This, therefore, motivates us even more to look for additional couplings to other SM particles, specifically fermions. For the rest of the sensitivity plots, we fix $g_{\phi \gamma \gamma}=1.6 \times 10^{-4}~$ GeV$^{-1}$, which is still allowed by other experimental bounds. This choice of coupling also demands that the relic abundance for the other models are governed by DM annihilations to the respective leptons.

Figure.~\ref{neutrinocouplingsensitivity} considers neutrinophilic scalars produced through three-body decays of charged mesons, with $\phi$ emitted from the neutrino leg. The signal scales as $(y_\nu g_{\phi \gamma \gamma})^2$, reducing the reliance on the $g_{\phi \gamma \gamma}$ coupling compared to the photophilic case. For the fixed value of $g_{\phi\gamma\gamma}$, we show the sensitivity in terms of $y_\nu$. The sensitivity plots include the strongest of the existing bounds as derived in Ref.~\cite{Dev:2025tdv, Dev:2024ygx}, which included bounds from the three-body decay branching ratio measurements from NA62~\cite{NA62:2021bji} (for $K^{\pm}$) and PIENU~\cite{PIENU:2021clt} (for $\pi^{\pm}$) at 90\% C.L., and the corrections to the $K^{\pm} \to e^{\pm} + \mybracketbar{\nu_e}$ decay width from the scalar-incuded loops. Additionally, we also show the bounds from MiniBooNE for a 10-event threshold. Although the kinematically accessible scalar masses can reach $m_{K^+} - m_e = 0.494~\text{GeV}$ for DUNE, SBND, and ICARUS, and $m_{\pi^+} - m_e = 0.139~\text{GeV}$ for CCM200, we find that our sensitivities fall short of these mass thresholds. This is due to the suppressed meson flux at higher energies, which results in DM with a lesser kinetic energy reaching the detector. This, in turn, leads to a suppressed $2 \rightarrow 3$ cross section (as seen in Fig. \ref{fig:CrossSectionvsEnergy}) and consequently to a lower sensitivity. 

Figure.~\ref{muoncouplingsensitivity} shows the muonphilic scalar mediator sensitivities. Here, we find that DUNE ND has the potential to probe couplings beyond those probed by the NA64$\mu$~\cite{NA64:2024klw} for $m_\phi < 40~\rm{MeV}$. For reference, we include a purple band representing parameters preferring the previous discrepancy in $(g-2)_\mu$ at $90\%$ C.L.. In the black dashed line, we show the $1\sigma$ limit based on the current measurement of $(g-2)_\mu$~\cite{Muong-2:2025xyk}, assuming that this muonphilic scalar is the only new BSM contribution. We also show the regions excluded by the BaBar$4\mu$~\cite{BaBar:2016sci} searches for an invisibly decaying scalar~\cite{Capdevilla:2021kcf}. Since the CMS$4\mu$~\cite{CMS:2024jyb} bounds are mostly prevalent for $m_{\phi} > 6~\text{GeV}$~\cite{Capdevilla:2021kcf}, and subdominant for the invisible decaying scalar, we do not present them in our sensitivity plot. 

Figure.\ref{electroncouplingsensitivity} shows the sensitivity for an electrophilic scalar. In this case, stringent NA64 constraints~\cite{NA64:2020qwq, NA64:2023wbi} leave no uncovered parameter space for the above experiments. Notably, CCM200 shows enhanced reach due to significant contributions from electron-positron annihilation, Compton-like scattering, and electron bremsstrahlung, which dominate in this setup. In black dashed lines, we show the limits from direct detection experiments such as DAMIC-M~\cite{DAMIC-M:2025luv}, DarkSide-50~\cite{DarkSide:2022knj}, PandaX-4T~\cite{Zhang:2025ajc}, and Xenon-1T (solar reflected Ref.~\cite{Emken:2024nox}). These limits are calculated assuming that the 100\% ambient dark matter density is composed of the candidate under consideration, whereas our analysis considers a light dark matter candidate that may not saturate the thermal dark matter abundance. It is also possible that the origin of dark matter is non-thermal. If the candidate accounts for only a fraction of the total dark matter, the corresponding bounds would scale accordingly.

In Fig. \ref{upquarkcouplingsensitivity}, up-philic scalar mediator $\phi$ production proceeds via two mechanisms: $K^+ \rightarrow \pi^+ + \phi$ decays (Fig. \ref{fig:Kptopipupcoupling}) and proton bremsstrahlung (Fig. \ref{fig:protonbremscalar}). The former dominates for $m_\phi \leq m_{K^+} - m_{\pi^+} = 0.35~\text{GeV}$, while the latter becomes important at higher mediator masses $m_{\phi} > 0.35~\text{GeV}$. Here, we assume that the scalar mediator $\phi$ couples exclusively to up quarks, which induces an effective scalar-proton interaction \cite{Batell:2018fqo}. This, in turn, leads to constraints from direct detection experiments at 90\% C.L., as shown in the sensitivity plot. The current direct detection limits of ambient DM consist of results from the $\nu$-cleus~\cite{CRESST:2017ues}, CRESST~\cite{CRESST:2015txj, CRESST:2017cdd}, CDMSlite~\cite{SuperCDMS:2015eex}, PICO~\cite{PICO:2017tgi}, and XENON1T~\cite{XENON:2017vdw} experiments. 
Additional constraints include the MiniBooNE bound at 90\% C.L., which comes from the null result in the nucleon elastic scattering channel~\cite{MiniBooNE:2017nqe, MiniBooNE:2012jpi, MiniBooNEDM:2018cxm}. That result was obtained from the dedicated beam dump run using $1.86 \times 10^{20}$ POT directed onto a steel beam dump. This setup differs from the configuration considered in our analysis, where we consider the target mode of the BNB, in which protons are directed at a beryllium target rather than an iron dump. Another important constraint arises from the E787/E949 experiments~\cite{E787:2002qfb, E787:2004ovg, E949:2007xyy} conducted at Brookhaven National Laboratory, which searched for the rare invisible decay of Kaons: $K \rightarrow \pi+\text{invisible}$.  
\par 

Among all the scenarios, the neutrino-scalar coupling provides the broadest access to unexplored parameter space, primarily due to the weaker existing constraints on $y_\nu$. We further observe in all plots, from Fig.~\ref{neutrinocouplingsensitivity} to Fig.~\ref{upquarkcouplingsensitivity}, that neither the existing bounds nor the sensitivities of our benchmark experiments can probe the entirety of the thermal relic line (thick purple line) across the mass scale considered. This, therefore, motivates future efforts to dedicate searches via the proposed $2\to3$ scattering mechanism.

\section{\label{sec:conclusion}Summary and Conclusions}

In this work, we have analyzed in detail several scalar mediator DM models-namely, photophilic, neutrinophilic, electro/muon-philic, and quarkphilic (up-quark only), within the context of a $2 \rightarrow 3$ scattering process that features a high-energy photon in the final state. We have shown that this distinctive photon serves as a powerful experimental probe of scalar portal DM in neutrino experiments. 
\par

Our study considers four experiments: SBND, ICARUS-NuMI, DUNE ND, and CCM200, where, for the first three experiments, we considered DM produced from both the target and absorber (from escaped protons). In addition, we analyzed the completed MicroBooNE and MiniBooNE experiments to place constraints on the respective model parameter spaces. Among all models, only the neutrinophilic scenario allows MiniBooNE to place new constraints; in all other cases, the bounds remain within previously excluded regions.
\par 

For ongoing and upcoming experiments, we find that the neutrinophilic model probes the most previously unexplored parameter space due to weaker existing constraints on the coupling $y_\nu$, followed by the quarkphilic scenario (assuming the scalar couples only to up quarks). In terms of experimental reach, DUNE ND offers the strongest sensitivity due to its high beam energy and on-axis positioning, followed by ICARUS-NuMI and then SBND. Due to stringent constraints from NA64, we find that this process is not quite effective in probing unexplored parameter space for purely photophilic and electro/muon-philic models. 
\par

We discussed how various observables, such as timing spectra, final-state photon energies, and angular spectra, can be used to discriminate signal from background. In particular, we show that timing information plays a crucial role in separating DM-induced events from neutrino-induced backgrounds, especially when applying timing cuts. These cuts retain most of the signal while significantly reducing the background. 

\par
Additionally, we analyzed the energy and angular spectra of the final-state photon, which tends to be more energetic and forward than the SM background photons, offering another clear discriminator. For instance, in the case of MicroBooNE's measured background comparison, we demonstrated that the signal photon energy peaks well above the dominant background region and is much more forward-directed. Other potential DM signals, such as electron and nuclear recoils, are not as efficient due to larger backgrounds and smaller efficiencies. Overall, we find that the $2\to 3$ scattering mechanism results in monophoton signals that stand out from all SM neutrino backgrounds that dominantly occur via $2\to 2$ scattering. This distinction is due to better extraction of energy into the final state photon via the $2\to 3$ mechanism, making this feature universal throughout all the models considered and therefore all DM production mechanisms. We also highlighted the role of geometric configuration: off-axis and more distant detectors (e.g., DUNE PRISM and ICARUS-NuMI) help temporally isolate signal events more effectively. 
\par

We further investigated energy spectra of DM particles for a given mass, as well as the DM counts as a function of mass, that enter each detector across production mechanisms (Primakoff, two and three body meson decays, proton and electron bremsstrahlung). We also looked into the spatial distributions of the DM particles at the face of the SBND detector. These distributions provide key insights into the production dynamics for each model, dependence on the beam energy, and the geometric configuration of the experiment, and can serve as valuable inputs for future DM simulations or cross-checks of flux predictions.
\par

This analysis presents a novel methodology to search for scalar portal DM scenarios with photon couplings. Interesting phenomenology can be derived from scalar portal dark matter that couples to gluons as well, arising from an effective Lagrangian of the form $\phi G_{\mu\nu}G^{\mu\nu}$. These couplings would lead to interesting signatures in both the production side of dark matter as well as in the detection. While this has not been the focus of this paper, we are investigating these possibilities in future studies. 

This analysis can be readily extended to explore other types of light mediator models or to refine experimental strategies to isolate new physics signals in high-background environments.  Although we demonstrate them in the context of neutrino experiments, this can be adapted to other types of beam dump experiments and colliders. Beyond the mentioned searches for lab-produced DM, the $2\to 3$ processes can be used to search for cosmic-ray-boosted dark matter at experiments such as DUNE FD, T2K, and HyperK, and blazar-produced/boosted dark matter at IceCube and KM3NeT~\cite{Dev:2025czz}, etc.
\par

\section*{Acknowledgments}
We thank Wooyoung Jang and Hyunyong Kim for their work on \texttt{GEANT4} simulations. We also thank Adrian Thompson for his work on the \texttt{RKHorn} package, and Doojin Kim, Kevin Wood, and Jaehoon Yu for useful discussions. This work of BD, DG, and AK is supported by the U.S. Department of Energy
Grant~DE-SC0010813. For facilitating portions of this research, AK wishes to acknowledge the Center for Theoretical Underground Physics and Related Areas (CETUP), The Institute for Underground Science at Sanford Underground Research Facility (SURF), and the South Dakota Science and Technology Authority for hospitality and financial support, as well as for providing a stimulating environment.

\appendix

\section{\label{app:threebody}Three body decays}

The branching ratio of a three-body decay process can be evaluated from the following expression:

\begin{equation}
    \frac{d^2 \text{BR}_{\phi_{l/\nu}}({\texttt{m} \to l^{\pm}\nu_l\phi})}{dm_{12}^2 dm_{23}^2} = \frac{1}{(2\pi)^3 32 m_\texttt{m}^3 \Gamma_{\texttt{m}}} |\mathcal{M}_{\phi_{l/\nu}}({\texttt{m} \to l^{\pm}\nu_l\phi})|^2
    \label{eq:diffdecay}
\end{equation}

The matrix element squared for $\texttt{m} \to l^{\pm}\nu_l\phi$ for a neutrinophilic scalar is:

\begin{equation}
    \begin{aligned}
        |\mathcal{M}_{\phi_\nu}({\texttt{m} \to l^{\pm}\nu_l\phi})|^2 &=  \frac{4G_F^2 V_{uq}^2 f_{\texttt{m}}^2 y_{\nu}^2}{(-m_{12}^2 -m_{23}^2 + m_{\mu}^2 + m_{\texttt{m}}^2 + m_{\phi}^2)^2}\bigg((m_{12}^6 + m_{12}^4(m_{23}^2 - 2(m_{\phi}^2 + m_{\texttt{m}}^2) - m_l^2) \\&+ m_{12}^2(m_{23}^4 - m_{23}^2(2(m_{\phi}^2 + m_{\texttt{m}}^2) + m_l^2) + (m_\phi^2 + m_{\texttt{m}}^2)^2+2m_{\phi}^2m_l^2)\\ & m_l^2 (m_\phi^2(m_{23}^2 - m_{\texttt{m}}^2) + m_{\texttt{m}}^2(m_{\texttt{m}}^2 + m_l^2 - m_{23}^2) - m_{\phi}^4) \bigg)
    \end{aligned}
    \label{eq:neuphilic_matel}
\end{equation}

For the electrophilic(muonphilic) scalar where the scalar emanates from the electron(muon) leg, the matrix element squared is,
\begin{equation}
    \begin{aligned}
        |\mathcal{M}_{\phi_l}({\texttt{m} \to l^{\pm}\nu_l\phi})|^2 &= y_{22}^2 \frac{4G_F^2 V_{uq}^2 f_{\texttt{m}}^2}{(m_{23}^2 - m_{l}^2)^2} \bigg(m_{23}^2 (m_{12}^2 + m_{23}^2 - m_{\phi}^2) 
        (m_{23}^2 - m_{l}^2) +   
        2 m_{23}^2 m_{l}^2 (-m_{23}^2 + m_{\texttt{m}}^2) \\&- (m_{23}^2 - m_{\phi}^2 + m_{l}^2) (m_{23}^4 - m_{l}^2 m_{\texttt{m}}^2) \bigg)
    \end{aligned}
    \label{eq:lepphilic_matel}
\end{equation}
where the limits of integration are,
\begin{equation}
    \begin{aligned}
        m_{l}^2 \leq &m_{12}^2 \leq (m_\texttt{m} - m_{\phi})^2\\
        (E_{l}^* + E_{\phi}^*)^2 - (p_{l}^* + p_{\phi}^*)^2 \leq &m_{23}^2 \leq (E_{l}^* + E_{\phi}^*)^2 - (p_{l}^* - p_{\phi}^*)^2
    \end{aligned}
    \label{eq:m12lims}
\end{equation}

To integrate over $m_{23}^2$ while fixing the value of $m_{12}^2$, we choose a frame where $p_1 + p_2$ is at rest. The quantities with the star describe those in the rest frame of $p_1+p_2$. These quantities are as follows,

\begin{equation}
    \begin{aligned}
        E_l^* = \frac{m_{12}^2 + m_l^2}{2\sqrt{m_{12}}}&; \quad
        E_\phi^* = \frac{m_{\texttt{m}}^2 - m_{12}^2 - m_\phi^2}{2\sqrt{m_{12}}} \\
        p_{l,\phi}^* &= \sqrt{E_{l,\phi}^{*2} - m_{l,\phi}^2}
    \end{aligned}
    \label{eq:m23energies}
\end{equation}

The energy of the scalar in the rest frame of the decaying meson can be found for a given Dalitz variable.

\begin{equation}
    E_{\phi} = \frac{m_{\texttt{m}}^2 + m_{\phi}^2 - m_{12}^2}{2m_{\texttt{m}}} 
    \label{eq:energyphi}
\end{equation}

\begin{figure}[!htbp]
    \centering
        \includegraphics[scale=0.55]{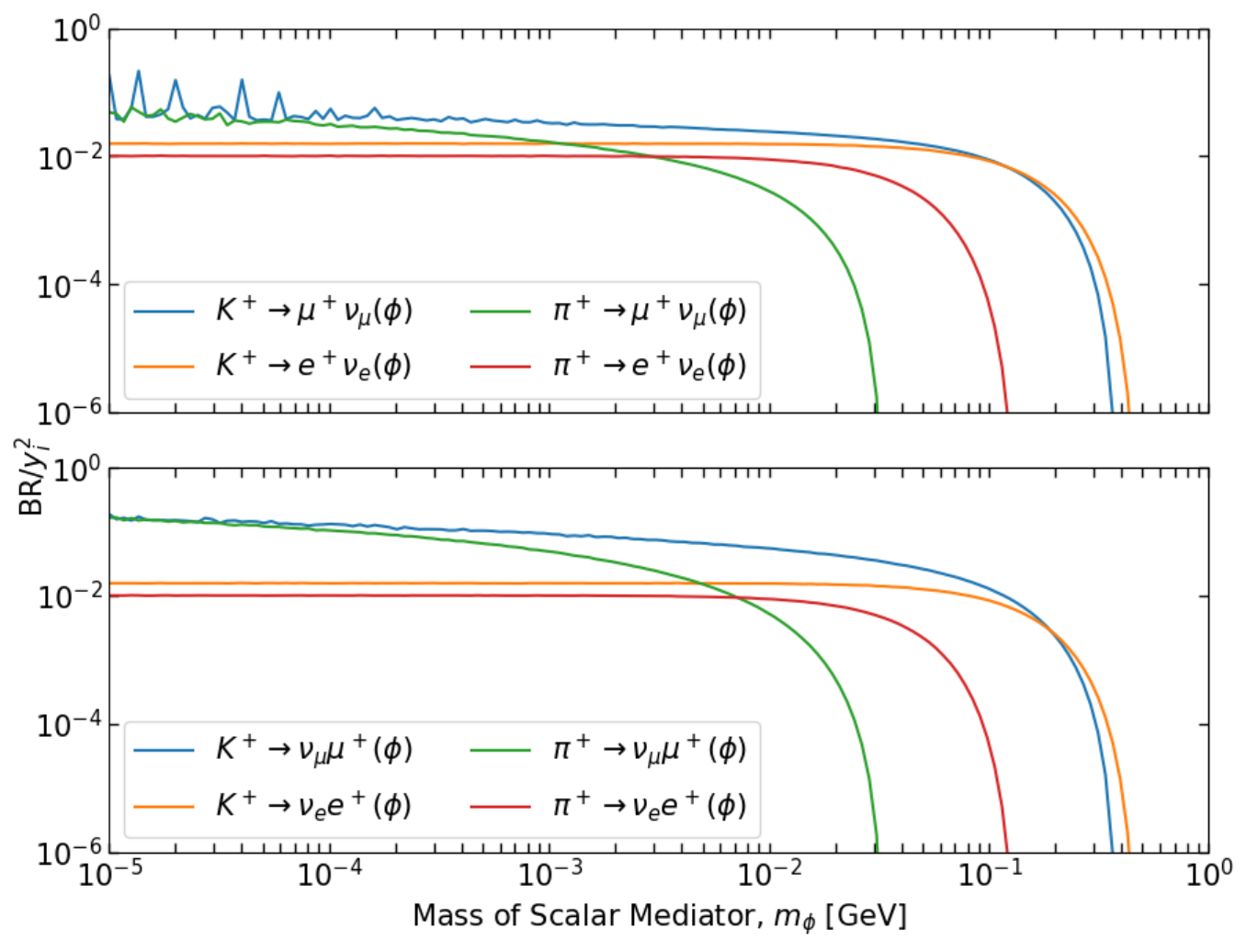}
    \captionsetup{justification=Justified, singlelinecheck=false}
    \caption{Branching ratios of three body meson decays involving a light scalar mediator $\phi$, as a function of its mass. The branching ratios are normalized by the square of the effective coupling $y_i^2$.}
    \label{fig:BRplot}
\end{figure}

Figure.~\ref{fig:BRplot} shows these branching ratios with the scalar mediator $\phi$ emitted from the neutrino leg (top panel) and from the charged lepton leg (bottom panel). The kinematic threshold for each decay channel is determined by the condition $m_\phi < m_{{\texttt{m}}^+} - m_{l^+}$, beyond which the branching ratio drops to zero. The $y$-axis shows the branching ratio normalized to unit coupling squared, while the $x$-axis corresponds to the scalar mediator mass.

\section{\label{app:2to3} $2 \rightarrow 3$ Scattering Process: Cross-section and Phase Space Limits} 
For completeness, here we also outline the procedure for calculating the cross-section and kinematics for the $2 \rightarrow 3$ process as can also be found in Refs.~\cite{Byckling:1971vca, Dutta:2025fgz}. The $2 \rightarrow 3$ process, $\chi(p_\chi) + N(p_N) \rightarrow \chi(k_\chi) + \gamma(k_\gamma) + N(k_N)$, is described in terms of five independent Mandelstam variables: 
\begin{itemize}
\item $s=(p_\chi+p_N)^2$ 
\item $t_1=(p_\chi-k_\chi)^2$ 
\item $t_2=(p_N-k_N)^2$ 
\item $s_1=(k_\chi+k_\gamma)^2$  
\item $s_2=(k_\gamma+k_N)^2$ 
\end{itemize}
Here, $p_i$ and $k_i$ denote the four-momenta of the incoming and outgoing particles, respectively. The relevant scalar products in terms of these Mandelstam variables and the particle masses are given by:
\begin{itemize}
\item $p_\chi.p_N$=$\frac{1}{2}(s-m_\chi^2-m_N^2)$ 
\item $p_\chi.k_\chi$=$\frac{1}{2}(2m_\chi^2-t_1)$
\item $p_\chi.k_\gamma$=$\frac{1}{2}(s_1+t_1-t_2-m_\chi^2)$
\item $p_\chi.k_N$=$\frac{1}{2}(s-s_1+t_2-m_N^2)$
\item $p_N.k_\chi$=$\frac{1}{2}(s-s_2+t_1-m_\chi^2)$
\item $p_N.k_\gamma$=$\frac{1}{2}(s_2+t_2-t_1-m_N^2)$
\item $p_N.k_N$=$\frac{1}{2}(2m_N^2-t_2)$
\item $k_\chi.k_\gamma$=$\frac{1}{2}(s_1-m_\chi^2)$
\item $k_\chi.k_N$=$\frac{1}{2}(s-s_1-s_2)$
\item $k_\gamma.k_N$=$\frac{1}{2}(s_2-m_N^2)$
\end{itemize}

The three-body phase space is given by:
\begin{equation}\label{equationA1}
dR_3(s)=\int_{}^{}\frac{d^3k_\chi}{2E^{'}_\chi}\frac{d^3k_\gamma}{2E^{'}_\gamma}\frac{d^3k_N}{2E^{'}_p}\delta^4(p_\chi+p_N-k_\chi-k_\gamma-k_N) = \frac{\pi}{16\lambda^{1/2}(s,m_\chi^2,m_N^2)}\frac{dt_1ds_2dt_2ds_1}{(-\Delta_4)^{1/2}}
\end{equation}

Here, $\Delta_4$ denotes the symmetric $4\times 4$ Gram determinant, defined as:

\begin{equation}\label{equationA2}
\Delta_4=\frac{1}{16}\Bigg[\begin{vmatrix}
2s_2 & s_2-t_1+m_N^2 & s+s_2-m_\chi^2 & s_2+m_N^2 \\
s_2-t_1+m_N^2 & 2m_N^2 & s-m_\chi^2+m_N^2 & 2m_N^2-t_2 \\
s+s_2-m_\chi^2 & s-m_\chi^2+m_N^2 & 2s & s-s_1+m_N^2 \\
s_2+m_N^2 & 2m_N^2-t_2 & s-s_1+m_N^2 & 2m_N^2 \\
\end{vmatrix}
\Bigg]
\end{equation}

In the second expression, the integration variables have been transformed into four independent Mandelstam variables, converting the phase space into a Lorentz-invariant quantity. The corresponding integration limits for these variables are given below:

\begin{itemize}
\item $s_2$:\Bigg($m_N^2$, $(\sqrt{s}-m_\chi)^2$\Bigg) 
\item $t_1^{\pm}=2m_\chi^2 - \frac{1}{2s}\Bigg((s+m_\chi^2-m_N^2)(s-s_2+m_\chi^2)\mp \lambda^{1/2}(s, m_\chi^2, m_N^2)\lambda^{1/2}(s, s_2, m_\chi^2)\Bigg)$
\item $t_2^{\pm}=2m_N^2-\frac{1}{2s_2}(s_2+m_N^2-t_1)(s_2+m_N^2)\pm \frac{1}{2s_2}\lambda^{1/2}(s_2, m_N^2, t_1)\lambda^{1/2}(s_2, m_N^2, 0)$
\item \begin{align} 
s_1^{\pm}=s+m_N^2-\frac{1}{\lambda(s_2, t_1, m_N^2)}\Biggl[\begin{vmatrix}
2m_N^2 & s_2-t_1+m_N^2 & 2m_N^2-t_2\\
s_2-t_1+m_N^2 & 2s_2 & s_2+m_N^2\\
s-m_\chi^2+m_N^2 & s+s_2-m_\chi^2 & 0
\end{vmatrix} \nonumber \\
 \mp  2\{G(s, t_1, s_2, m_\chi^2, m_N^2, m_\chi^2)G(s_2, t_2, m_N^2, t_1, m_N^2, 0)\}^{1/2}\Biggr]
\end{align}
\end{itemize}

where, 

\begin{equation}\label{equationA3}
G(x,y,z,u,v,w)=-\frac{1}{2}\begin{vmatrix}
0 & 1 & 1 & 1 & 1 \\
1 & 0 & v & x & z \\
1 & v & 0 & u & y \\
1 & x & u & 0 & w \\
1 & z & y & w & 0 \\
\end{vmatrix}
\end{equation}

Within the physically allowed phase space, the following conditions must be satisfied: $G(x, y, z, u, v, w) \leq 0$, $\Delta_4 \leq 0$ and $\lambda(x, y, z) \geq 0$. Under these constraints, the total cross-section is given as:

\begin{equation}\label{equationA4}
\sigma(s)=\int_{s_2^-}^{s_2^+} \int_{t_1^-}^{t_1^+} \int_{t_2^-}^{t_2^+} \int_{s_1^+}^{s_1^-} <|\mathcal{M}|^2>\frac{1}{(2\pi)^5}\frac{\pi}{32\lambda(s,m_\chi^2,m_N^2)}\frac{1}{(-\Delta_4)^{1/2}}ds_1dt_2dt_1ds_2
\end{equation}

The integration must be performed in the specified order. Altering the order of integration requires adjusting the corresponding limits accordingly.

To determine the kinematics of the outgoing photon, we consider the rest frame of the photon and the recoiled nucleus. Let $\theta_{p_Nk_N}^{R\gamma N}$ and $\phi^{R\gamma N}_{p_N k_N}$ denote the polar and azimuthal angles of particle 3 (the outgoing nucleus) measured with respect to the direction of the incoming nucleus, which defines the $z$-axis. In this frame, the outgoing photon's polar and azimuthal angles are given by $\pi - \theta_{p_Nk_N}^{R\gamma N}$ and $\pi + \phi^{R\gamma N}_{p_N k_N}$, respectively. In terms of the variables $t_1$, $s_2$, $\phi^{R\gamma N}_{p_N k_N}$ and $\theta_{p_Nk_N}^{R\gamma N}$, the total cross-section is given by:
\begin{equation}\label{equationA5}
\sigma(s)=\int_{s_2^-}^{s_2^+} \int_{t_1^-}^{t_1^+} \int_{-1}^{1} \int_{0}^{2\pi}\frac{\sqrt{\lambda(s, 0, m_N^2)}}{32\pi^4\lambda(s, m_\chi^2, m_N^2)}\frac{<|\mathcal{M}|^2>}{8s_2}d\cos(\theta_{p_Nk_N}^{R\gamma N})d\phi_{p_Nk_N}^{R\gamma N}d{t_1}d{s_2} 
\end{equation}

By treating the outgoing dark matter particle and the recoiled nucleus as an effective two-body system with total invariant mass squared $s_{\chi N}$, the energy of the outgoing photon in the center-of-mass (COM) frame of the incoming dark matter and nucleus is given by:
\begin{equation}\label{equationA6}
E_{\gamma}^{CM}=\frac{s-s_{\chi N}}{2\sqrt{s}}
\end{equation}
where, $s_{\chi N}=s+m_\chi^2+m_N^2-s_1-s_2$. Once the polar and azimuthal angles are randomly sampled within their allowed ranges, the corresponding values of the Mandelstam variables $t_2$ and $s_1$ can be determined using the following expressions:

\begin{align}\label{equationA7}
t_2=2m_N^2-\frac{1}{2s_2}(s_2+m_N^2-t_1)(s_2+m_N^2) + \cos(\theta_{p_Nk_N}^{R\gamma N})\frac{1}{2s_2}\lambda^{1/2}(s_2, m_N^2, t_1)\lambda^{1/2}(s_2, m_N^2, 0)
\end{align}

\begin{align}\label{equationA8}
s_1=s+m_N^2-\frac{1}{\lambda(s_2, t_1, m_N^2)}\Biggl[\begin{vmatrix}
2m_N^2 & s_2-t_1+m_N^2 & 2m_N^2-t_2\\
s_2-t_1+m_N^2 & 2s_2 & s_2+m_N^2\\
s-m_\chi^2+m_N^2 & s+s_2-m_\chi^2 & 0
\end{vmatrix} \nonumber \\
 +  2\{G(s, t_1, s_2, m_\chi^2, m_N^2, m_\chi^2)G(s_2, t_2, m_N^2, t_1, m_N^2, 0)\}^{1/2}\cos(\phi^{R\gamma N}_{p_N k_N})\Biggr]
 \end{align}

\begin{figure}[!htbp]
    \centering
        \includegraphics[scale=0.45]{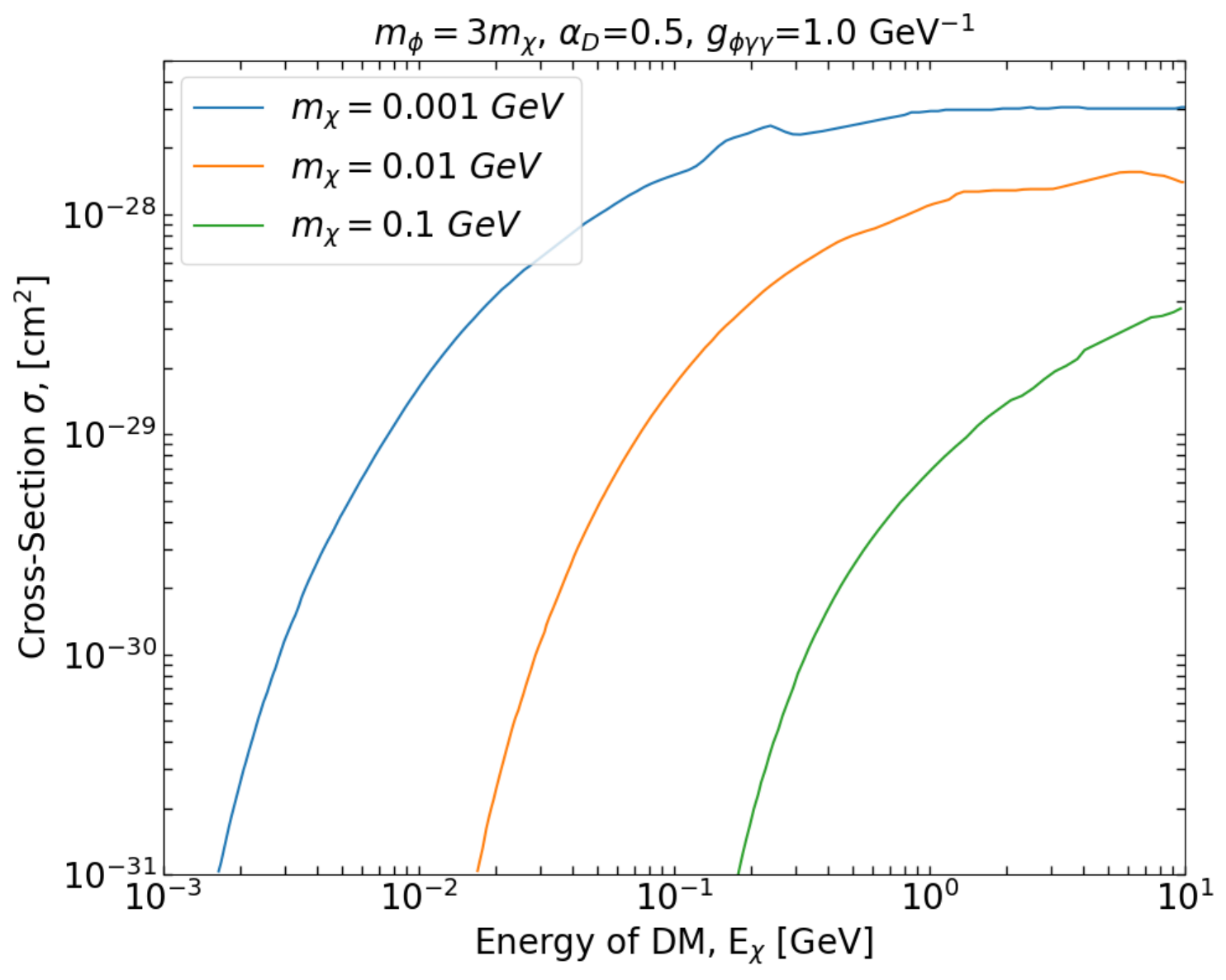}
    \captionsetup{justification=Justified, singlelinecheck=false}
    \caption{Cross-section $\sigma_{2 \rightarrow 3}$ of $\chi + N \rightarrow \chi + \gamma + N$ shown as a function of incoming dark matter energy $E_\chi$ for three different dark matter masses: $m_\chi=0.001, 0.01~\text{and}~0.1~$GeV.}
    \label{fig:CrossSectionvsEnergy}
\end{figure}

Upon obtaining the energy and angles in the COM frame of the incoming DM and nucleus, we calculate the boost matrix needed to stop the incoming nucleus to switch to the lab frame. Applying the calculated boost matrix on the outgoing photon's four vector gives us its energy and momentum in the lab frame. In Fig.~\ref{fig:CrossSectionvsEnergy}, we show the dependence of the $2 \rightarrow 3$ scattering cross section on the incoming dark matter (DM) energy for different DM masses: $m_\chi=0.001, 0.01, 0.1~$GeV. The benchmark parameters used for this plot are: $m_\phi=3m_\chi$, $\alpha_D=0.5$ and $g_{\phi \gamma \gamma}=1.0~$GeV$^{-1}$.

\section{\label{FormFactor}Details on Form Factor}

In this analysis, we consider both the coherent and incoherent scattering regimes. For momentum transfer squared $Q^2 < 0.1~\text{GeV}^2$, we assume coherent scattering, where the incoming DM particle interacts with the entire nucleus. In this regime, we adopt the Helm form factor, parameterized as~\cite{Helm:1956zz, Engel:1991wq}
\begin{equation}\label{10a}
F^{\text{Helm}}(E_R) = \frac{3}{(Q r)^3} e^{-Q^2 s^2 / 2} \left( \sin(Q r) - Q r \cos(Q r) \right),
\end{equation}
where $s = 0.9~\text{fm}$ is the skin thickness, $r = 3.9 \times (A/40)^{1/3}~\text{fm}$ is the effective nuclear radius, and $Q = \sqrt{2 m_N E_R}$ is the momentum transfer. Here, $A$ denotes the mass number of the nucleus. We neglect the nuclear magnetic form factor, as it lacks the $Z^2$ enhancement relevant in this context~\cite{Brdar:2020quo}.
\par

In the incoherent regime ($Q^2 > 0.1~\text{GeV}^2$), the DM particle scatters off individual nucleons. We model this interaction using a dipole form factor given by \cite{Perdrisat:2006hj}
\begin{equation}\label{10b}
F^{\text{dipole}}(E_R) = \frac{1}{1 + \frac{Q^2}{M_D^2}},
\end{equation}
where $M_D = 1.18 + 0.83 \, A^{1/3}$ GeV, with $A$ again denoting the nuclear mass number.

\section{\label{RecoilEnergy}Final State $\gamma$ Energy Spectra vs Nuclear Recoil Energy}

\begin{figure}[!htbp]
    \centering
        \includegraphics[scale=0.45]{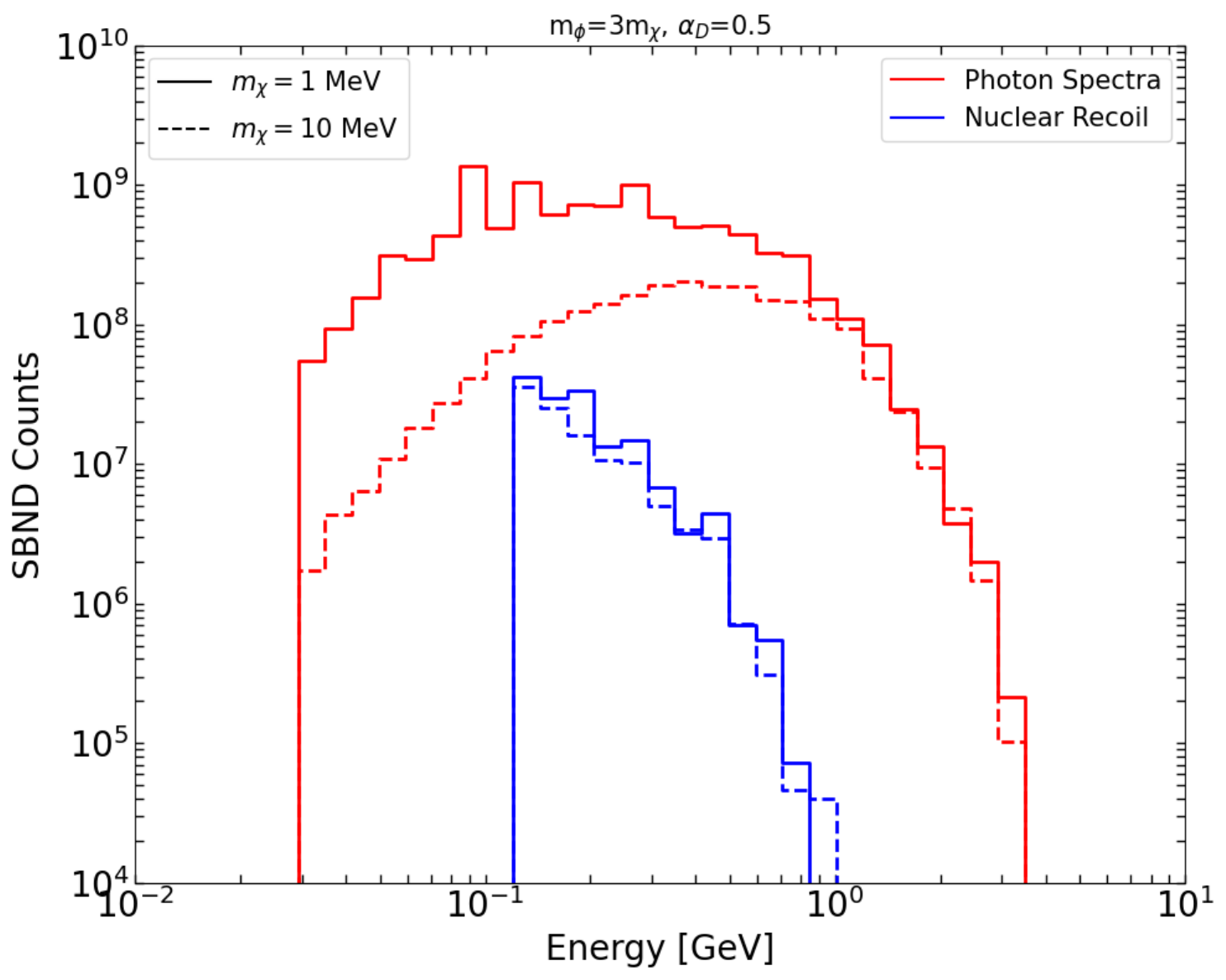}
    \captionsetup{justification=Justified, singlelinecheck=false} 
    \caption{Energy spectra of the final state photon (red) and the nucleus from recoils (blue) that can be observed at SBND for the neutrinophilic model. Here, we consider 1 MeV (solid) and 10 MeV (dashed) incoming DM masses. For sake of demonstration, we fix $g_{\phi\gamma\gamma} = 1.0~\text{GeV}^{-1}$, $\alpha_D=0.5$ and $g_\nu=5 \times 10^{-3}$ and $g_\nu=10^{-2}$ respectively. Since, uniform binning is considered for the x-axis, the y-axis denotes counts/bin and not counts per bin width. Hence, in order to obtain the total number of photons one can sum over the values for each bin and not explicitly multiply with bin-widths.}
    \label{fig:recoilenergyspectra}
\end{figure}

Figure.~\ref{fig:recoilenergyspectra} compares the energy spectra of the final-state photon and the recoil energy of the outgoing nucleus for the SBND experiment, under the neutrinophilic scalar model. In this analysis, a photon detection threshold of 30 MeV and a nuclear recoil threshold of 120 MeV are considered. The benchmark parameters used are: $m_\phi=3m_\chi$, $\alpha_D=0.5$, $g_{\phi \gamma \gamma}=1.0~$GeV$^{-1}$, and the maximum allowed value of $y_\nu$ consistent with the chosen scalar mass. 
\par 

As shown in the figure, the recoil energy spectrum is significantly softer. In contrast, the single-photon ($1\gamma$) final state provides a much cleaner and more distinctive signal, enhancing its utility in dark matter searches.

\bibliography{DMScalar}

@article{Emken:2024nox,
    author = "Emken, Timon and Essig, Rouven and Xu, Hailin",
    title = "{Solar reflection of dark matter with dark-photon mediators}",
    eprint = "2404.10066",
    archivePrefix = "arXiv",
    primaryClass = "hep-ph",
    doi = "10.1088/1475-7516/2024/07/023",
    journal = "JCAP",
    volume = "07",
    pages = "023",
    year = "2024"
}

@database{NIST:2019,
        author = "Berger, M.J. and others",
        title = "{XCOM: Photon Cross Section Database (version 1.5)}",
        collaboration = "NIST",
        year = "2010",
        doi = "10.18434/T48G6X",
        url = "ttp://physics.nist.gov/xcom"
}

@article{DarkSide:2022knj,
    author = "Agnes, P. and others",
    collaboration = "DarkSide",
    title = "{Search for Dark Matter Particle Interactions with Electron Final States with DarkSide-50}",
    eprint = "2207.11968",
    archivePrefix = "arXiv",
    primaryClass = "hep-ex",
    reportNumber = "FERMILAB-PUB-22-566-ND-PPD-SCD",
    doi = "10.1103/PhysRevLett.130.101002",
    journal = "Phys. Rev. Lett.",
    volume = "130",
    number = "10",
    pages = "101002",
    year = "2023"
}

@article{DAMIC-M:2025luv,
    author = "Aggarwal, K. and others",
    collaboration = "DAMIC-M",
    title = "{Probing Benchmark Models of Hidden-Sector Dark Matter with DAMIC-M}",
    eprint = "2503.14617",
    archivePrefix = "arXiv",
    primaryClass = "hep-ex",
    doi = "10.1103/2tcc-bqck",
    journal = "Phys. Rev. Lett.",
    volume = "135",
    number = "7",
    pages = "071002",
    year = "2025"
}

@article{Zhang:2025ajc,
    author = "Zhang, Minzhen and others",
    title = "{Search for Light Dark Matter with 259-day data in PandaX-4T}",
    eprint = "2507.11930",
    archivePrefix = "arXiv",
    primaryClass = "hep-ex",
    month = "7",
    year = "2025"
}

@article{NA64:2024klw,
    author = "Andreev, Yu. M. and others",
    collaboration = "NA64",
    title = "{First Results in the Search for Dark Sectors at NA64 with the CERN SPS High Energy Muon Beam}",
    eprint = "2401.01708",
    archivePrefix = "arXiv",
    primaryClass = "hep-ex",
    reportNumber = "CERN-EP-2023-306",
    doi = "10.1103/PhysRevLett.132.211803",
    journal = "Phys. Rev. Lett.",
    volume = "132",
    number = "21",
    pages = "211803",
    year = "2024"
}

@article{BaBar:2016sci,
    author = "Lees, J. P. and others",
    collaboration = "BaBar",
    title = "{Search for a muonic dark force at BABAR}",
    eprint = "1606.03501",
    archivePrefix = "arXiv",
    primaryClass = "hep-ex",
    reportNumber = "BABAR-PUB-16-003, SLAC-PUB-16549",
    doi = "10.1103/PhysRevD.94.011102",
    journal = "Phys. Rev. D",
    volume = "94",
    number = "1",
    pages = "011102",
    year = "2016"
}

@article{CMS:2024jyb,
    author = "Hayrapetyan, Aram and others",
    collaboration = "CMS",
    title = "{Model-independent search for pair production of new bosons decaying into muons in proton-proton collisions at $\sqrt{s}$ = 13 TeV}",
    eprint = "2407.20425",
    archivePrefix = "arXiv",
    primaryClass = "hep-ex",
    reportNumber = "CMS-HIG-21-004, CERN-EP-2024-199",
    doi = "10.1007/JHEP12(2024)172",
    journal = "JHEP",
    volume = "12",
    pages = "172",
    year = "2024"
}

@article{Chen:2018vkr,
    author = "Chen, Chien-Yi and Kozaczuk, Jonathan and Zhong, Yi-Ming",
    title = "{Exploring leptophilic dark matter with NA64-$\mu$}",
    eprint = "1807.03790",
    archivePrefix = "arXiv",
    primaryClass = "hep-ph",
    doi = "10.1007/JHEP10(2018)154",
    journal = "JHEP",
    volume = "10",
    pages = "154",
    year = "2018"
}

@article{Rella:2022len,
    author = {Rella, Claudia and D{\"o}brich, Babette and Yu, Tien-Tien},
    title = "{Searching for muonphilic dark sectors with proton beams}",
    eprint = "2205.09870",
    archivePrefix = "arXiv",
    primaryClass = "hep-ph",
    doi = "10.1103/PhysRevD.106.035023",
    journal = "Phys. Rev. D",
    volume = "106",
    number = "3",
    pages = "035023",
    year = "2022"
}

@article{Tsai:1986tx,
    author = "Tsai, Yung-Su",
    editor = "Loken, S. C.",
    title = "{AXION BREMSSTRAHLUNG BY AN ELECTRON BEAM}",
    reportNumber = "SLAC-PUB-3926",
    doi = "10.1103/PhysRevD.34.1326",
    journal = "Phys. Rev. D",
    volume = "34",
    pages = "1326",
    year = "1986"
}

@article{Tsai:1989vw,
    author = "Tsai, Yung-Su",
    title = "{PRODUCTION OF NEUTRAL BOSONS BY AN ELECTRON BEAM}",
    reportNumber = "SLAC-PUB-4877",
    doi = "10.1103/PhysRevD.40.760",
    journal = "Phys. Rev. D",
    volume = "40",
    pages = "760",
    year = "1989"
}

@book{Byckling:1971vca,
    author = "Byckling, Eero and Kajantie, K.",
    title = "{Particle Kinematics}: {(Chapters I-VI, X)}",
    publisher = "University of Jyvaskyla",
    address = "Jyvaskyla, Finland",
    year = "1971"
}

@article{Dutta:2025fgz,
    author = "Dutta, Bhaskar and Karthikeyan, Aparajitha and Kim, Doojin and Thompson, Adrian and Van de Water, Richard G.",
    title = "{Photon Excess from Dark Matter and Neutrino Scattering at MiniBooNE and MicroBooNE}",
    eprint = "2504.08071",
    archivePrefix = "arXiv",
    primaryClass = "hep-ph",
    reportNumber = "MI-HET-855",
    month = "4",
    year = "2025"
}

@article{PhysRevD.55.1233,
  title = {Muon bremsstrahlung on heavy atoms},
  author = {Andreev, Yu. M. and Bugaev, E. V.},
  journal = {Phys. Rev. D},
  volume = {55},
  issue = {3},
  pages = {1233--1243},
  numpages = {0},
  year = {1997},
  month = {Feb},
  publisher = {American Physical Society},
  doi = {10.1103/PhysRevD.55.1233},
  url = {https://link.aps.org/doi/10.1103/PhysRevD.55.1233}
}

@article{Dent:2020jhf,
    author = "Dent, James B. and Dutta, Bhaskar and Newstead, Jayden L. and Thompson, Adrian",
    title = "{Inverse Primakoff Scattering as a Probe of Solar Axions at Liquid Xenon Direct Detection Experiments}",
    eprint = "2006.15118",
    archivePrefix = "arXiv",
    primaryClass = "hep-ph",
    reportNumber = "MI-TH-2018",
    doi = "10.1103/PhysRevLett.125.131805",
    journal = "Phys. Rev. Lett.",
    volume = "125",
    number = "13",
    pages = "131805",
    year = "2020"
}

@article{SOLAX:1997lpz,
    author = "Avignone, III, F. T. and others",
    collaboration = "SOLAX",
    title = "{Experimental search for solar axions via coherent Primakoff conversion in a germanium spectrometer}",
    eprint = "astro-ph/9708008",
    archivePrefix = "arXiv",
    doi = "10.1103/PhysRevLett.81.5068",
    journal = "Phys. Rev. Lett.",
    volume = "81",
    pages = "5068--5071",
    year = "1998"
}

@article{Creswick:1997pg,
    author = "Creswick, R. J. and Avignone, III, F. T. and Farach, H. A. and Collar, J. I. and Gattone, A. O. and Nussinov, S. and Zioutas, K.",
    title = "{Theory for the direct detection of solar axions by coherent Primakoff conversion in germanium detectors}",
    eprint = "hep-ph/9708210",
    archivePrefix = "arXiv",
    doi = "10.1016/S0370-2693(98)00183-X",
    journal = "Phys. Lett. B",
    volume = "427",
    pages = "235--240",
    year = "1998"
}

@article{Avignone:1988bv,
    author = "Avignone, F. T. and Baktash, C. and Barker, W. C. and Calaprice, F. P. and Dunford, R. W. and Haxton, W. C. and Kahana, D. and Kouzes, R. T. and Miley, H. S. and Moltz, D. M.",
    title = "{Search for Axions From the 1115-kev Transition of $^{65}$Cu}",
    doi = "10.1103/PhysRevD.37.618",
    journal = "Phys. Rev. D",
    volume = "37",
    pages = "618--630",
    year = "1988"
}

@inproceedings{PRISM,
  author = {Marco Del Tutto},
  title = {SBND-PRISM: Sampling Multiple Off-Axis Neutrino Fluxes with the Same Detector},
  booktitle = {APS April Meeting 2021, Volume 66, Number 5},
  year = {2021},
  organization = {American Physical Society},
  address = {Online},
  note = {Presentation at APS April Meeting 2021}
}

@article{Steigman:1979kw,
    author = "Steigman, G.",
    title = "{Cosmology Confronts Particle Physics}",
    doi = "10.1146/annurev.ns.29.120179.001525",
    journal = "Ann. Rev. Nucl. Part. Sci.",
    volume = "29",
    pages = "313--338",
    year = "1979"
}

@article{Wolfram:1978gp,
    author = "Wolfram, Stephen",
    title = "{Abundances of Stable Particles Produced in the Early Universe}",
    reportNumber = "CALT-68-689, PRINT-78-0529 (OXFORD)",
    doi = "10.1016/0370-2693(79)90426-X",
    journal = "Phys. Lett. B",
    volume = "82",
    pages = "65--68",
    year = "1979"
}

@article{Lee:1977ua,
    author = "Lee, Benjamin W. and Weinberg, Steven",
    editor = "Srednicki, M. A.",
    title = "{Cosmological Lower Bound on Heavy Neutrino Masses}",
    reportNumber = "FERMILAB-PUB-77-041-T",
    doi = "10.1103/PhysRevLett.39.165",
    journal = "Phys. Rev. Lett.",
    volume = "39",
    pages = "165--168",
    year = "1977"
}

@article{Dutta:2019fxn,
    author = "Dutta, Bhaskar and Ghosh, Sumit and Kumar, Jason",
    title = "{A sub-GeV dark matter model}",
    eprint = "1905.02692",
    archivePrefix = "arXiv",
    primaryClass = "hep-ph",
    reportNumber = "MI-TH-1921, UH511-1305-2019",
    doi = "10.1103/PhysRevD.100.075028",
    journal = "Phys. Rev. D",
    volume = "100",
    pages = "075028",
    year = "2019"
}

@article{Dutta:2020scq,
    author = "Dutta, Bhaskar and Ghosh, Sumit and Li, Tianjun",
    title = "{Explaining $(g-2)_{\mu,e}$, the KOTO anomaly and the MiniBooNE excess in an extended Higgs model with sterile neutrinos}",
    eprint = "2006.01319",
    archivePrefix = "arXiv",
    primaryClass = "hep-ph",
    reportNumber = "MI-TH-2012",
    doi = "10.1103/PhysRevD.102.055017",
    journal = "Phys. Rev. D",
    volume = "102",
    number = "5",
    pages = "055017",
    year = "2020"
}

@article{Escudero:2019gvw,
    author = "Escudero, Miguel and Witte, Samuel J.",
    title = "{A CMB search for the neutrino mass mechanism and its relation to the Hubble tension}",
    eprint = "1909.04044",
    archivePrefix = "arXiv",
    primaryClass = "astro-ph.CO",
    reportNumber = "KCL-2019-71",
    doi = "10.1140/epjc/s10052-020-7854-5",
    journal = "Eur. Phys. J. C",
    volume = "80",
    number = "4",
    pages = "294",
    year = "2020"
}

@article{Batell:2018fqo,
    author = "Batell, Brian and Freitas, Ayres and Ismail, Ahmed and Mckeen, David",
    title = "{Probing Light Dark Matter with a Hadrophilic Scalar Mediator}",
    eprint = "1812.05103",
    archivePrefix = "arXiv",
    primaryClass = "hep-ph",
    reportNumber = "PITT-PACC-1817",
    doi = "10.1103/PhysRevD.100.095020",
    journal = "Phys. Rev. D",
    volume = "100",
    number = "9",
    pages = "095020",
    year = "2019"
}

@article{XENON:2017vdw,
    author = "Aprile, E. and others",
    collaboration = "XENON",
    title = "{First Dark Matter Search Results from the XENON1T Experiment}",
    eprint = "1705.06655",
    archivePrefix = "arXiv",
    primaryClass = "astro-ph.CO",
    doi = "10.1103/PhysRevLett.119.181301",
    journal = "Phys. Rev. Lett.",
    volume = "119",
    number = "18",
    pages = "181301",
    year = "2017"
}

@article{PICO:2017tgi,
    author = "Amole, C. and others",
    collaboration = "PICO",
    title = "{Dark Matter Search Results from the PICO-60 C$_3$F$_8$ Bubble Chamber}",
    eprint = "1702.07666",
    archivePrefix = "arXiv",
    primaryClass = "astro-ph.CO",
    reportNumber = "FERMILAB-PUB-17-058-AE-PPD",
    doi = "10.1103/PhysRevLett.118.251301",
    journal = "Phys. Rev. Lett.",
    volume = "118",
    number = "25",
    pages = "251301",
    year = "2017"
}

@article{SuperCDMS:2015eex,
    author = "Agnese, R. and others",
    collaboration = "SuperCDMS",
    title = "{New Results from the Search for Low-Mass Weakly Interacting Massive Particles with the CDMS Low Ionization Threshold Experiment}",
    eprint = "1509.02448",
    archivePrefix = "arXiv",
    primaryClass = "astro-ph.CO",
    reportNumber = "IPPP-15-56, DCTP-15-112, FERMILAB-PUB-15-394-AE",
    doi = "10.1103/PhysRevLett.116.071301",
    journal = "Phys. Rev. Lett.",
    volume = "116",
    number = "7",
    pages = "071301",
    year = "2016"
}

@article{CRESST:2017cdd,
    author = "Petricca, F. and others",
    editor = "Clark, Ken and Jillings, Chris and Kraus, Christine and Saffin, Jenna and Scorza, Silvia",
    collaboration = "CRESST",
    title = "{First results on low-mass dark matter from the CRESST-III experiment}",
    eprint = "1711.07692",
    archivePrefix = "arXiv",
    primaryClass = "astro-ph.CO",
    doi = "10.1088/1742-6596/1342/1/012076",
    journal = "J. Phys. Conf. Ser.",
    volume = "1342",
    number = "1",
    pages = "012076",
    year = "2020"
}

@article{CRESST:2015txj,
    author = "Angloher, G. and others",
    collaboration = "CRESST",
    title = "{Results on light dark matter particles with a low-threshold CRESST-II detector}",
    eprint = "1509.01515",
    archivePrefix = "arXiv",
    primaryClass = "astro-ph.CO",
    doi = "10.1140/epjc/s10052-016-3877-3",
    journal = "Eur. Phys. J. C",
    volume = "76",
    number = "1",
    pages = "25",
    year = "2016"
}

@article{CRESST:2017ues,
    author = "Angloher, G. and others",
    collaboration = "CRESST",
    title = "{Results on MeV-scale dark matter from a gram-scale cryogenic calorimeter operated above ground}",
    eprint = "1707.06749",
    archivePrefix = "arXiv",
    primaryClass = "astro-ph.CO",
    doi = "10.1140/epjc/s10052-017-5223-9",
    journal = "Eur. Phys. J. C",
    volume = "77",
    number = "9",
    pages = "637",
    year = "2017"
}

@article{E787:2002qfb,
    author = "Adler, S and others",
    collaboration = "E787",
    title = "{Search for the decay K+ ---{\ensuremath{>}} pi+ nu anti-nu in the momentum region P(pi) less than 195-MeV/c}",
    eprint = "hep-ex/0201037",
    archivePrefix = "arXiv",
    reportNumber = "BNL-68967",
    doi = "10.1016/S0370-2693(02)01911-1",
    journal = "Phys. Lett. B",
    volume = "537",
    pages = "211--216",
    year = "2002"
}

@article{E787:2004ovg,
    author = "Adler, S. and others",
    collaboration = "E787",
    title = "{Further search for the decay K+ ---{\ensuremath{>}} pi+ nu anti-nu in the momentum region P {\ensuremath{<}} 195-MeV/c}",
    eprint = "hep-ex/0403034",
    archivePrefix = "arXiv",
    reportNumber = "BNL-72163-2004-JA, KEK-PREPRINT-2004-2",
    doi = "10.1103/PhysRevD.70.037102",
    journal = "Phys. Rev. D",
    volume = "70",
    pages = "037102",
    year = "2004"
}

@article{E949:2007xyy,
    author = "Adler, S. and others",
    collaboration = "E949, E787",
    title = "{Measurement of the $K^+ \to \pi^+ \nu \bar{\nu}$ Branching Ratio}",
    eprint = "0709.1000",
    archivePrefix = "arXiv",
    primaryClass = "hep-ex",
    reportNumber = "BNL-79257-2007-JA, KEK-2007-34, TRI-PP-07-18, TUHEP-EX-07-002, FERMILAB-PUB-08-065-CD-E",
    doi = "10.1103/PhysRevD.77.052003",
    journal = "Phys. Rev. D",
    volume = "77",
    pages = "052003",
    year = "2008"
}

@article{MiniBooNE:2017nqe,
    author = "Aguilar-Arevalo, A. A. and others",
    collaboration = "MiniBooNE",
    title = "{Dark Matter Search in a Proton Beam Dump with MiniBooNE}",
    eprint = "1702.02688",
    archivePrefix = "arXiv",
    primaryClass = "hep-ex",
    reportNumber = "FERMILAB-PUB-17-059-AD-E-ND",
    doi = "10.1103/PhysRevLett.118.221803",
    journal = "Phys. Rev. Lett.",
    volume = "118",
    number = "22",
    pages = "221803",
    year = "2017"
}

@article{MiniBooNE:2012jpi,
    author = "Dharmapalan, R. and others",
    collaboration = "MiniBooNE",
    title = "{Light Mass WIMP Searches with a Neutrino Experiment: A Proposal for Further MiniBooNE Running}",
    eprint = "1211.2258",
    archivePrefix = "arXiv",
    primaryClass = "hep-ex",
    reportNumber = "FERMILAB-PROPOSAL-1032",
    month = "9",
    year = "2012"
}

@article{MiniBooNEDM:2018cxm,
    author = "Aguilar-Arevalo, A. A. and others",
    collaboration = "MiniBooNE DM",
    title = "{Dark Matter Search in Nucleon, Pion, and Electron Channels from a Proton Beam Dump with MiniBooNE}",
    eprint = "1807.06137",
    archivePrefix = "arXiv",
    primaryClass = "hep-ex",
    reportNumber = "LA-UR-18-26421, FERMILAB-PUB-18-334-ND",
    doi = "10.1103/PhysRevD.98.112004",
    journal = "Phys. Rev. D",
    volume = "98",
    number = "11",
    pages = "112004",
    year = "2018"
}

@article{MicroBooNE:2025rsd,
    author = "Abratenko, P. and others",
    collaboration = "MicroBooNE",
    title = "{First Search for Neutral Current Coherent Single-Photon Production in MicroBooNE}",
    eprint = "2502.06091",
    archivePrefix = "arXiv",
    primaryClass = "hep-ex",
    reportNumber = "FERMILAB-PUB-25-0056",
    month = "2",
    year = "2025"
}

@article{MiniBooNE:2008hfu,
    author = "Aguilar-Arevalo, A. A. and others",
    collaboration = "MiniBooNE",
    title = "{The Neutrino Flux Prediction at MiniBooNE}",
    eprint = "0806.1449",
    archivePrefix = "arXiv",
    primaryClass = "hep-ex",
    reportNumber = "FERMILAB-PUB-08-161-AD-E",
    doi = "10.1103/PhysRevD.79.072002",
    journal = "Phys. Rev. D",
    volume = "79",
    pages = "072002",
    year = "2009"
}

@article{MicroBooNE:2016pwy,
    author = "Acciarri, R. and others",
    collaboration = "MicroBooNE",
    title = "{Design and Construction of the MicroBooNE Detector}",
    eprint = "1612.05824",
    archivePrefix = "arXiv",
    primaryClass = "physics.ins-det",
    reportNumber = "FERMILAB-PUB-16-613-ND",
    doi = "10.1088/1748-0221/12/02/P02017",
    journal = "JINST",
    volume = "12",
    number = "02",
    pages = "P02017",
    year = "2017"
}

@article{Perdrisat:2006hj,
    author = "Perdrisat, C. F. and Punjabi, V. and Vanderhaeghen, M.",
    title = "{Nucleon Electromagnetic Form Factors}",
    eprint = "hep-ph/0612014",
    archivePrefix = "arXiv",
    reportNumber = "WM-06-115, JLAB-THY-06-595",
    doi = "10.1016/j.ppnp.2007.05.001",
    journal = "Prog. Part. Nucl. Phys.",
    volume = "59",
    pages = "694--764",
    year = "2007"
}

@article{Helm:1956zz,
    author = "Helm, Richard H.",
    title = "{Inelastic and Elastic Scattering of 187-Mev Electrons from Selected Even-Even Nuclei}",
    doi = "10.1103/PhysRev.104.1466",
    journal = "Phys. Rev.",
    volume = "104",
    pages = "1466--1475",
    year = "1956"
}

@article{Engel:1991wq,
    author = "Engel, J.",
    title = "{Nuclear form-factors for the scattering of weakly interacting massive particles}",
    doi = "10.1016/0370-2693(91)90712-Y",
    journal = "Phys. Lett. B",
    volume = "264",
    pages = "114--119",
    year = "1991"
}

@article{Brdar:2020quo,
    author = "Brdar, Vedran and Greljo, Admir and Kopp, Joachim and Opferkuch, Toby",
    title = "{The Neutrino Magnetic Moment Portal: Cosmology, Astrophysics, and Direct Detection}",
    eprint = "2007.15563",
    archivePrefix = "arXiv",
    primaryClass = "hep-ph",
    reportNumber = "CERN-TH-2020-130, MITP/20-041",
    doi = "10.1088/1475-7516/2021/01/039",
    journal = "JCAP",
    volume = "01",
    pages = "039",
    year = "2021"
}

@article{Dutta:2025npn,
    author = "Dutta, Bhaskar and Goswami, Debopam and Karthikeyan, Aparajitha and Kelly, Kevin J.",
    title = "{Dirt/Detector/Dump: complementary BSM production at Short-Baseline Neutrino Facilities}",
    eprint = "2501.09840",
    archivePrefix = "arXiv",
    primaryClass = "hep-ph",
    reportNumber = "MI-HET-842",
    doi = "10.1007/JHEP05(2025)240",
    journal = "JHEP",
    volume = "05",
    pages = "240",
    year = "2025"
}

@article{Foroughi-Abari:2020gju,
    author = "Foroughi-Abari, Saeid and Ritz, Adam",
    title = "{LSND Constraints on the Higgs Portal}",
    eprint = "2004.14515",
    archivePrefix = "arXiv",
    primaryClass = "hep-ph",
    doi = "10.1103/PhysRevD.102.035015",
    journal = "Phys. Rev. D",
    volume = "102",
    number = "3",
    pages = "035015",
    year = "2020"
}

@article{Batell:2019nwo,
    author = "Batell, Brian and Berger, Joshua and Ismail, Ahmed",
    title = "{Probing the Higgs Portal at the Fermilab Short-Baseline Neutrino Experiments}",
    eprint = "1909.11670",
    archivePrefix = "arXiv",
    primaryClass = "hep-ph",
    reportNumber = "PITT-PACC-1906, OSU-HEP-19-07",
    doi = "10.1103/PhysRevD.100.115039",
    journal = "Phys. Rev. D",
    volume = "100",
    number = "11",
    pages = "115039",
    year = "2019"
}

@article{Gehrlein:2025tko,
    author = "Gehrlein, Julia and Shoemaker, Ian M. and Thapa, Anil",
    title = "{Monophotons at Neutrino Experiments from Neutrino Polarizability}",
    eprint = "2506.14881",
    archivePrefix = "arXiv",
    primaryClass = "hep-ph",
    month = "6",
    year = "2025"
}

@misc{SearchesforBSM,
title={Searches for Beyond Standard Model
Physics in the SBND neutrino experiment},
author={Crespo-Anadón, J. I.},
collaboration={SBND Collaboration},
year={2023}
}

@misc{SearchesforBSM1,
title={Searches for Beyond Standard Model
Physics in the SBND experiment},
author={Luo, X.},
collaboration={SBND Collaboration},
year={2024}
}

@article{SBND:2024vgn,
    author = "Abratenko, P. and others",
    collaboration = "SBND",
    title = "{Scintillation light in SBND: simulation, reconstruction, and expected performance of the photon detection system}",
    eprint = "2406.07514",
    archivePrefix = "arXiv",
    primaryClass = "physics.ins-det",
    reportNumber = "FERMILAB-PUB-24-0303-PPD",
    doi = "10.1140/epjc/s10052-024-13306-3",
    journal = "Eur. Phys. J. C",
    volume = "84",
    number = "10",
    pages = "1046",
    year = "2024"
}

@article{Hasnip:2025gyi,
    author = "Hasnip, Ciaran",
    collaboration = "DUNE",
    title = "{DUNE-PRISM: Reducing neutrino interaction model dependence with a movable neutrino detector}",
    eprint = "2501.14811",
    archivePrefix = "arXiv",
    primaryClass = "hep-ex",
    month = "1",
    year = "2025"
}

@phdthesis{Hasnip:2023ygr,
    author = "Hasnip, C.",
    title = "{DUNE-PRISM - a new method to measure neutrino oscillations}",
    reportNumber = "FERMILAB-THESIS-2023-21",
    doi = "10.5287/ora-9opn82v9m",
    school = "Oxford University, Oxford U.",
    year = "2023"
}

@article{CCM:2021jmk,
    author = "Aguilar-Arevalo, A. A. and others",
    collaboration = "CCM",
    title = "{Prospects for detecting axionlike particles at the Coherent CAPTAIN-Mills experiment}",
    eprint = "2112.09979",
    archivePrefix = "arXiv",
    primaryClass = "hep-ph",
    reportNumber = "LA-UR-21-28474",
    doi = "10.1103/PhysRevD.107.095036",
    journal = "Phys. Rev. D",
    volume = "107",
    number = "9",
    pages = "095036",
    year = "2023"
}

@article{CCM:2021leg,
    author = "Aguilar-Arevalo, A. A. and others",
    collaboration = "CCM",
    title = "{First dark matter search results from Coherent CAPTAIN-Mills}",
    eprint = "2105.14020",
    archivePrefix = "arXiv",
    primaryClass = "hep-ex",
    reportNumber = "LA-UR-21-24983",
    doi = "10.1103/PhysRevD.106.012001",
    journal = "Phys. Rev. D",
    volume = "106",
    number = "1",
    pages = "012001",
    year = "2022"
}

@article{DUNE:2016hlj,
    author = "Acciarri, R. and others",
    collaboration = "DUNE",
    title = "{Long-Baseline Neutrino Facility (LBNF) and Deep Underground Neutrino Experiment (DUNE)}: {Conceptual Design Report, Volume 1: The LBNF and DUNE Projects}",
    eprint = "1601.05471",
    archivePrefix = "arXiv",
    primaryClass = "physics.ins-det",
    reportNumber = "FERMILAB-DESIGN-2016-01",
    month = "1",
    year = "2016"
}

@article{DUNE:2021tad,
    author = "Hewes, V. and others",
    collaboration = "DUNE",
    title = "{Deep Underground Neutrino Experiment (DUNE) Near Detector Conceptual Design Report}",
    eprint = "2103.13910",
    archivePrefix = "arXiv",
    primaryClass = "physics.ins-det",
    reportNumber = "FERMILAB-PUB-21-067-E-LBNF-PPD-SCD-T",
    doi = "10.3390/instruments5040031",
    journal = "Instruments",
    volume = "5",
    number = "4",
    pages = "31",
    year = "2021"
}

@article{MicroBooNE:2015bmn,
    author = "Acciarri, R. and others",
    collaboration = "MicroBooNE, LAr1-ND, ICARUS-WA104",
    title = "{A Proposal for a Three Detector Short-Baseline Neutrino Oscillation Program in the Fermilab Booster Neutrino Beam}",
    eprint = "1503.01520",
    archivePrefix = "arXiv",
    primaryClass = "physics.ins-det",
    reportNumber = "FERMILAB-PUB-15-0726-AD-PPD",
    month = "3",
    year = "2015"
}

@article{ICARUS:2004wqc,
    author = "Amerio, S. and others",
    collaboration = "ICARUS",
    title = "{Design, construction and tests of the ICARUS T600 detector}",
    doi = "10.1016/j.nima.2004.02.044",
    journal = "Nucl. Instrum. Meth. A",
    volume = "527",
    pages = "329--410",
    year = "2004"
}

@article{Carlson:2012pc,
    author = "Carlson, Carl E. and Rislow, Benjamin C.",
    title = "{New Physics and the Proton Radius Problem}",
    eprint = "1206.3587",
    archivePrefix = "arXiv",
    primaryClass = "hep-ph",
    doi = "10.1103/PhysRevD.86.035013",
    journal = "Phys. Rev. D",
    volume = "86",
    pages = "035013",
    year = "2012"
}

@article{Barger:2011mt,
    author = "Barger, Vernon and Chiang, Cheng-Wei and Keung, Wai-Yee and Marfatia, Danny",
    title = "{Constraint on parity-violating muonic forces}",
    eprint = "1109.6652",
    archivePrefix = "arXiv",
    primaryClass = "hep-ph",
    doi = "10.1103/PhysRevLett.108.081802",
    journal = "Phys. Rev. Lett.",
    volume = "108",
    pages = "081802",
    year = "2012"
}

@article{Laha:2013xua,
    author = "Laha, Ranjan and Dasgupta, Basudeb and Beacom, John F.",
    title = "{Constraints on New Neutrino Interactions via Light Abelian Vector Bosons}",
    eprint = "1304.3460",
    archivePrefix = "arXiv",
    primaryClass = "hep-ph",
    doi = "10.1103/PhysRevD.89.093025",
    journal = "Phys. Rev. D",
    volume = "89",
    number = "9",
    pages = "093025",
    year = "2014"
}

@article{Foroughi-Abari:2021zbm,
    author = "Foroughi-Abari, Saeid and Ritz, Adam",
    title = "{Dark sector production via proton bremsstrahlung}",
    eprint = "2108.05900",
    archivePrefix = "arXiv",
    primaryClass = "hep-ph",
    doi = "10.1103/PhysRevD.105.095045",
    journal = "Phys. Rev. D",
    volume = "105",
    number = "9",
    pages = "095045",
    year = "2022"
}

@article{CCM:2023itc,
    author = "Aguilar-Arevalo, A. A. and others",
    collaboration = "CCM",
    title = "{Testing meson portal dark sector solutions to the MiniBooNE anomaly at the Coherent CAPTAIN Mills experiment}",
    eprint = "2309.02599",
    archivePrefix = "arXiv",
    primaryClass = "hep-ph",
    reportNumber = "LA-UR-23-29529",
    doi = "10.1103/PhysRevD.109.095017",
    journal = "Phys. Rev. D",
    volume = "109",
    number = "9",
    pages = "095017",
    year = "2024",
    note = "[Addendum: Phys.Rev.D 111, 035030 (2025)]"
}

@article{Freese:2017idy,
    author = "Freese, Katherine",
    editor = "Bianchi, Massimo and Jantzen, Robert T. and Ruffini, Remo",
    title = "{Status of Dark Matter in the Universe}",
    eprint = "1701.01840",
    archivePrefix = "arXiv",
    primaryClass = "astro-ph.CO",
    doi = "10.1142/S0218271817300129",
    journal = "Int. J. Mod. Phys.",
    volume = "1",
    number = "06",
    pages = "325--355",
    year = "2017"
}

@article{babcock1939rotation,
  title={The rotation of the Andromeda Nebula},
  author={Babcock, Horace W},
  journal={Lick Observatory bulletin; no. 498; Lick Observatory bulletins; no. 498., Berkeley: University of California Press,[1939], p. 41-51,[2] leaves of plates; 31 cm.},
  volume={19},
  pages={41--51},
  year={1939}
}

@article{Markevitch:2001ri,
    author = "Markevitch, M. and Gonzalez, A. H. and David, L. and Vikhlinin, A. and Murray, S. and Forman, W. and Jones, C. and Tucker, W.",
    title = "{A Textbook example of a bow shock in the merging galaxy cluster 1E0657-56}",
    eprint = "astro-ph/0110468",
    archivePrefix = "arXiv",
    doi = "10.1086/339619",
    journal = "Astrophys. J. Lett.",
    volume = "567",
    pages = "L27",
    year = "2002"
}

@article{Markevitch:2004qk,
    author = "Markevitch, Maxim and Govoni, F. and Brunetti, G. and Jerius, D.",
    title = "{Bow shock and radio halo in the merging cluster A520}",
    eprint = "astro-ph/0412451",
    archivePrefix = "arXiv",
    doi = "10.1086/430695",
    journal = "Astrophys. J.",
    volume = "627",
    pages = "733--738",
    year = "2005"
}

@article{Zwicky:1933gu,
    author = "Zwicky, F.",
    title = "{Die Rotverschiebung von extragalaktischen Nebeln}",
    doi = "10.1007/s10714-008-0707-4",
    journal = "Helv. Phys. Acta",
    volume = "6",
    pages = "110--127",
    year = "1933"
}

@article{Zwicky:1937zza,
    author = "Zwicky, F.",
    title = "{On the Masses of Nebulae and of Clusters of Nebulae}",
    doi = "10.1086/143864",
    journal = "Astrophys. J.",
    volume = "86",
    pages = "217--246",
    year = "1937"
}

@article{Neto:2007vq,
    author = "Neto, Angelo F. and Gao, Liang and Bett, Philip and Cole, Shaun and Navarro, Julio F. and Frenk, Carlos S. and White, Simon D. M. and Springel, Volker and Jenkins, Adrian",
    title = "{The statistics of lambda CDM Halo Concentrations}",
    eprint = "0706.2919",
    archivePrefix = "arXiv",
    primaryClass = "astro-ph",
    doi = "10.1111/j.1365-2966.2007.12381.x",
    journal = "Mon. Not. Roy. Astron. Soc.",
    volume = "381",
    pages = "1450--1462",
    year = "2007"
}

@article{Maccio:2008pcd,
    author = "Maccio', Andrea V. and Dutton, Aaron A. and Bosch, Frank C. van den",
    title = "{Concentration, Spin and Shape of Dark Matter Haloes as a Function of the Cosmological Model: WMAP1, WMAP3 and WMAP5 results}",
    eprint = "0805.1926",
    archivePrefix = "arXiv",
    primaryClass = "astro-ph",
    doi = "10.1111/j.1365-2966.2008.14029.x",
    journal = "Mon. Not. Roy. Astron. Soc.",
    volume = "391",
    pages = "1940--1954",
    year = "2008"
}

@article{Springel:2008cc,
    author = "Springel, Volker and Wang, Jie and Vogelsberger, Mark and Ludlow, Aaron and Jenkins, Adrian and Helmi, Amina and Navarro, Julio F. and Frenk, Carlos S. and White, Simon D. M.",
    title = "{The Aquarius Project: the subhalos of galactic halos}",
    eprint = "0809.0898",
    archivePrefix = "arXiv",
    primaryClass = "astro-ph",
    doi = "10.1111/j.1365-2966.2008.14066.x",
    journal = "Mon. Not. Roy. Astron. Soc.",
    volume = "391",
    pages = "1685--1711",
    year = "2008"
}

@article{XENON:2018voc,
    author = "Aprile, E. and others",
    collaboration = "XENON",
    title = "{Dark Matter Search Results from a One Ton-Year Exposure of XENON1T}",
    eprint = "1805.12562",
    archivePrefix = "arXiv",
    primaryClass = "astro-ph.CO",
    doi = "10.1103/PhysRevLett.121.111302",
    journal = "Phys. Rev. Lett.",
    volume = "121",
    number = "11",
    pages = "111302",
    year = "2018"
}

@article{XENON:2019gfn,
    author = "Aprile, E. and others",
    collaboration = "XENON",
    title = "{Light Dark Matter Search with Ionization Signals in XENON1T}",
    eprint = "1907.11485",
    archivePrefix = "arXiv",
    primaryClass = "hep-ex",
    doi = "10.1103/PhysRevLett.123.251801",
    journal = "Phys. Rev. Lett.",
    volume = "123",
    number = "25",
    pages = "251801",
    year = "2019"
}

@article{XENON:2020fgj,
    author = "Aprile, E. and others",
    collaboration = "XENON",
    title = "{Search for inelastic scattering of WIMP dark matter in XENON1T}",
    eprint = "2011.10431",
    archivePrefix = "arXiv",
    primaryClass = "hep-ex",
    doi = "10.1103/PhysRevD.103.063028",
    journal = "Phys. Rev. D",
    volume = "103",
    number = "6",
    pages = "063028",
    year = "2021"
}

@article{DarkSide:2018ppu,
    author = "Agnes, P. and others",
    collaboration = "DarkSide",
    title = "{Constraints on Sub-GeV Dark-Matter\textendash{}Electron Scattering from the DarkSide-50 Experiment}",
    eprint = "1802.06998",
    archivePrefix = "arXiv",
    primaryClass = "astro-ph.CO",
    reportNumber = "FERMILAB-PUB-18-052-AD-AE-CD-E",
    doi = "10.1103/PhysRevLett.121.111303",
    journal = "Phys. Rev. Lett.",
    volume = "121",
    number = "11",
    pages = "111303",
    year = "2018"
}

@article{SuperCDMS:2020aus,
    author = "Alkhatib, I. and others",
    collaboration = "SuperCDMS",
    title = "{Light Dark Matter Search with a High-Resolution Athermal Phonon Detector Operated Above Ground}",
    eprint = "2007.14289",
    archivePrefix = "arXiv",
    primaryClass = "hep-ex",
    doi = "10.1103/PhysRevLett.127.061801",
    journal = "Phys. Rev. Lett.",
    volume = "127",
    pages = "061801",
    year = "2021"
}

@article{PandaX:2023xgl,
    author = "Huang, Di and others",
    collaboration = "PandaX",
    title = "{Search for Dark-Matter\textendash{}Nucleon Interactions with a Dark Mediator in PandaX-4T}",
    eprint = "2308.01540",
    archivePrefix = "arXiv",
    primaryClass = "hep-ex",
    doi = "10.1103/PhysRevLett.131.191002",
    journal = "Phys. Rev. Lett.",
    volume = "131",
    number = "19",
    pages = "191002",
    year = "2023"
}

@article{SENSEI:2024yyt,
    author = "Bloch, Itay M. and others",
    collaboration = "SENSEI",
    title = "{SENSEI at SNOLAB: Single-Electron Event Rate and Implications for Dark Matter}",
    eprint = "2410.18716",
    archivePrefix = "arXiv",
    primaryClass = "astro-ph.CO",
    reportNumber = "FERMILAB-PUB-24-0767-PPD, YITP-SB-2024-25",
    doi = "10.1103/PhysRevLett.134.161002",
    journal = "Phys. Rev. Lett.",
    volume = "134",
    number = "16",
    pages = "161002",
    year = "2025"
}

@article{XENON:2019zpr,
    author = "Aprile, E. and others",
    collaboration = "XENON",
    title = "{Search for Light Dark Matter Interactions Enhanced by the Migdal Effect or Bremsstrahlung in XENON1T}",
    eprint = "1907.12771",
    archivePrefix = "arXiv",
    primaryClass = "hep-ex",
    doi = "10.1103/PhysRevLett.123.241803",
    journal = "Phys. Rev. Lett.",
    volume = "123",
    number = "24",
    pages = "241803",
    year = "2019"
}

@article{Berlin:2018bsc,
    author = "Berlin, Asher and Blinov, Nikita and Krnjaic, Gordan and Schuster, Philip and Toro, Natalia",
    title = "{Dark Matter, Millicharges, Axion and Scalar Particles, Gauge Bosons, and Other New Physics with LDMX}",
    eprint = "1807.01730",
    archivePrefix = "arXiv",
    primaryClass = "hep-ph",
    reportNumber = "FERMILAB-PUB-18-310-A, SLAC-PUB-17297",
    doi = "10.1103/PhysRevD.99.075001",
    journal = "Phys. Rev. D",
    volume = "99",
    number = "7",
    pages = "075001",
    year = "2019"
}

@article{Banerjee:2019pds,
    author = "Banerjee, D. and others",
    title = "{Dark matter search in missing energy events with NA64}",
    eprint = "1906.00176",
    archivePrefix = "arXiv",
    primaryClass = "hep-ex",
    reportNumber = "CERN-EP-2019-116",
    doi = "10.1103/PhysRevLett.123.121801",
    journal = "Phys. Rev. Lett.",
    volume = "123",
    number = "12",
    pages = "121801",
    year = "2019"
}

@article{Andreev:2024lps,
    author = "Andreev, Yu. M. and others",
    title = "{First constraints on the $L_\mu-L_\tau$ explanation of the muon $g-2$ anomaly from NA64-$e$ at CERN}",
    eprint = "2404.06982",
    archivePrefix = "arXiv",
    primaryClass = "hep-ex",
    reportNumber = "CERN-EP-2024-103",
    month = "4",
    year = "2024"
}

@article{COHERENT:2021pvd,
    author = "Akimov, D. and others",
    collaboration = "COHERENT",
    title = "{First Probe of Sub-GeV Dark Matter beyond the Cosmological Expectation with the COHERENT CsI Detector at the SNS}",
    eprint = "2110.11453",
    archivePrefix = "arXiv",
    primaryClass = "hep-ex",
    doi = "10.1103/PhysRevLett.130.051803",
    journal = "Phys. Rev. Lett.",
    volume = "130",
    number = "5",
    pages = "051803",
    year = "2023"
}

@article{LSND:2001akn,
    author = "Auerbach, L. B. and others",
    collaboration = "LSND",
    title = "{Measurement of electron - neutrino - electron elastic scattering}",
    eprint = "hep-ex/0101039",
    archivePrefix = "arXiv",
    doi = "10.1103/PhysRevD.63.112001",
    journal = "Phys. Rev. D",
    volume = "63",
    pages = "112001",
    year = "2001"
}

@article{SENSEI:2020dpa,
    author = "Barak, Liron and others",
    collaboration = "SENSEI",
    title = "{SENSEI: Direct-Detection Results on sub-GeV Dark Matter from a New Skipper-CCD}",
    eprint = "2004.11378",
    archivePrefix = "arXiv",
    primaryClass = "astro-ph.CO",
    reportNumber = "YITP-SB-2020-6, FERMILAB-PUB-20-158-AE-E",
    doi = "10.1103/PhysRevLett.125.171802",
    journal = "Phys. Rev. Lett.",
    volume = "125",
    number = "17",
    pages = "171802",
    year = "2020"
}

@article{Rott:2018rlw,
    author = "Rott, Carsten",
    collaboration = "JSNS\$\textasciicircum{}2\$",
    title = "{Status and Prospects of the JSNS$^2$ Experiment}",
    eprint = "1811.03321",
    archivePrefix = "arXiv",
    primaryClass = "physics.ins-det",
    doi = "10.22323/1.340.0185",
    journal = "PoS",
    volume = "ICHEP2018",
    pages = "185",
    year = "2019"
}

@article{Lattaud:2022jnq,
    author = "Lattaud, Hugues",
    collaboration = "EDELWEISS",
    title = "{Sub-GeV dark matter searches with EDELWEISS: New results and prospects}",
    eprint = "2211.04176",
    archivePrefix = "arXiv",
    primaryClass = "astro-ph.GA",
    doi = "10.21468/SciPostPhysProc.12.012",
    journal = "SciPost Phys. Proc.",
    volume = "12",
    pages = "012",
    year = "2023"
}

@article{PandaX:2024muv,
    author = "Bo, Zihao and others",
    collaboration = "PandaX",
    title = "{First Measurement of Solar $^8$B Neutrino Flux through Coherent Elastic Neutrino-Nucleus Scattering in PandaX-4T}",
    eprint = "2407.10892",
    archivePrefix = "arXiv",
    primaryClass = "hep-ex",
    month = "7",
    year = "2024"
}

@article{XENON:2024wpa,
    author = "Aprile, E. and others",
    collaboration = "XENON",
    title = "{The XENONnT dark matter experiment}",
    eprint = "2402.10446",
    archivePrefix = "arXiv",
    primaryClass = "physics.ins-det",
    doi = "10.1140/epjc/s10052-024-12982-5",
    journal = "Eur. Phys. J. C",
    volume = "84",
    number = "8",
    pages = "784",
    year = "2024"
}

@article{LZ:2023lvz,
    author = "Aalbers, J. and others",
    collaboration = "LZ",
    title = "{First constraints on WIMP-nucleon effective field theory couplings in an extended energy region from LUX-ZEPLIN}",
    eprint = "2312.02030",
    archivePrefix = "arXiv",
    primaryClass = "hep-ex",
    doi = "10.1103/PhysRevD.109.092003",
    journal = "Phys. Rev. D",
    volume = "109",
    number = "9",
    pages = "092003",
    year = "2024"
}

@article{SuperCDMS:2024yiv,
    author = "Albakry, M. F. and others",
    collaboration = "SuperCDMS",
    title = "{Light dark matter constraints from SuperCDMS HVeV detectors operated underground with an anticoincidence event selection}",
    eprint = "2407.08085",
    archivePrefix = "arXiv",
    primaryClass = "hep-ex",
    reportNumber = "FERMILAB-PUB-24-0376-PPD",
    doi = "10.1103/PhysRevD.111.012006",
    journal = "Phys. Rev. D",
    volume = "111",
    number = "1",
    pages = "012006",
    year = "2025"
}

@article{Dutta:2023fij,
    author = "Dutta, Bhaskar and Huang, Wei-Chih and Newstead, Jayden L.",
    title = "{Probing the Dark Sector with Nuclear Transition Photons}",
    eprint = "2302.10250",
    archivePrefix = "arXiv",
    primaryClass = "hep-ph",
    doi = "10.1103/PhysRevLett.131.111801",
    journal = "Phys. Rev. Lett.",
    volume = "131",
    number = "11",
    pages = "111801",
    year = "2023"
}

@article{Dutta:2024kuj,
    author = "Dutta, Bhaskar and Huang, Wei-Chih and Kim, Doojin and Newstead, Jayden L. and Park, Jong-Chul and Ali, Iman Shaukat",
    title = "{Exciting Prospects for Dark Matter at Large-Volume Neutrino Detectors}",
    eprint = "2402.04184",
    archivePrefix = "arXiv",
    primaryClass = "hep-ph",
    month = "2",
    year = "2024"
}

@article{deGouvea:2018cfv,
    author = "de Gouv\^ea, Andr\'e and Fox, Patrick J. and Harnik, Roni and Kelly, Kevin J. and Zhang, Yue",
    title = "{Dark Tridents at Off-Axis Liquid Argon Neutrino Detectors}",
    eprint = "1809.06388",
    archivePrefix = "arXiv",
    primaryClass = "hep-ph",
    reportNumber = "FERMILAB-PUB-18-433-T, NUHEP-TH/18-10",
    doi = "10.1007/JHEP01(2019)001",
    journal = "JHEP",
    volume = "01",
    pages = "001",
    year = "2019"
}

@article{Adrian:2022nkt,
    author = "Adrian, P. H. and others",
    title = "{Searching for prompt and long-lived dark photons in electroproduced e+e- pairs with the heavy photon search experiment at JLab}",
    eprint = "2212.10629",
    archivePrefix = "arXiv",
    primaryClass = "hep-ex",
    reportNumber = "JLAB-PHY-23-3738, FERMILAB-PUB-22-983-PPD",
    doi = "10.1103/PhysRevD.108.012015",
    journal = "Phys. Rev. D",
    volume = "108",
    number = "1",
    pages = "012015",
    year = "2023"
}

@article{Foot:1990mn,
    author = "Foot, Robert",
    title = "{New Physics From Electric Charge Quantization?}",
    reportNumber = "MAD/TH/90-14",
    doi = "10.1142/S0217732391000543",
    journal = "Mod. Phys. Lett. A",
    volume = "6",
    pages = "527--530",
    year = "1991"
}

@article{He:1990pn,
    author = "He, X. G. and Joshi, Girish C. and Lew, H. and Volkas, R. R.",
    title = "{NEW Z-prime PHENOMENOLOGY}",
    reportNumber = "UM-P-90/42, OZ-P-90/16",
    doi = "10.1103/PhysRevD.43.R22",
    journal = "Phys. Rev. D",
    volume = "43",
    pages = "22--24",
    year = "1991"
}

@article{He:1991qd,
    author = "He, Xiao-Gang and Joshi, Girish C. and Lew, H. and Volkas, R. R.",
    title = "{Simplest Z-prime model}",
    reportNumber = "CERN-TH-6084-91, UM-P-91-32, OZ-91-07",
    doi = "10.1103/PhysRevD.44.2118",
    journal = "Phys. Rev. D",
    volume = "44",
    pages = "2118--2132",
    year = "1991"
}

@article{Heeck:2011wj,
    author = "Heeck, Julian and Rodejohann, Werner",
    title = "{Gauged  $L_\mu  -  L_\tau$  Symmetry  at  the  Electroweak  Scale}",
    eprint = "1107.5238",
    archivePrefix = "arXiv",
    primaryClass = "hep-ph",
    doi = "10.1103/PhysRevD.84.075007",
    journal = "Phys. Rev. D",
    volume = "84",
    pages = "075007",
    year = "2011"
}

@article{Patt:2006fw,
    author = "Patt, Brian and Wilczek, Frank",
    title = "{Higgs-field portal into hidden sectors}",
    eprint = "hep-ph/0605188",
    archivePrefix = "arXiv",
    reportNumber = "MIT-CTP-3745",
    month = "5",
    year = "2006"
}

@article{Dutta:2022fdt,
    author = "Dutta, Bhaskar and Ghosh, Sumit and Li, Tianjun and Thompson, Adrian and Verma, Ankur",
    title = "{Non-standard neutrino interactions in light mediator models at reactor experiments}",
    eprint = "2209.13566",
    archivePrefix = "arXiv",
    primaryClass = "hep-ph",
    reportNumber = "MI-HET-775, KIAS-P22043",
    doi = "10.1007/JHEP03(2023)163",
    journal = "JHEP",
    volume = "03",
    pages = "163",
    year = "2023"
}

@article{Dutta:2024hqq,
    author = "Dutta, Bhaskar and Ghosh, Sumit and Kelly, Kevin J. and Li, Tianjun and Thompson, Adrian and Verma, Ankur",
    title = "{Non-standard neutrino interactions mediated by a light scalar at DUNE}",
    eprint = "2401.02107",
    archivePrefix = "arXiv",
    primaryClass = "hep-ph",
    doi = "10.1007/JHEP07(2024)213",
    journal = "JHEP",
    volume = "07",
    pages = "213",
    year = "2024"
}

@article{Muong-2:2021ojo,
    author = "Abi, B. and others",
    collaboration = "Muon g-2",
    title = "{Measurement of the Positive Muon Anomalous Magnetic Moment to 0.46 ppm}",
    eprint = "2104.03281",
    archivePrefix = "arXiv",
    primaryClass = "hep-ex",
    reportNumber = "FERMILAB-PUB-21-132-E",
    doi = "10.1103/PhysRevLett.126.141801",
    journal = "Phys. Rev. Lett.",
    volume = "126",
    number = "14",
    pages = "141801",
    year = "2021"
}

@article{Muong-2:2021vma,
    author = "Albahri, T. and others",
    collaboration = "Muon g-2",
    title = "{Measurement of the anomalous precession frequency of the muon in the Fermilab Muon $g−2$ Experiment}",
    eprint = "2104.03247",
    archivePrefix = "arXiv",
    primaryClass = "hep-ex",
    reportNumber = "FERMILAB-PUB-21-183-E",
    doi = "10.1103/PhysRevD.103.072002",
    journal = "Phys. Rev. D",
    volume = "103",
    number = "7",
    pages = "072002",
    year = "2021"
}

@article{Dutta:2023fnl,
    author = "Dutta, Bhaskar and Karthikeyan, Aparajitha and Kim, Doojin",
    title = "{Longer-lived mediators from charged mesons and photons at neutrino experiments}",
    eprint = "2308.01491",
    archivePrefix = "arXiv",
    primaryClass = "hep-ph",
    reportNumber = "MI-HET-809",
    doi = "10.1103/PhysRevD.109.075029",
    journal = "Phys. Rev. D",
    volume = "109",
    number = "7",
    pages = "075029",
    year = "2024"
}

@article{Cesarotti:2023udo,
    author = "Cesarotti, Cari and Kahn, Yonatan and Krnjaic, Gordan and Rocha, Duncan and Spitz, Joshua",
    title = "{New {\ensuremath{\mu}} forces from {\ensuremath{\nu}}{\ensuremath{\mu}} sources}",
    eprint = "2311.10829",
    archivePrefix = "arXiv",
    primaryClass = "hep-ph",
    reportNumber = "FERMILAB-PUB-23-539-T, MIT-CTP/5649",
    doi = "10.1103/PhysRevD.110.055032",
    journal = "Phys. Rev. D",
    volume = "110",
    number = "5",
    pages = "055032",
    year = "2024"
}

@article{Blinov:2024gcw,
    author = "Blinov, Nikita and Gori, Stefania and Hamer, Nick",
    title = "{Diphoton signals of muon-philic scalars at DarkQuest}",
    eprint = "2405.17651",
    archivePrefix = "arXiv",
    primaryClass = "hep-ph",
    doi = "10.1103/PhysRevD.110.075006",
    journal = "Phys. Rev. D",
    volume = "110",
    number = "7",
    pages = "075006",
    year = "2024"
}

@article{Planck:2015fie,
    author = "Ade, P. A. R. and others",
    collaboration = "Planck",
    title = "{Planck 2015 results. XIII. Cosmological parameters}",
    eprint = "1502.01589",
    archivePrefix = "arXiv",
    primaryClass = "astro-ph.CO",
    doi = "10.1051/0004-6361/201525830",
    journal = "Astron. Astrophys.",
    volume = "594",
    pages = "A13",
    year = "2016"
}

@article{Slatyer:2009yq,
    author = "Slatyer, Tracy R. and Padmanabhan, Nikhil and Finkbeiner, Douglas P.",
    title = "{CMB Constraints on WIMP Annihilation: Energy Absorption During the Recombination Epoch}",
    eprint = "0906.1197",
    archivePrefix = "arXiv",
    primaryClass = "astro-ph.CO",
    doi = "10.1103/PhysRevD.80.043526",
    journal = "Phys. Rev. D",
    volume = "80",
    pages = "043526",
    year = "2009"
}

@article{deNiverville:2011it,
    author = "deNiverville, Patrick and Pospelov, Maxim and Ritz, Adam",
    title = "{Observing a light dark matter beam with neutrino experiments}",
    eprint = "1107.4580",
    archivePrefix = "arXiv",
    primaryClass = "hep-ph",
    doi = "10.1103/PhysRevD.84.075020",
    journal = "Phys. Rev. D",
    volume = "84",
    pages = "075020",
    year = "2011"
}

@article{deNiverville:2015mwa,
    author = "deNiverville, Patrick and Pospelov, Maxim and Ritz, Adam",
    title = "{Light new physics in coherent neutrino-nucleus scattering experiments}",
    eprint = "1505.07805",
    archivePrefix = "arXiv",
    primaryClass = "hep-ph",
    doi = "10.1103/PhysRevD.92.095005",
    journal = "Phys. Rev. D",
    volume = "92",
    number = "9",
    pages = "095005",
    year = "2015"
}

@article{Batell:2009di,
    author = "Batell, Brian and Pospelov, Maxim and Ritz, Adam",
    title = "{Exploring Portals to a Hidden Sector Through Fixed Targets}",
    eprint = "0906.5614",
    archivePrefix = "arXiv",
    primaryClass = "hep-ph",
    doi = "10.1103/PhysRevD.80.095024",
    journal = "Phys. Rev. D",
    volume = "80",
    pages = "095024",
    year = "2009"
}

@article{deNiverville:2012ij,
    author = "deNiverville, Patrick and McKeen, David and Ritz, Adam",
    title = "{Signatures of sub-GeV dark matter beams at neutrino experiments}",
    eprint = "1205.3499",
    archivePrefix = "arXiv",
    primaryClass = "hep-ph",
    doi = "10.1103/PhysRevD.86.035022",
    journal = "Phys. Rev. D",
    volume = "86",
    pages = "035022",
    year = "2012"
}

@article{Kahn:2014sra,
    author = "Kahn, Yonatan and Krnjaic, Gordan and Thaler, Jesse and Toups, Matthew",
    title = "{DAE{\ensuremath{\delta}}ALUS and dark matter detection}",
    eprint = "1411.1055",
    archivePrefix = "arXiv",
    primaryClass = "hep-ph",
    reportNumber = "MIT-CTP-4591, FERMILAB-PUB-14-608-ND",
    doi = "10.1103/PhysRevD.91.055006",
    journal = "Phys. Rev. D",
    volume = "91",
    number = "5",
    pages = "055006",
    year = "2015"
}

@article{Kumar:2013iva,
    author = "Kumar, Jason and Marfatia, Danny",
    title = "{Matrix element analyses of dark matter scattering and annihilation}",
    eprint = "1305.1611",
    archivePrefix = "arXiv",
    primaryClass = "hep-ph",
    doi = "10.1103/PhysRevD.88.014035",
    journal = "Phys. Rev. D",
    volume = "88",
    number = "1",
    pages = "014035",
    year = "2013"
}

@article{Dutta:2018qei,
    author = "Dutta, Bhaskar and Ghosh, Sumit and Gogoladze, Ilia and Li, Tianjun",
    title = "{Three-loop neutrino masses via new massive gauge bosons from $E_6$ GUT}",
    eprint = "1805.01866",
    archivePrefix = "arXiv",
    primaryClass = "hep-ph",
    reportNumber = "MI-TH-1882",
    doi = "10.1103/PhysRevD.98.055028",
    journal = "Phys. Rev. D",
    volume = "98",
    number = "5",
    pages = "055028",
    year = "2018"
}

@article{Chauhan:2023faf,
    author = "Chauhan, Garv and Dev, P. S. Bhupal and Dubovyk, Ievgen and Dziewit, Bartosz and Flieger, Wojciech and Grzanka, Krzysztof and Gluza, Janusz and Karmakar, Biswajit and Zi{\k{e}}ba, Szymon",
    title = "{Phenomenology of lepton masses and mixing with discrete flavor symmetries}",
    eprint = "2310.20681",
    archivePrefix = "arXiv",
    primaryClass = "hep-ph",
    doi = "10.1016/j.ppnp.2024.104126",
    journal = "Prog. Part. Nucl. Phys.",
    volume = "138",
    pages = "104126",
    year = "2024"
}

@article{Dev:2021axj,
    author = "Dev, P. S. Bhupal and Dutta, Bhaskar and Ghosh, Tathagata and Han, Tao and Qin, Han and Zhang, Yongchao",
    title = "{Leptonic scalars and collider signatures in a UV-complete model}",
    eprint = "2109.04490",
    archivePrefix = "arXiv",
    primaryClass = "hep-ph",
    reportNumber = "PITT-PACC-2108, MI-TH-2112, HRI-RECAPP-2021-007",
    doi = "10.1007/JHEP03(2022)068",
    journal = "JHEP",
    volume = "03",
    pages = "068",
    year = "2022"
}

@article{Dev:2025czz,
    author = "Dev, P. S. Bhupal and Dutta, Bhaskar and Karthikeyan, Aparajitha and Maitra, Writasree and Strigari, Louis E. and Verma, Ankur",
    title = "{`Dark' Matter Effect as a Novel Solution to the KM3-230213A Puzzle}",
    eprint = "2505.22754",
    archivePrefix = "arXiv",
    primaryClass = "hep-ph",
    reportNumber = "MI-HET-859",
    month = "5",
    year = "2025"
}

@article{DELPHI:2008uka,
    author = "Abdallah, J. and others",
    collaboration = "DELPHI",
    title = "{Search for one large extra dimension with the DELPHI detector at LEP}",
    eprint = "0901.4486",
    archivePrefix = "arXiv",
    primaryClass = "hep-ex",
    reportNumber = "CERN-PH-EP-2008-013",
    doi = "10.1140/epjc/s10052-009-0874-9",
    journal = "Eur. Phys. J. C",
    volume = "60",
    pages = "17--23",
    year = "2009"
}

@article{BaBar:2017tiz,
    author = "Lees, J. P. and others",
    collaboration = "BaBar",
    title = "{Search for Invisible Decays of a Dark Photon Produced in ${e}^{+}{e}^{-}$ Collisions at BaBar}",
    eprint = "1702.03327",
    archivePrefix = "arXiv",
    primaryClass = "hep-ex",
    reportNumber = "BABAR-PUB-17-001, SLAC-PUB-16923",
    doi = "10.1103/PhysRevLett.119.131804",
    journal = "Phys. Rev. Lett.",
    volume = "119",
    number = "13",
    pages = "131804",
    year = "2017"
}

@article{NA64:2020qwq,
    author = "Banerjee, D. and others",
    collaboration = "NA64",
    title = "{Search for Axionlike and Scalar Particles with the NA64 Experiment}",
    eprint = "2005.02710",
    archivePrefix = "arXiv",
    primaryClass = "hep-ex",
    reportNumber = "CERN-EP-2020-068",
    doi = "10.1103/PhysRevLett.125.081801",
    journal = "Phys. Rev. Lett.",
    volume = "125",
    number = "8",
    pages = "081801",
    year = "2020"
}

@article{Darme:2020sjf,
    author = "Darm{\'e}, Luc and Giacchino, Federica and Nardi, Enrico and Raggi, Mauro",
    title = "{Invisible decays of axion-like particles: constraints and prospects}",
    eprint = "2012.07894",
    archivePrefix = "arXiv",
    primaryClass = "hep-ph",
    doi = "10.1007/JHEP06(2021)009",
    journal = "JHEP",
    volume = "06",
    pages = "009",
    year = "2021"
}

@article{PIENU:2021clt,
    author = "Aguilar-Arevalo, A. and others",
    collaboration = "PIENU",
    title = "{Search for three body pion decays ${\pi}^+{\to}l^+{\nu}X$}",
    eprint = "2101.07381",
    archivePrefix = "arXiv",
    primaryClass = "hep-ex",
    doi = "10.1103/PhysRevD.103.052006",
    journal = "Phys. Rev. D",
    volume = "103",
    number = "5",
    pages = "052006",
    year = "2021"
}

@article{NA62:2021bji,
    author = "Cortina Gil, Eduardo and others",
    collaboration = "NA62",
    title = "{Search for $K^+$ decays to a muon and invisible particles}",
    eprint = "2101.12304",
    archivePrefix = "arXiv",
    primaryClass = "hep-ex",
    reportNumber = "CERN-EP-2021-018",
    doi = "10.1016/j.physletb.2021.136259",
    journal = "Phys. Lett. B",
    volume = "816",
    pages = "136259",
    year = "2021"
}

@article{NA64:2023wbi,
    author = "Andreev, Yu. M. and others",
    collaboration = "NA64",
    title = "{Search for Light Dark Matter with NA64 at CERN}",
    eprint = "2307.02404",
    archivePrefix = "arXiv",
    primaryClass = "hep-ex",
    reportNumber = "CERN-EP-2023-130",
    doi = "10.1103/PhysRevLett.131.161801",
    journal = "Phys. Rev. Lett.",
    volume = "131",
    number = "16",
    pages = "161801",
    year = "2023"
}

@article{Dutta:2020vop,
    author = "Dutta, Bhaskar and Kim, Doojin and Liao, Shu and Park, Jong-Chul and Shin, Seodong and Strigari, Louis E. and Thompson, Adrian",
    title = "{Searching for dark matter signals in timing spectra at neutrino experiments}",
    eprint = "2006.09386",
    archivePrefix = "arXiv",
    primaryClass = "hep-ph",
    reportNumber = "MI-TH-2014",
    doi = "10.1007/JHEP01(2022)144",
    journal = "JHEP",
    volume = "01",
    pages = "144",
    year = "2022"
}

@misc{RKHorn2024,
  author       = {Adrian Thompson},
  title        = {{athompson-git/RKHorn: RKHorn-alpha}},
  note = {{Available at \href{https://github.com/athompson-git/RKHorn}{https://github.com/athompson-git/RKHorn}. Archived as [\href{https://doi.org/10.5281/zenodo.14219233}{DOI:10.5281/zenodo.14219233}].}},
  year         = {2024},
  publisher    = {Zenodo},
  doi          = {10.5281/zenodo.14219233},
  url          = {https://doi.org/10.5281/zenodo.14219233}
}

@article{Allison:2016lfl,
    author = "Allison, J. and others",
    title = "{Recent developments in Geant4}",
    reportNumber = "FERMILAB-PUB-16-447-CD",
    doi = "10.1016/j.nima.2016.06.125",
    journal = "Nucl. Instrum. Meth. A",
    volume = "835",
    pages = "186--225",
    year = "2016"
}

@article{GEANT4:2002zbu,
    author = "Agostinelli, S. and others",
    collaboration = "GEANT4",
    title = "{GEANT4--a simulation toolkit}",
    reportNumber = "SLAC-PUB-9350, FERMILAB-PUB-03-339, CERN-IT-2002-003",
    doi = "10.1016/S0168-9002(03)01368-8",
    journal = "Nucl. Instrum. Meth. A",
    volume = "506",
    pages = "250--303",
    year = "2003"
}

@article{Allison:2006ve,
    author = "Allison, John and others",
    title = "{Geant4 developments and applications}",
    reportNumber = "SLAC-PUB-11870",
    doi = "10.1109/TNS.2006.869826",
    journal = "IEEE Trans. Nucl. Sci.",
    volume = "53",
    pages = "270",
    year = "2006"
}

@article{Kim:1979if,
    author = "Kim, Jihn E.",
    title = "{Weak Interaction Singlet and Strong CP Invariance}",
    reportNumber = "UPR-0120T",
    doi = "10.1103/PhysRevLett.43.103",
    journal = "Phys. Rev. Lett.",
    volume = "43",
    pages = "103",
    year = "1979"
}

@misc{MicroBooNE:2025ntu,
    author          = "Abratenko, P. and others",
    collaboration   = "MicroBooNE",
    title           = "{Inclusive Search for Anomalous Single-Photon Production in MicroBooNE}",
    eprint          = "2502.06064",
    archivePrefix   = "arXiv",
    primaryClass    = "hep-ex",
    reportNumber    = "FERMILAB-PUB-25-0055-PPD",
    year            = "2025",
    note            = "arXiv:2502.06064 [hep-ex]"
}

@article{Aliberti:2025beg,
    author = "Aliberti, R. and others",
    title = "{The anomalous magnetic moment of the muon in the Standard Model: an update}",
    eprint = "2505.21476",
    archivePrefix = "arXiv",
    primaryClass = "hep-ph",
    reportNumber = "CERN-TH-2025-101, FERMILAB-PUB-25-0344-T, INT-PUB-25-015, IPARCOS-UCM-25-029, KEK Preprint 2025-22, LTH 1403, MITP-25-037, UWThPh 2025-15, UWThPh
  2025-15, ZU-TH 37/25, IPARCOS-UCM-25-029",
    month = "5",
    year = "2025"
}

@article{Muong-2:2025xyk,
    author = "Aguillard, D. P. and others",
    collaboration = "Muon g-2",
    title = "{Measurement of the Positive Muon Anomalous Magnetic Moment to 127 ppb}",
    eprint = "2506.03069",
    archivePrefix = "arXiv",
    primaryClass = "hep-ex",
    reportNumber = "FERMILAB-PUB-25-0364-PPD",
    month = "6",
    year = "2025"
}

@online{numirookiebook,
  author = {Bruce Baller},
  title = {NuMI Rookie Book},
  url = {https://operations.fnal.gov/rookie_books/NuMI_v1.pdf},
  urldate = {2005-04-26}
}

@misc{kevinwood,
  note = {Private communication with Kevin Wood}
}

@article{AristizabalSierra:2020rom,
    author = "Aristizabal Sierra, D. and De Romeri, V. and Flores, L. J. and Papoulias, D. K.",
    title = "{Axionlike particles searches in reactor experiments}",
    eprint = "2010.15712",
    archivePrefix = "arXiv",
    primaryClass = "hep-ph",
    doi = "10.1007/JHEP03(2021)294",
    journal = "JHEP",
    volume = "03",
    pages = "294",
    year = "2021"
}

@article{Nardi:2018cxi,
    author = "Nardi, Enrico and Carvajal, Cristian D. R. and Ghoshal, Anish and Meloni, Davide and Raggi, Mauro",
    title = "{Resonant production of dark photons in positron beam dump experiments}",
    eprint = "1802.04756",
    archivePrefix = "arXiv",
    primaryClass = "hep-ph",
    doi = "10.1103/PhysRevD.97.095004",
    journal = "Phys. Rev. D",
    volume = "97",
    number = "9",
    pages = "095004",
    year = "2018"
}

@article{Dutta:2024nhg,
    author = "Dutta, Bhaskar and Karthikeyan, Aparajitha and Rai, Mudit and Kim, Hyunyong",
    title = "{Dark Matter Internal Pair Production: A Novel Direct Detection Mechanism}",
    eprint = "2410.07624",
    archivePrefix = "arXiv",
    primaryClass = "hep-ph",
    reportNumber = "MI-HET-84",
    doi = "10.1103/ry3x-dw48",
    journal = "Phys. Rev. Lett.",
    volume = "135",
    number = "1",
    pages = "011804",
    year = "2025"
}

@article{PhysRev.81.899,
  title = {Photo-Production of Neutral Mesons in Nuclear Electric Fields and the Mean Life of the Neutral Meson},
  author = {Primakoff, H.},
  journal = {Phys. Rev.},
  volume = {81},
  issue = {5},
  pages = {899--899},
  numpages = {0},
  year = {1951},
  month = {Mar},
  publisher = {American Physical Society},
  doi = {10.1103/PhysRev.81.899},
  url = {https://link.aps.org/doi/10.1103/PhysRev.81.899}
}

@article{Dev:2024ygx,
    author = "Dev, P. S. Bhupal and Kim, Doojin and Sathyan, Deepak and Sinha, Kuver and Zhang, Yongchao",
    title = "{New Laboratory Constraints on Neutrinophilic Mediators}",
    eprint = "2407.12738",
    archivePrefix = "arXiv",
    primaryClass = "hep-ph",
    reportNumber = "CETUP-2024-005",
    month = "7",
    year = "2024"
}

@article{Dev:2025tdv,
    author = "Dev, P. S. Bhupal and Kim, Doojin and Sathyan, Deepak and Sinha, Kuver and Zhang, Yongchao",
    title = "{New Constraints on Neutrino-Dark Matter Interactions: A Comprehensive Analysis}",
    eprint = "2507.01000",
    archivePrefix = "arXiv",
    primaryClass = "hep-ph",
    reportNumber = "CETUP-2023-022",
    month = "7",
    year = "2025"
}

@article{Capdevilla:2021kcf,
    author = "Capdevilla, Rodolfo and Curtin, David and Kahn, Yonatan and Krnjaic, Gordan",
    title = "{Systematically testing singlet models for $(g-2)_{\mu}$}",
    eprint = "2112.08377",
    archivePrefix = "arXiv",
    primaryClass = "hep-ph",
    reportNumber = "FERMILAB-PUB-21-737-T",
    doi = "10.1007/JHEP04(2022)129",
    journal = "JHEP",
    volume = "04",
    pages = "129",
    year = "2022"
}

\end{document}